%% file: Born21UA.tex
\documentclass[aps,prc,showpacs,nofootinbib]{revtex4-1}\usepackage{amsmath}
\usepackage{amssymb}
\usepackage{epsfig}
\usepackage{graphicx} 
\usepackage{subfigure}

\newcommand\nn{\nonumber} 
\newcommand\ba{\begin{eqnarray}}
\newcommand\ea{\end{eqnarray}} 
\newcommand\alb{\begin{align}}
\newcommand\ale{\end{align}} 
\newcommand\be{\begin{equation}}
\newcommand\ee{\end{equation}}

\begin{document}

%%%%%%%%%%%%%%%%%%%%%%%%%%%%%%%%%%%%
\title{Contribution of the non-resonant mechanism to the double and single differential distributions over the invariant variables in the reaction $e^+ + e^- \to N + \bar N +\pi^0.$ }

%%%%%%%%%%%%%%%%%%%%%%%%%%%%%%%%%%%%

\author{G. I. Gakh}
\email{gakh@kipt.kharkov.ua}
\affiliation{\it National Science Centre, Kharkov Institute of
Physics and Technology, Akademicheskaya 1, and V. N. Karazin
Kharkov National University, Dept. of
Physics and Technology, 31 Kurchatov, 61108 Kharkov, Ukraine}
\author{M.I. Konchatnij}
 \email{konchatnij@kipt.kharkov.ua}
%% \cortext[cor2]{}
\affiliation{\it National Science Centre, Kharkov Institute of
Physics and Technology, Akademicheskaya 1, and V. N. Karazin
Kharkov National University, Dept. of
Physics and Technology, 31 Kurchatov, 61108 Kharkov, Ukraine}
\author{N.P. Merenkov}
\email{merenkov@@kipt.kharkov.ua}
\affiliation{\it National Science Centre, Kharkov Institute of
Physics and Technology, Akademicheskaya 1, and V. N. Karazin
Kharkov
National University, Dept. of
Physics and Technology, 31 Kurchatov, 61108 Kharkov, Ukraine}

\author{E. Tomasi--Gustafsson}
\email{egle.tomasi@cea.fr}
\affiliation{\it IRFU, CEA, Universit\'e Paris-Saclay, 91191
Gif-sur-Yvette, France}

\begin{abstract}
The general analysis of the differential cross section and various polarization observables is performed
for the process $e^+ + e^- \to N + \bar N +\pi^0 $ assuming that the annihilation occurs through the exchange of one virtual photon. The dependence of the differential distributions over invariant variables is derived for the reaction $e^+ + e^- \to N + \bar N +\pi^0 $ in the so-called non-resonant mechanism, applying the conservation of the hadron electromagnetic currents and the P-invariance of the hadron electromagnetic interaction. The detection in an exclusive experimental set up where the nucleon (or antinucleon) and pion are detected in coincidence is considered. A number of single and double differential distributions have been calculated analytically and numerical estimates are given for the $p\bar p\pi^0$ and $n\bar n\pi^0$ channels, in the Born (non-resonant) approximation, in the energy range from threshold up to $s=16$ GeV$^2$.
\end{abstract}

\maketitle

\section{Introduction}

The aim of  hadronic physics is to understand hadrons as composite systems of 
strongly interacting quarks and gluons. The existing theory of the strong
interactions, QCD, does not apply in the GeV energy region. The charge and magnetic distributions that are fundamental characteristics of the hadrons,  are formalized in terms of elastic and transition electromagnetic form factors. These form factors characterize the internal structure of the hadrons and they are
the analytical functions of one kinematical variable $q^2$, the square of the four-momentum of the virtual photon. This variable can be negative (in the scattering type
experiments, the space-like region) or positive (in the annihilation type
experiments, the time-like region). In the time-like region these form factors are
complex functions, whereas in the space-like region they are real ones. The study of
the hadron form factors in both kinematical regions is important to get complementary
information about the hadron structure.

The importance of a good experimental knowledge of the nucleon electromagnetic form
factors in a wide $q^2$ range is quite clear since QCD predictions from
non-perturbative (low $q^2$ values) to perturbative (high $q^2$ values) regime can
then be tested according to their capability to reproduce the form factor
measurements for any $q^2$ value.,Let us note that any model where the interaction is based on the valence quarks can hardly foresee a
neutron magnetic form factor bigger than the proton one. In this connection, one can  mention that earlier
predictions based on the vector meson dominance \cite{Iachello:1972nu,Bijker:2004yu} and Skyrme \cite{Holzwarth:1996xq} based models give
$|G^{n}_{M}|\geq |G^p_{M}|$.

In the space-like region the charge and magnetic form factors were determined separately, both for the proton and neutron, in a wide $q^2$ range (see, for example, review \cite{Pacetti:2015iqa} and references therein).  In the time-like region, the annihilation cross section for $e^+ + e^- \to p + \bar p $ was measured at Novosibirsk in the threshold region \cite{Akhmetshin:2015ifg,CMD-3:2018kql}, by the BaBar collaboration at SLAC  \cite{Lees:2013xe, Lees:2013uta} and by the BESIII collaboration at Beijing in several works, using initial state radiation \cite{BESIII:2015axk, BESIII:2021rqk} and beam scan method \cite{BESIII:2019hdp}  providing  the first separation between electric and magnetic form factors. Precise data on the neutron  effective form factors  have been also published \cite{BESIII:2021tbq}.  Unexpected features  where highlighted, that deserve more accurate investigations. Among them: 
i) in the explored energy range ($q^2< 6$ GeV$^2$), the inequality $\,G^{n}_{M} > G^p_{M}$ takes place; ii)  the differential cross section measurement
suggests that the $G_E^{n}$ and $G_E^{p}$ are comparable, at threshold; 
iii) the
steep decrease of the proton form factor near the threshold, and the presence of a dip in the
$e^+e^-\to $ hadrons in the same region, suggest the presence of a narrow resonance just below
the threshold (this state is consistent with an $N\bar N$ bound state, the so-called baryonium) \cite{Wycech:2015qra};
iv)  the cross section and effective proton form factor show oscillating behavior \cite{Lees:2013uta,Bianconi:2015owa} (BaBar collaboration) confirmed by the BESIII data. The neutron data from BESIII confirm that similar oscillations exist, but are shifted by a phase Ref. \cite{BESIII:2021tbq}.
A review of the form factor data data collected by the BESIII collaboration can be found in Ref. \cite{Huang:2021xte}.

In this work we focus of the 'inelastic' annihilation reaction $e^+ + e^- \to p + \bar p +\pi $ that is related by crossing symmetry to the reactions i)-$\pi+N\to e^++e^-+N$  containing information on time-like from factors as suggested in Ref. \cite{Rekalo:1965}, and investigated by the HADES collaboration \cite{HADES:2020kwd}; ii)- $\bar N+N\to e^++e^-+\pi$  that allows to determine FFs in the physical and even unphysical  region, see (\cite{Adamuscin:2007iv} and References therein) and will be investigated in PANDA, FAIR; iii)- as well as $e^-+N\to e^-+N+\pi $. Current measurements at electron-beam facilities accumulated a considerable amount of
precise data of the meson photo- and electro-production reactions on the nucleon target, opening the 
opportunity to make quantitative study of the $N^*$ structure \cite{Burkert:2004sk} and extract electromagnetic transition form factors.
Recent experiments have suggested new $N^*$ states which
strongly couple to various reaction channels but not to the dominant $\pi N$ channels \cite{McNabb:2003nf,Sumihama:2005er}.
The study of the $e^-+N\to e^-+N+\pi $ reaction is a privileged channel to investigate $N^*$ spectroscopy. The description of the
experimental results in the framework of  constituent quark models is not satisfactory in terms of 'missing resonances'. For example, the prediction of a substantial number of  N$^*$ resonances around 2 $GeV/c^2$, has not been confirmed so far \cite{Capstick:2000qj}. The reason may be the weak coupling of these $N^*$ states to $\pi N$ and $\gamma N$ states, stressing the need to investigate other reactions.

The study of the different transition form factors in the
time-like region is very important. The measurement of the $e^+e^-$ reactions allows one to study also the excited hyperon states, such as
$\Lambda^*, \Sigma^*$ and $\Xi^*$ \cite{Zou:2000wg,Zou:2007mub,Asner:2009zz}. The corresponding  experiments at the Beijing Electron-Positron Collider (BEPC) \cite{Li:1999uwc,Zou:2000nu} started already about 20 years ago. Up to now, the N$^*$ production from $e^+e^-$  annihilations has been
studied only around the charmonium region. The experimental results on N$^*$ from $e^+e^-$ annihilations and their phenomenological
implications can be found in the review \cite{Zou:2018lse}.

BES/BESII/BESIII Collaborations have published their results on N$^*$ production from the decays of the charmonium states \cite{Zou:2018lse}.
Some interesting results on the N$^*$s production have been obtained. The N(1440) peak was observed for the first time directly
from $\pi$N invariant mass spectrum (due to the absence of the strong $\Delta$ peak). Besides several well known N$^*$ resonances
around 1520 MeV and 1670 MeV, three new N$^*$ resonances above 2 GeV were found using partial wave analyses. The measurement of the
$\psi(2S) \to \bar pp\pi^0$ channel (by CLEO Collaboration) found a similar strong N$^*$(1440) peak \cite{Alexander:2010vd}. There is no obvious
N$^*$(1440) peak for $e^+e^- \to \bar pp\pi^0$ in the vicinity of the $\psi$(3770)\cite{Ablikim:2014kxa}.

The time-like region became accessible with the advent of high-precision, high-intensity $e^+e^-$ colliders at intermediate energies.
New data from BESIII, collected in a high-precision energy scan in 2015, will offer improved precision over a large q$^2$ range.
The coming upgrade of the BEPCII collider up to c.m.s. energies of 4.9 GeV will allow to study more details of the $N^*$ production.
The topics which planned to study at BESIII in the near future can be found in \cite{Yuan:2019zfo}.

The process $e^+e^-\to p\bar p \pi^0 $ has been investigated at the BEPCII collider (China) at the vicinity of the
$\psi (3770)$ resonance \cite{Ablikim:2014kxa}. The cross section of the decay $\psi (3770)\to p\bar p \pi^0 $ is measured
taking into account the interference between the continuum and resonant production amplitudes. The continuum cross
section was described by a function $C/s^{\lambda}$ with unknown exponent $\lambda$. The aim of the experiment was
to determine the width of the decay $\psi (3770)\to p\bar p \pi^0 $ since these data are required for the preparing
$\bar P$ANDA experiment \cite{Lundborg:2005am} in which planned to investigate, in particular, the charmonium and charmonium
hybrid states \cite{Lutz:2009ff}. Later \cite{Ablikim:2017gtb}, the BESIII Collaboration has measured this reaction in the vicinity of the
$Y(4260)$ resonance, more precisely in the energy range $\sqrt{s}=4.008 - 4.600$ GeV. No resonant structure is observed
in the shape of the cross section $e^+e^-\to p\bar p \pi^0 $.

In this paper, we open the series of works devoted to the general analysis of the differential cross section and polarization observables
in the process $e^+ + e^- \to N + \bar N +\pi^0 $, where $N\,(\bar N)$ is proton\,\,(antiproton) or neutron\,\,(antineutron) in the one-photon-annihilation approximation. We intend to account for the continuum (non-resonant) and resonance
(with different possible vector mesons  or excited baryons in intermediate virtual states of Feynman diagrams) contributions and concentrate on invariant variables distributions.
In this part of our work we consider the general analysis and investigate in details the non-resonant contribution.

The paper is organized as follows. The general structure of the hadronic tensor for the case of unpolarized final hadrons
and polarized nucleon is given in Sec. IIA. The invariant amplitudes of the process $e^+ + e^- \to N + \bar N +\pi^0 $ are
introduced in Sec. IIB. Sec. IIC contains the description of the nucleon polarization 4-vector in terms of the
4-momenta of the final particles. Sec. III contains the discussion of the kinematics.  The model for the $e^+ + e^- \to N + \bar N +\pi^0 $
reaction mechanism is given in Sec. IV. The discussion of the obtained results is given in Sec. V. Conclusions are set  in Sec. VI.

\section{Formalism}

The reaction
\begin{equation}\label{eq:1}
e^- (k_1) + e^+(k_2) \to N(p_1)+\bar N(p_2)+\pi^0(k),
\end{equation}
in the one-photon-annihilation approximation for the non-resonant mechanism, is described by two Feynman diagrams of Fig.~1. The notation of
the particle four-momenta is indicated in the parenthesis. Here $q=k_1+k_2=p_1+p_2+k$ is the  four-momentum of
the virtual photon and $k^2=m^2, \, p_1^2=p_2^2=M^2$, $m (M)$ is the pion (nucleon) mass. Further, we neglect the electron mass
where it is possible.

The matrix element, in this approximation, can be written as a contraction of the leptonic $(e l_{\mu})$ and hadronic $(e J_{\mu})$ currents
\begin{equation}\label{eq:2}
\mathcal{M}=\frac{e^2}{q^2}l^{\mu}J_{\mu}, \ \ l^{\mu}=\bar v(k_2)\gamma^{\mu}u(k_1).
\end{equation}

Then, the square of the matrix element is 
\begin{equation}\label{eq:3}
|\mathcal{M}|^2=\frac{16\pi^2\alpha^2}{q^4}L^{\mu\nu}H_{\mu\nu}, \ \
L^{\mu\nu}=l^{\mu}l^{\nu*}, \ \  H_{\mu\nu}=J_{\mu}J_{\nu}^*.
\end{equation}

\begin{figure}
\centering
\includegraphics[width=0.4\textwidth]{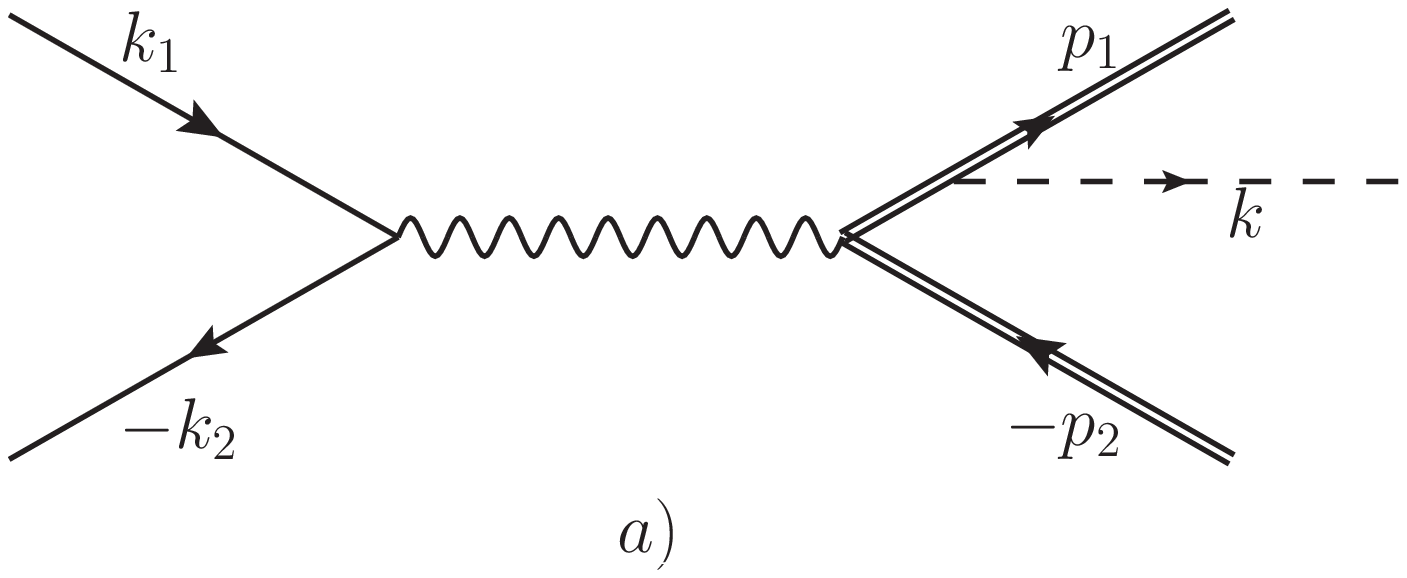}
\includegraphics[width=0.4\textwidth]{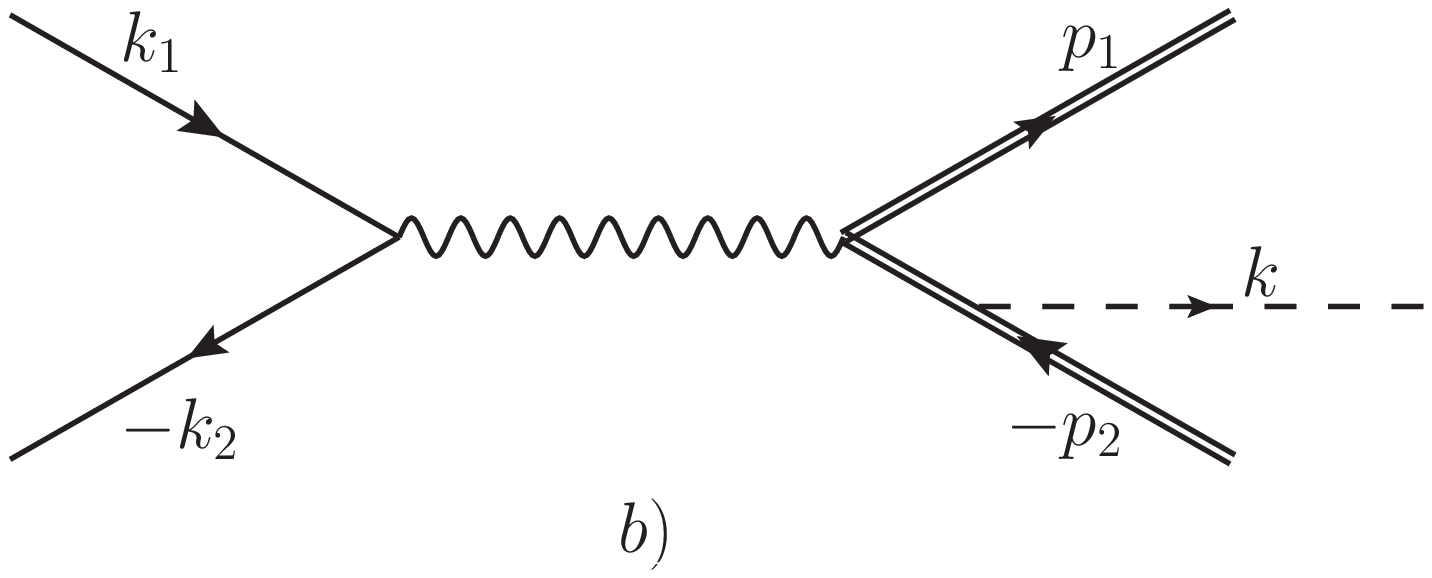}
%\hspace{1cm}
%\includegraphics[width=0.3\textwidth]{gena2019c.eps}
 \parbox[t]{0.9\textwidth}{\caption{The simplest Feynman diagrams which describe the continuum (non-resonant) contribution to the process (\ref{eq:1}); (a)$-$ with intermediate nucleon,
 (b)$-$ with intermediate antinucleon.}\label{fig.1}}
\end{figure}

\subsection{The structure of the hadronic tensor}

The hadronic tensor $H_{\mu\nu}$ has the following general form for the case when the polarizations of the
final particles are not measured
\begin{equation}\label{eq:4}
H_{\mu\nu}(0)=H_1\tilde g_{\mu\nu}+H_2\tilde k_{\mu}\tilde k_{\nu}+H_3\tilde
p_{\mu}\tilde p_{\nu}+H_4(\tilde p_{\mu}\tilde k_{\nu}+\tilde p_{\nu}\tilde
k_{\mu})+iH_5(\tilde p_{\mu}\tilde k_{\nu}-\tilde p_{\nu}\tilde
k_{\mu}),
\end{equation}
where $\tilde g_{\mu\nu}=g_{\mu\nu}-q_{\mu}q_{\nu}/q^2$, $\tilde k_{\mu}=k_{\mu}-(k\cdot q/q^2)q_{\mu}$
and $\tilde p_{\mu}=p_{\mu}-(p_\cdot q/q^2)q_{\mu},\, p=p_1-p_2$. $H_i$ (i=1-5) are the so-called
structure functions depending on three invariant variables $s_1,\,s_2$ and $s\equiv q^2$ (see below).

The leptonic tensor $L_{\mu\nu}$ has the following form in the case when electron beam is polarized
\begin{equation}\label{eq:5}
L_{\mu\nu}=-q^2g_{\mu\nu}+2(k_{1\mu}k_{2\nu}+k_{1\nu}k_{2\mu})+2im_e (\mu\nu\eta q),
\end{equation}
where $(\mu\nu ab)=\epsilon_{\mu\nu\varrho\sigma}a^{\varrho}b^{\sigma}$ and $\eta_{\mu}$ is the
spin four-vector of the electron (we chose $\epsilon^{0123}=-~\epsilon_{0123}=+1$), $m_e$ is the
electron mass.

At chosen normalization, the differential cross section of the process (\ref{eq:1}), in terms of the leptonic and hadronic tensors,
has the following form (further we use $\eta^\mu = k_1^\mu/m_e$ and neglect the electron mass for the initial particles electromagnetic current)
\begin{equation}\label{eq:6difsec}
d\,\sigma =\frac{\alpha^2}{8\pi^3\,q^6}L^{\mu\nu}\,H_{\mu\nu}\,d\,R_3, \ \ d\,R_3
=\frac{d^3p_1}{2\,E_1}\,\frac{d^3p_2}{2\,E_2}\,\frac{d^3k}{2\,E}\,\delta(k_1+k_2-p_1-p_2-k),
\end{equation}
where $E_1\,(E_2)$ is the nucleon (antinucleon) energy and $E$ is the pion one.

In the case when the nucleon polarization is measured, we can use the following form of the hadronic tensor
$$H_{\mu\nu}= \frac{1}{2}H_{\mu\nu}(0)+ T_{\mu\nu}\,,$$
where the tensor $T_{\mu\nu}$ depends on the nucleon polarization 4-vector $S^\mu$ and can be written as the
sum of the symmetrical $T^{(s)}_{\mu\nu}$ and antisymmetrical $T^{(a)}_{\mu\nu}$ parts. The symmetrical
part can be written as follows

\ba\label{eq:7}
T^{(s)}_{\mu\nu}(S)&=&Im \big\{T_{kk}\, \tilde{k}^{\mu\nu}_k + T_{kp}\, \tilde{k}^{\mu\nu}_p + T_{pk}\,
\tilde{p}^{\mu\nu}_k + T_{pp}\, \tilde {p}^{\mu\nu}_p + T_G\, \tilde{G}^{\mu\nu} +\nn\\
&&T_K\, \tilde{K}^{\mu\nu} + T_P\, \tilde{P}^{\mu\nu} + T_{KP}\, \widetilde{KP}^{\mu\nu} \big\}\,,
\ea
where
\ba
\tilde{k}^{\mu\nu}_ka&=&a\tilde{k}^{\mu}(\nu k q S)+\tilde{k}^{\nu}(\mu k q S),\,\,\tilde{k}^{\mu\nu}_p=\tilde{k}^{\mu}(\nu p q S)+\tilde{k}^{\nu}(\mu p q S)\,,\nn\\
\tilde{p}^{\mu\nu}_k &=& \tilde{p}^{\mu}(\nu k q S)+\tilde{p}^{\nu}(\mu k q S),\,\,\tilde{p}^{\mu\nu}_p=\tilde{p}^{\mu}(\nu p q S)+\tilde{p}^{\nu}(\mu p q S)\,,\nn\\
\tilde{G}^{\mu\nu}&=&\tilde{g}^{\mu\nu}(k p q S),\,\,\tilde{K}^{\mu\nu}=\tilde{k}^\mu\tilde{k}^\nu(k p q S),\,\, \tilde{P}^{\mu\nu}=\tilde{p}^\mu\tilde{p}^\nu(k p q S)\,,\nn\\
\widetilde{KP}^{\mu\nu}&=&[\tilde{k}^\mu\tilde{p}^\nu +\tilde{p}^\mu\tilde{k}^\nu](k p q S),\,\, (\mu a b c)=\epsilon_{\mu\nu\varrho\sigma}a^\nu b^{\varrho}c^{\sigma},\,\, (a b c d)=
\epsilon_{\mu\nu\varrho\sigma} a^\mu b^\nu c^{\varrho}d^{\sigma}. 
\nn
\ea

The antisymmetrical part is
\begin{equation}\label{eq:8}
T^{(a)}_{\mu\nu}(S)=i\,Re \big\{T_{s}(\mu \nu q S) + T_{pps}(p S)(\mu \nu p q) + T_{pqs}(q S)(\mu \nu p q) +  T_{kps}(p S)(\mu \nu k q) + T_{kqs}(q S)(\mu \nu k q) \big\}.
\end{equation}
Note that the form of the tensor  $T^{(s)}_{\mu\nu}$ given by Eq.\,(\ref{eq:7}) is not unique, and this point
is discussed in Appendix A.

\subsection{Invariant amplitudes}

The general form of the matrix element (\ref{eq:2}) can be chosen by analogy with the
process of the pion electroproduction on the nucleons \cite{Adler:1968tw}. If the gauge invariance and the space parity conservation take place,
we have
\begin{equation}\label{eq:9}
\mathcal{M}=\frac{e^2}{q^2}\varphi^+_{\pi}\,\Sigma_{i=1}^6\,\bar u(p_1)\,\gamma_5\,M_i\, v(p_2)\,A_i, \ \ \gamma_5=i \gamma^0 \gamma^1 \gamma^2 \gamma^3,
\end{equation}
where $\varphi_\pi$ is the pion wave function and the $M_i$ structures have the following form
\ba\label{eq:10}
M_1&=&\frac{1}{2}\gamma^{\mu}\gamma^{\nu}F_{\mu\nu}, \ \
M_2=p^{\mu}k^{\nu}F_{\mu\nu}, \ \
M_3=\gamma^{\mu}k^{\nu}F_{\mu\nu}, \ \
M_4=(\gamma^{\mu}p^{\nu}-2M\gamma^{\mu}\gamma^{\nu})F_{\mu\nu},
\\
M_5&=&q^{\mu}k^{\nu}F_{\mu\nu}, \ \
M_6=q^{\mu}\gamma^{\nu}F_{\mu\nu}, \ \
F_{\mu\nu}=l_{\mu}q_{\nu}-l_{\nu}q_{\mu}. \nn
\ea
The invariant amplitudes $A_i$ (i=1-6) are the complex functions of three independent variables: for example, $q^2-$ the square of the total invariant mass of the final hadrons, and
$s_{1,2}=(p_{1,2}+k)^2$ -the square of the invariant masses of the $N\,\pi^0 $  and $\bar{N}\,\pi^0$ systems.

Equations (\ref{eq:9}) and (\ref{eq:10}) mean that, in general case, $ J_{\mu}$ can be written as follows
\begin{equation}\label{eq:11}
J_{\mu}=\varphi^+_{\pi}\,\bar u(p_1)\,\gamma_5\,\hat O_{\mu}\,v(p_2),
\end{equation}
where the matrix $\hat O_{\mu}$ has the form
\ba\label{eq:12}
\hat O_{\mu}&=&(k\cdot q\,p_{\mu}-p\cdot q\,k_{\mu})A_2-q^2\tilde k_{\mu}A_5+(k\cdot q\,A_3+p\cdot q\,A_4-q^2\,A_6)\gamma_{\mu}+
\nn\\
&&
+(A_6\,q_{\mu}-A_4\,p_{\mu}-A_3\,k_{\mu})\hat q+(A_1-4M\,A_4)(\gamma_{\mu}\,\hat q -q_{\mu}). 
\ea
%%%%%%%%%%%%%%%%%%%%%%%%%%%
\subsection{The nucleon polarization 4-vector}
%%%%%%%%%%%%%%%%%%%%%%%%%%%%
In the rest frame of the nucleon $({\bf{p}_1}=0)$ its polarization four-vector has the form $S_r^\mu = (0, {\bf{n}}), \ \
{\bf{n}}^2=1,$  and, in general case, 3-vector ${\bf{n}}$ has three independent components: two in the
plane $({\bf{q}},\,{\bf{k}})$ and one along 3-vector $\big[{\bf{k}}\times{\bf{q}}\big].$ It means that
in arbitrary Lorentz system 4-vector $S^\mu$ can be expressed by means of 4-vectors of the particle momenta
and expanded by three independent 4-vectors: longitudinal $S^\mu_L,$ transversal $S^\mu_T$ and normal
$S^\mu_N.$

Let us choose the longitudinal polarization such that in the rest frame ${\bf{n}}=-{\bf{q}}/|{\bf{q}}|.$
It can be expressed in terms of 4-vectors $p_1^{\mu}$ and $q^\mu,$ and has the following form
\begin{equation}\label{eq:SL}
S^\mu_L=\frac{q\cdot p_1\,p_1^{\mu}-M^2\,q^\mu}{M\, K}\,, \ \ K=\sqrt{(q\cdot p_1)^2-q^2\,M^2}\,, \ \
S_L\cdot p_1=0, \ \ S_L^2 = - 1 .
\end{equation}
Note also that in c.\,m.\,s. of the process (\ref{eq:1}), where ${\bf q}=0,$
$$S^\mu_L=\bigg(\,\frac{|{\bf{p}_1}|}{M}, \, \frac{E_1\,{\bf{p}_1}}{M\,|{\bf{p}_1}|}\,\bigg).$$

The transversal polarization was chosen to be orthogonal to the longitudinal one, that is
$$S^\mu_T\cdot S^\mu_L = 0, \ \to S_T\cdot p_1 = 0, \ S_T\cdot q = 0, \ S^2_T= -1. $$
The relation $S_T\cdot q = 0$ indicates that the polarization 4-vector $S^\mu_T$ is expressed in terms of
the 4-vectors $\tilde{p}_1,\,\,\tilde{p}_2$ and $\tilde{k}.$ Only two 4-vectors are independent since we
have the following relation $\tilde{p}_1+\tilde{p}_2+\tilde{k}=0.$ Choosing $\tilde{p}_1$ and $\tilde{k}$
one can obtain
\begin{equation}\label{ST}
S^\mu_T=\frac{(q^2\,k\cdot p_1-q\cdot p_1\,k\cdot q)\,\tilde{p}_1^{\mu} + [(q\cdot p_1)^2-q^2M^2]\,\tilde{k}^\mu}{K\, N}\,,
\end{equation}
where
$$N=\sqrt{-(\mu k p_1 q)(\mu k p_1 q)}\,, \ N^2=2\,k\cdot q\,k\cdot p_1\,q\cdot p_1-q^2\,(k\cdot p_1)^2)-M^2(k\cdot q)^2-m^2(q\cdot p_1)^2+q^2 M^2 m^2\,.$$

In both coordinate system (the rest system and c.\,m.\,s.) the 4-vector $S^\mu_T$ has no time
component and its space component is
$$
\frac{[{\bf{p}_1 }\times[{\bf k}\times{\bf{p}_1}]]}{|[{\bf{p}_1 }\times[{\bf k}\times{\bf{p}_1}]]|}, \ \ 
\frac{[{\bf q}\times[{\bf k}\times{\bf{q}}]]}{|[{\bf q}\times[{\bf k}\times{\bf{q}}]]|},
$$
in and c.\,m.\,s. and rest frame, correspondingly.

It is clear that the normal polarization is
\begin{equation}\label{SN}
S^\mu_N=\frac{(\mu k p_1 q)}{N}\, = \Big(0,~\frac{[{\bf k}\times{\bf{p}_1}]}{|[{\bf k}\times{\bf{p}_1}]|}\Big) \ \mbox{(c.\,m.\,s.)} =
\Big(0,~\frac{[{\bf q}\times{\bf k}]}{|[{\bf q}\times{\bf k}]|}\Big) \ (\mbox{rest system})\,.
\end{equation}
%%%%%%%%%%%%%%%%
\section{Kinematics}
%%%%%%%%%%%%%%%%%%%%%%%
We define five independent invariant variables as follows
\ba\label{eq:invar}
s&=&(k_1+k_2)^2=(p_1+p_2+k)^2, \ s_1=(p_1+k)^2=(k_1+k_2-p_2)^2,
\\
s_2&=&(p_2+k)^2=(k_1+k_2-p_1)^2, \ t_1=(k_1-p_1)^2=(p_2+k-k_2)^2, \ t_2=(k_2-p_2)^2=(p_1+k-k_1)^2.
\nn
\ea

The scalar products of the 4-momenta in the process can be written in terms of these invariants as
\ba\label{scalar}
2k_1\cdot p_2& =& s-s_1+t_2-m_e^2, \ 2k_2\cdot p_1 = s-s_2+t_1-m_e^2, \ 2k_1\cdot k = s_1+t_1-t_2-M^2,\\
2k_2\cdot k& =& s_2+t_2-t_1-M^2, \ 2 p_1\cdot p_2 =s-s_1-s_2+m^2, \ 2 k_1\cdot k_2=s-2 m_e^2,\nn\\
2k_2\cdot p_2 &=&M^2+m_e^2-t_2, \ 2k_1\cdot p_1 =M^2+m_e^2-t_1, \ 2k\cdot p_1=s_1-M^2-m^2, \ 2k\cdot p_2=s_2-M^2-m^2.\nn
\ea
The kinematical regions allowed for the invariant variables can be obtained from the condition of the positivity of the quantity
$(-\Delta) = (k_1 k_2 p_1 p_2)^2,$ where $\Delta$ is the Gramian determinant. It has a form
\be
\Delta=\frac{1}{16}\left|
\begin{array}{l l l l } 
2 m_e^2               &\ s-2 m_e^2            & \ M^2+m_e^2-t_1   &\  s-s_1+t_2-m_e^2 \\
s-2m_e^2            &\ 2m_e^2                 & \ s-s_2+t_1-m_e^2 &\ M^2+m_e^2-t_2\\
M^2 + m_e^2-t_1&\ s-s_2+t_1 -m_e^2 & \ 2M^2                    & \ s-s_1-s_2+m^2\\
 s-s_1+t_2-m_e^2&\ M^2+m_e^2-t_2  & \ s-s_1-s_2+m^2    &\ 2M^2
\end{array}
\right|\,.
\nn
\ee
Taking into account the azimuthal symmetry relative to the line of the colliding electron-positron beams,
the phase space of the final particles can be written as \cite{Byckling:1971vca}
\begin{equation}\label{phasespace}
d\,R_3=\frac{\pi}{16(s-2m_e^2)}
\frac{dt_1\,dt_2\,ds_1\,ds_2}{\sqrt{-\Delta}}\,.
\end{equation}
Note that the electron mass can be neglected in our calculations, with a very high accuracy.

All the scalar products in hadronic part depend on the variables $s,~s_1$ and $s_2$
\ba
&&q^2\equiv s, \ p^2 = 2 M^2-m^2+s_1+s_2-s, \ q\cdot p= k\cdot p =\frac{s_1-s_2}{2}, \ k\cdot q= \frac{s_1+s_2}{2} - M^2,\nn\\
&&
d_1=s_2-M^2, \ d_2 = s_1-M^2.\nn
\label{eq:eqsch}
\ea
The double differential distributions are calculated as follows. To study the $(s_1,~ t_2)$ or $(s_2,~ t_1)$-distributions, it is sufficient  to measure one of the the 4-momenta 
$p_1$ or $p_2$, respectively. To investigate the $(s_1,~ s_2),~(s_1,~ t_1),~(s_2,~ t_2),~(t_1,~ t_2)$-distributions, both $p_1$ and $p_2$ have to be known.

Let us consider the range of the invariant variables and study the $(s_1,~ s_2)$ distribution. In this case, it is need to integrate over $t_1$ and $t_2$. From the positivity condition of the quantity $(- \Delta)$, we find
\be\label{eq:t1pm}
t_{1-}\leq t_1 \leq t_{1+}, \ \ t_{1\pm}=\frac{A(s, s_1, s_2, t_2) \pm 2 \sqrt{B(s, s_1, s_2)\,C(s, s_1, t_2)}}{(s+s_1-M^2)^2-4s s_1},
\ee
\ba
A(s, s_1, s_2, t_2)&= &m_e^2\big[2 M^4 -M^2(3 s_1+s_2)+s s_2- 2 m^2s -s_1(s-s_1-s_2)\big]- \nn\\
&&
 - M^2\big[m^2 s+s_1(s_2-2 s -t_2)+t_2(2 s-s_2)\big] - t_2\big[s(s_1+s_2-s -2 m^2) +s_1 s_2\big]+\nn\\
&&
+m^2 s(s-s_1)+M^6 - M^4(s+s_1+t_2)-s_1 s_2(s-s_1),\nn\\
B(s, s_1, s_2)&=& s_1 s_2(s_1+s_2-s)+2 M^6 -M^4(s+s_1+s_2+m^2)+\nn\\
&&
 +M^2\big[s s_2+s_1(s-2 s_2)+m^2(s_1+s_2-2 s)\big]+m^4 s +m^2\big[s(s-s_1-s_2)-s_1 s_2\big], \nn\\
C(s, s_1, t_2)& = &s\big[t_2(s-s_1+t_2-M^2) + M^2 s_1 \big] +m_e^2\big[M^4 -M^2(s+2s_1)-s(s_1+2 t_2)+s_1^2 +m_e^4 s\big]\nn.
\ea
The expression under the square root in Eq.~(\ref{eq:t1pm}) factorizes, and the limits on the variable $t_2~(s_2)$ can be found from the
condition $C(s, s_1, t_2)\geq 0 ~(B(s, s_1, s_2)\geq 0).$ For the variable $t_2$ they read
\begin{equation}\label{eq:t2pm}
t_{2-} \leq t_2  \leq t_{2+}, \ t_{2\pm}= \frac{1}{2}\left[M^2+2 m_e^2-s +s_1 \pm\sqrt{\left(1-\frac{4 m_e^2}{s}\right)\big[(s+s_1-M^2)^2 - 4 s s_1\big]}\right]
\end{equation}

The $s_2$ limits are 
\ba\label{eq:s2pm}
s_{2-} &\leq& s_2  \leq s_{2+}, \ s_{2\pm}= \frac{1}{2 s_1}\left(D(s,s_1)\pm \sqrt{F(s,s_1) G(s,s_1)}\right),
\\
D(s,s_1)&=& M^4 - M^2(s- 2 s_1+m^2)+m^2(s+s_1)+ s_1(s-s_1),\nn\\
F(s,s_1)&=&(s+s_1-M^2)^2-4 s s_1, \ \ G(s,s_1) = (s_1+m^2-M^2)^2- 4 m^2 s_1.\nn
\ea

Both expressions $F(s,s_1)$ and $G(s,s_1)$ have not to be negative, therefore
\begin{equation}\label{eq:s1pm}
(m+M)^2 \leq s_1  \leq (\sqrt{s}-M)^2.
\end{equation}
The inequalities (\ref{eq:t2pm}), (\ref{eq:s2pm}) and (\ref{eq:s1pm}) define the regions $(s_1,s_2)$ and  $(s_1,t_2)$
which are plotted in Figs.~2 a) and 2 b), correspondingly.
Because of the symmetry of the Gramian determinant with respect to the $(s_1\rightleftarrows s_2,~t_1\rightleftarrows t_2)-$permutations, one can apply these inequalities to limit also the region $(s_2,t_1)$.

It is interesting to investigate the distribution over the nucleon-antinucleon invariant mass squared $s_{12}= (p_1+p_2)^2 = 2M^2+m^2+s-s_1-s_2$. For this aim, we define firstly the region $(s_1,\,s_{12})$ and apply the inequality (see Eq. (\ref{eq:t1pm}))
$$B(s,\,s_1,\,s_2 = 2M^2+m^2 +s -s_1 -s_{12}) \geq 0$$
to obtain the limits on the $s_1$ variable at fixed values of the $s_{12}$ variable
\begin{equation}\label{eq:s1s12pm}
s_{1-} \leq s_1 \leq s_{1+},\ s_{1\pm}=\frac{1}{2}\bigg[2M^2+m^2+s-s_{12} \pm \sqrt{\Big(1-\frac{4M^2}{s_{12}}\Big)\big[(s+m^2-s_{12})^2 -4m^2 s\big]}\bigg].
\end{equation}
Taking into account that the expression under square root in Eq. (\ref{eq:s1s12pm}) has not to be negative, one finds the limits on the $s_{12}$ variable
$$4M^2 \leq s_{12} \leq(\sqrt{s}-m)^2.$$

As concerns the region $(t_1,~t_2),$ the corresponding boundaries are more complicated and the analytical
expressions for them require additional short notation. We introduce
\be
G(x,y,z,u,v,w)=-\frac{1}{2}\left|
\begin{array}{ccc}
 2 u & u-v+x & u+w-y \\
 u-v+x & 2 x & w+x-z \\
 u+w-y & w+x-z & 2 w \\
\end{array}
\right|\,,
\nn
\ee
with $s_{1}^-<s_1<s_{1}^+,$ and 
\ba
s_{1}^{\pm}&=&\frac{(a\pm b)}{(m_e^2-s_2^+)^2-4t_1s_2}, \
b=2 \sqrt{G\left(s,t_1,s_2,m_e^2,m_e^2,M^2\right) G\left(t_2,s_2,t_1,M^2,m_e^2,m^2\right)},\nn\\
a&=&s_2^-[s(t_1-m^2)+M^2(t_+-M^2)-s_2t_2]+s(t_2s_2^+-2M^2t_1)+m_e^2[m_e^2(s-2M^2)+\nn\\
&&+m^2(s-2M^2+2s_2)+M^2(M^2+t_++2s_2^+)+st_--s_2(t_++s_2)],\nn\\
&&\frac{-\lambda_{12}\lambda_2+a_1}{2t_2}<s_2<\frac{\lambda_{s}\lambda_1+b_1}{2m_e^2},\nn
\ea
where
\ba
a_1&=&t_-(t_2-m_e^2)+m^2(-M^2+m_e^2+t_2)+M^2t_+, \ b_1=s(t_1-M^2)+m_e^2(s+2M^2),\nn\\
s_2^{\pm}&=&s_2\pm t_1, \  t_{\pm}=t_1\pm t_2, \ \lambda (x,y,z)=x^2-2 x y-2 x z+y^2-2 y z+z^2,\nn\\
\lambda_{1,2}&=&\sqrt{\lambda \left(t_{1,2},m_e^2, M^2\right)}, \ \lambda_{12}=\sqrt{\lambda \left(t_1,t_2,m^2\right)}, \
\lambda_s=\sqrt{\lambda \left(s,m_e^2,m_e^2\right)}.\nn
\ea
The boundaries of the region $(t_1, t_2)$ are determined by the equation
\begin{equation}\label{eq:bt1t2}
\frac{-\lambda_{12}\lambda_2+a_1}{2t_2}=\frac{\lambda_{s}\lambda_1+b_1}{2m_e^2}.
\end{equation}
At such large energies, the electron mass cannot influence the kinematics. For the sake of simplicity, 
 the following formulas are derived in  the limit $m_e\, \to \,0.$. Eq. (\ref{eq:bt1t2}), in this limiting case, reads
 $$\frac{t_1 \left(M^2-s-t_1\right)}{M^2-t_1}=\frac{(t_2-M^2)(\lambda_{12}+m^2)
 +t_2t_-+M^2t_+}{2 t_2},$$
and gives
$$t_2^-<t_2<t_2^+$$
\ba 
t_2^{\pm}&=&\frac{a_2\pm b_2}{2 \left(M^2-t_1\right)
   \left(M^2-s-t_1\right)}, \ a_2=M^2[2t_1(s+t_1)-m^2s+2M^4-M^2(s+4t_1)]+\nn\\
&&
+st_1(m^2-s-t_1), \ b_2=s\bigl \{M^4[M^4+4t_1(t_1-M^2)+2t_1(s+t_1)]+\nn\\
&&
+t_1^2(s+t_1)(s+t_1-4M^2)+m^2(t_1-M^2)[m^2(t_1-M^2)+2M^4-2t_1(s+t_1)]\bigr\}^{\frac{1}{2}}, \nn\\
&&
\frac{a_3-b_3}{2}\leq t_1\leq \frac{a_3+b_3}{2}, \ a_3=2M(M+m)+m^2-s, \ b_3=
\sqrt{s-m^2} \sqrt{s-\left(m+2 M\right){}^2}.\nn
\ea
The regions $(t_2,\,t_1)$ and $(s_1,\,s_{12})$ are  plotted in the lower row in Fig.~2.

In addition, the dependence of the differential cross section on the invariant mass of the $N\,\bar{N}$-system is also of the utmost interest. It depends on the pion 4-momentum $k$ only, and allows, at least, investigations of the double distributions over invariants $\bar{t}_1 = (k_1-k)^2,~s_{12}=(p_1+p_2)^2$ or  $\bar{t}_2 =(k_2-k)^2,~s_{12}$.
To perform the corresponding calculations, it is necessary to investigate the Gramian determinant using $\bar{t}_1$ (or $\bar{t}_2$) and $s_{12}$ of five independent invariant variables.
In present paper, such kind of distributions are not considered and will be studied in a future publication.

\begin{figure}
\centering
\includegraphics[width=0.3\textwidth]{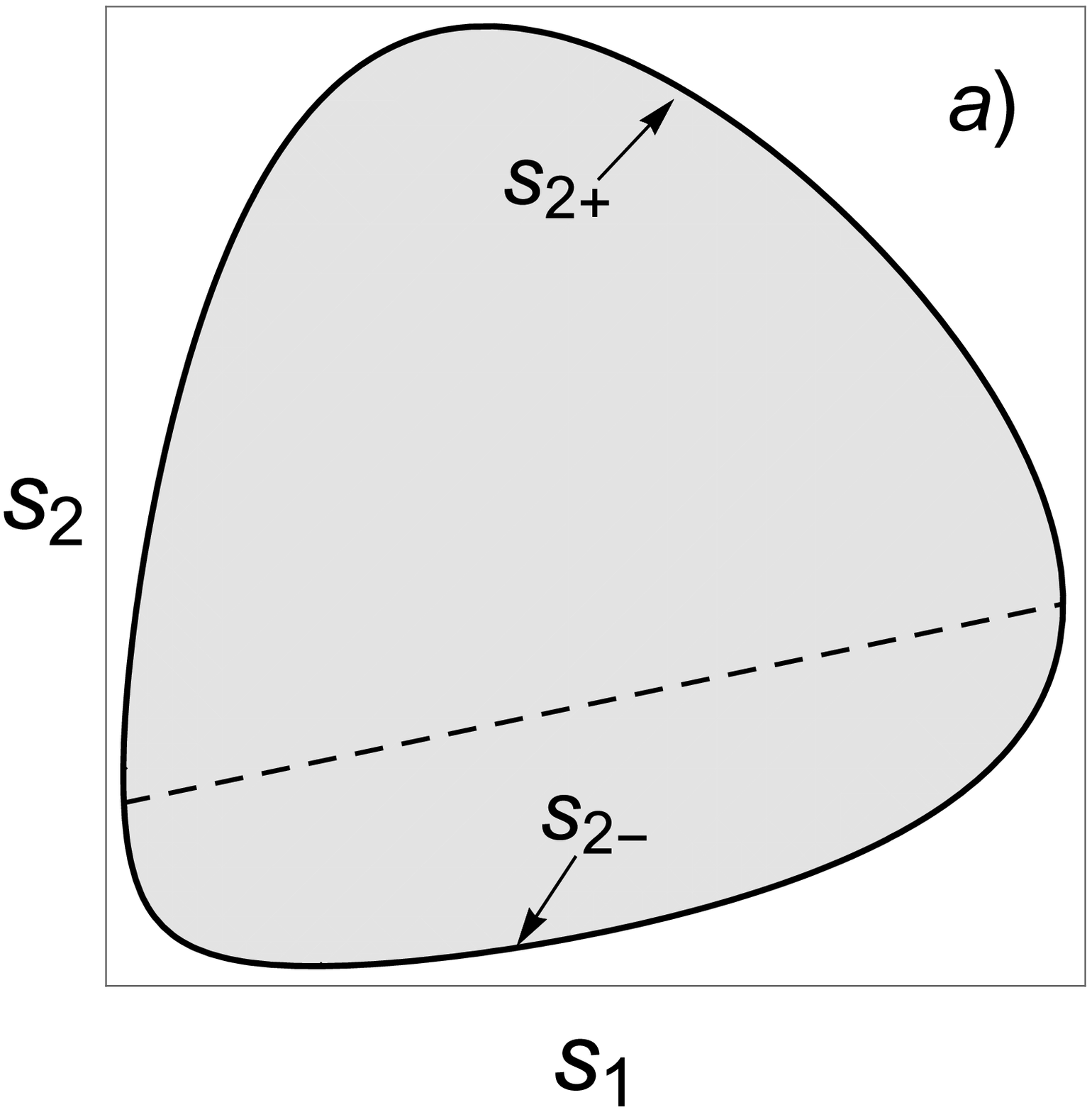}
\hspace{0.5cm}
\includegraphics[width=0.3\textwidth]{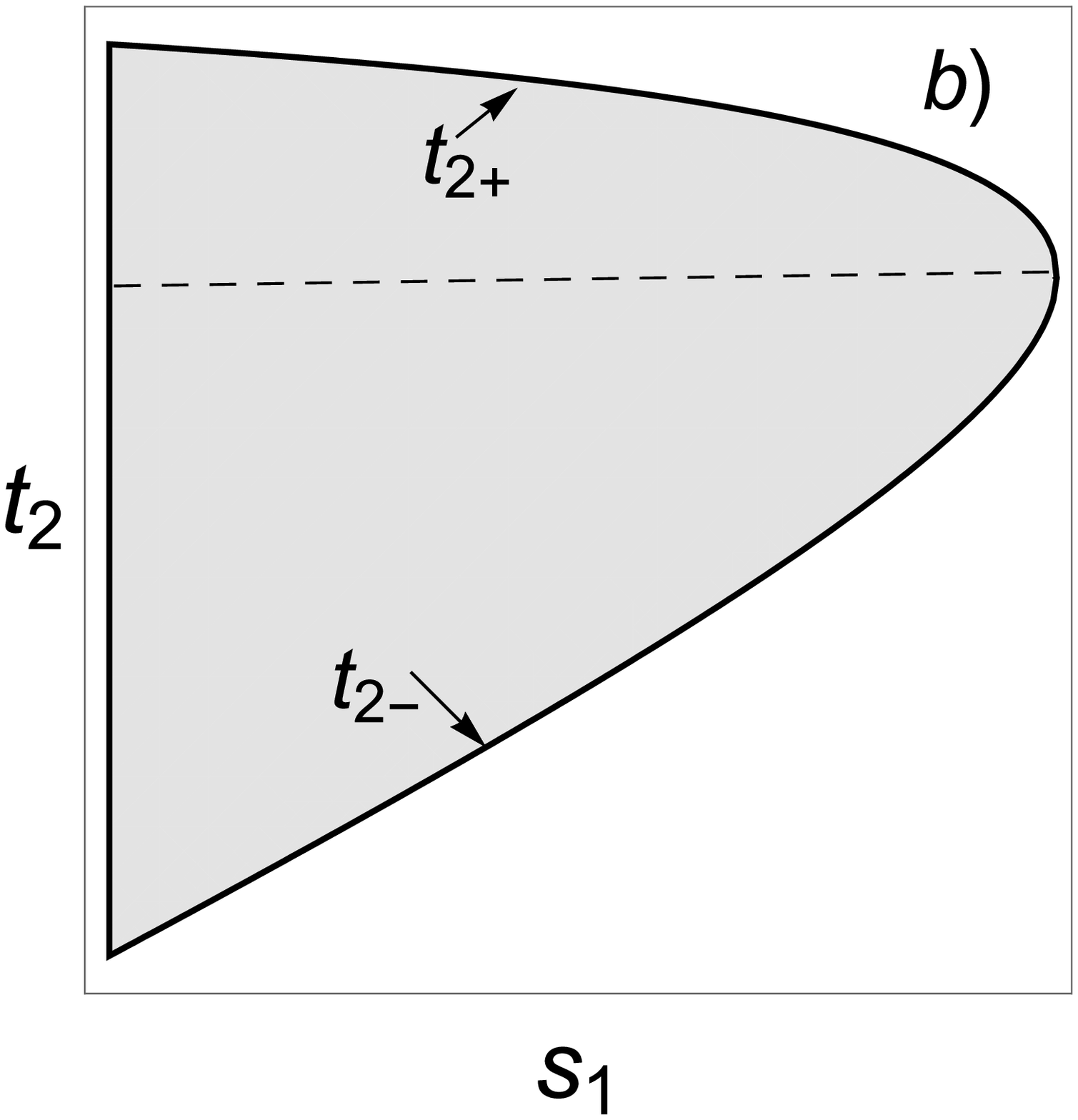}

\vspace{0.5cm}

\includegraphics[width=0.3\textwidth]{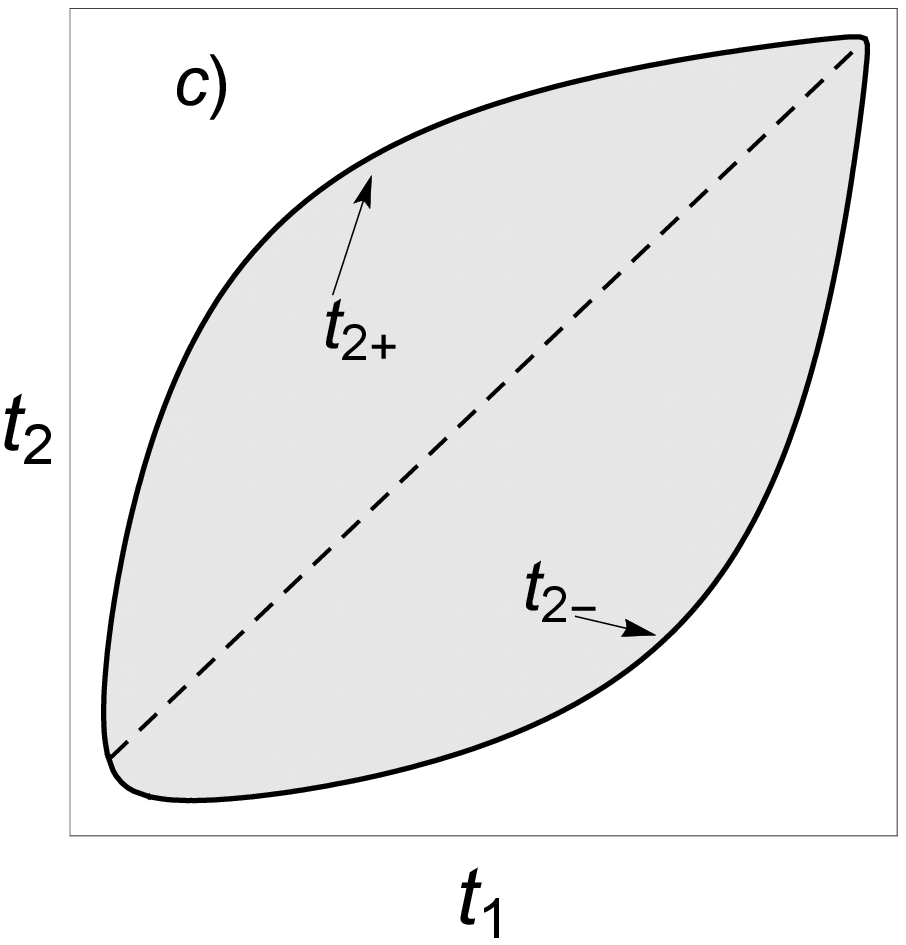}
\hspace{0.5cm}
\includegraphics[width=0.3\textwidth]{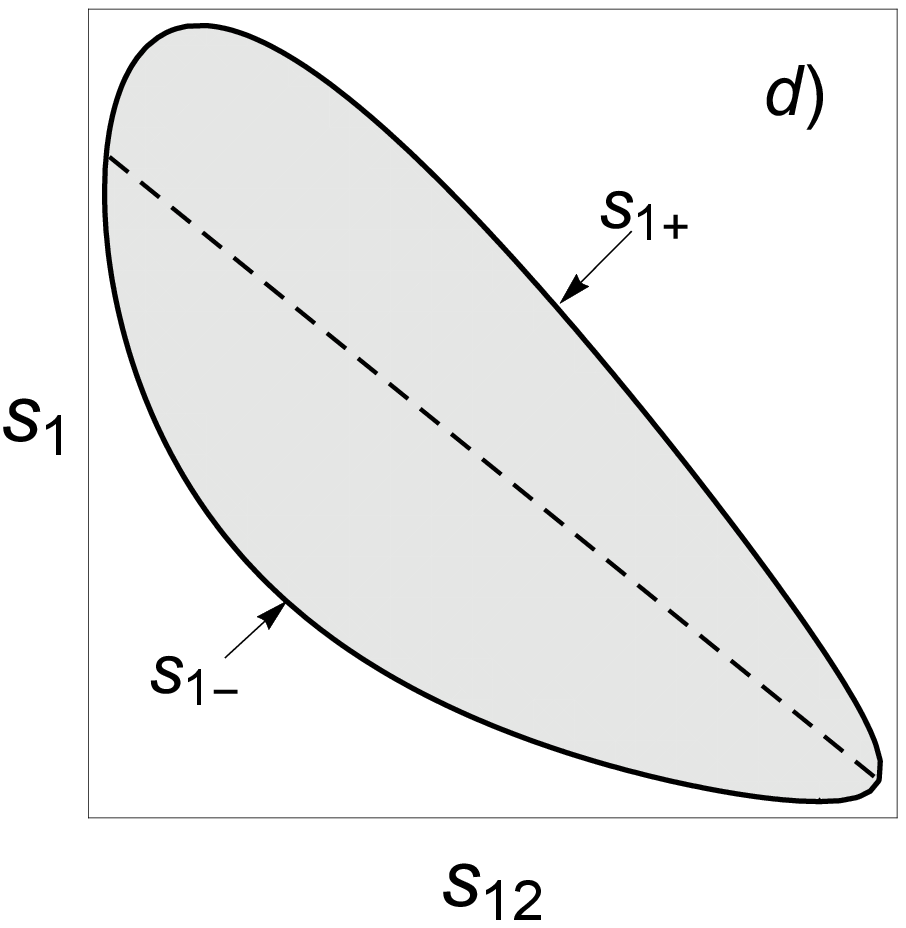}

\parbox[t]{0.9\textwidth}{\caption{The kinematical region  at $s=10~GeV^2$ for the double invariant variables : (a)$-(s_1,~s_2$), (b)$-(t_2,~s_1$),
(c) $-(t_2,~t_1$),  and (d) $-(s_1,\,s_{12})$.}\label{fig.2}}
\end{figure}
%%%%%%%%%%%%%%%%%%%%%%%%%%%%%%%%%%%%
\section{Continuum (non-resonant) contribution in process (\ref{eq:1})}
%%%%%%%%%%%%%%%%%%%%%%%%%%%%%%%%%%%%%
The non-resonant (continuum) contribution, to the reaction $e^+e^-\to N\bar N \pi^0$, 
is described by the diagrams given in Fig.~1~a),~b). The current, 
corresponding to the emission of the $\pi^0$ meson by the nucleon and antinucleon,
has the following form
\begin{equation}\label{eq:25}
J_{\mu}^B=g_{\pi^0 N N}\,\bar u(p_1)[\frac{1}{d_1}\Gamma_{\mu}^N(\hat p_1-\hat q+M)\gamma_5+
\frac{1}{d_2}\gamma_5(\hat q-\hat p_2+M)\Gamma_{\mu}^N]v(p_2),
\end{equation}
where $d_1=q^2-2q\cdot p_1$, $d_2=q^2-2q\cdot p_2$, $g_{\pi^0\,N\,N}$ is the coupling of the
$\pi^0\,N\,N$ interaction and
$$ \Gamma_{\mu}^N=F_1^N(q^2)\gamma_{\mu}-\frac{1}{2M}F_2^N(q^2)\sigma_{\mu\nu}q^{\nu}, $$
$\sigma_{\mu\nu}=(\gamma_{\mu}\gamma_{\nu}-\gamma_{\nu}\gamma_{\mu})/2$,
$F_1^N(q^2)$ and $F_2^N(q^2)$ are the Dirac and Pauli nucleon electromagnetic form factors which are
related to the Sachs magnetic and electric form factors by  $G_M^N(q^2)=F_1^N(q^2)+F_2^N(q^2)$,
$G_E^N(q^2)=F_1^N(q^2)+(q^2/4M^2)F_2^N(q^2)$.

The amplitudes $A_i (i=1-6)$, corresponding to the diagrams in Fig.~1, can be written as
\ba\label{eq:26}
A_1&=& - \frac{2g_{\pi^0 N N}}{d_1d_2}k\cdot q[F_1^N(q^2)-F_2^N(q^2)], \  A_2=\frac{2g_{\pi^0 N N}}{d_1d_2}F_1^N(q^2), 
A_3=-\frac{g_{\pi^0 N N}}{d_1d_2}\frac{q\cdot p}{M}F_2^N(q^2),\\
A_4&=&\frac{g_{\pi^0 N N}}{d_1d_2}\frac{k\cdot q}{M}F_2^N(q^2), \ A_5=A_6=0. \nn
\ea

The hadronic structure functions in Eq.~(\ref{eq:4}), which are independent on the nucleon polarization states, can be written, in general case, in terms of the invariant amplitudes as follows
\ba
H_1&=&2\Big\{[m^2 q^2 + (p\cdot q)^2 - (k\cdot q)^2]|A_{14}|^2+p^2\,|k\cdot q\,A_3+p\cdot q\,A_4-q^2\,A_6|^2- 
\nn\\
&&-4 M p\cdot q\, Re[k\cdot q\,A_3+p\cdot q\,A_4-q^2\,A_6]A_{14}^*\Big\}, \ \ A_{14}=A_1-4 M A_4,
\label{eq:27} \\
H_2&=&2\Bigl\{(p\cdot q)^2\,|A_4|\,^2 + [q^2(p^2+q^2) -(p\cdot q)^2]\,|A_3|\,^2 +(q-k)^2[(p\cdot q)^2\,|A_2|\,^2 +q^4\,|A_5|\,^2]\nn
\\&&
+q^2(q^2\,|A_6|\,^2 - |A_{14}|\,^2)\Bigr\} + 4\,Re\Bigl\{-p\cdot q\,(p\cdot q\,A_2 + q^2\,A_5 )A_{14}^* + p\cdot q\,(q^2\,A_3-2 M p\cdot q\,A_2) A_4^* +
\nn\\&&+q^2\,[2 M(p\cdot q\,A_2 +q^2\,A_5) - q^2\,A_3 - p\cdot q\,A_4]\,A_6^* - 2 M q^2 p\cdot q\,A_2\,A_3^* +\nn\\&&
+q^2[p\cdot q\,(q-k)^2\,A_2 - 2 M (q^2 A_3 + p\cdot q\,A_4)]\,A_5^* \Bigr\}\ \, \label{eq:28}\\
H_3 &=& 2\Bigl\{(k\cdot q)^2 [(q-k)^2 |A_2|\,^2 - |A_3|\,^2] + q^2 |A_{14}|\,^2 -q^4 |A_6|\,^2 +[(k\cdot q -q^2)^2 +q^2\,p^2]\,|A_4|\,^2\Bigr\} + \nn\\
&&
+ 4 Re\Bigl\{k\cdot q(q^2-k\cdot q)\,A_2 (A_{14}^* + 2 M A_4^*) +q^2 (2 M A_4 A_{14}^* + k\cdot q\,A_3\,A_6^*)\Bigr\}, \label{eq:29}\\\
H_4&=&2\Bigl\{k\cdot q\,p\cdot q (|A_3|\,^2 - (q-k)^2 |A_2|\,^2) + p\cdot q (q^2- k\cdot q) |A_4|\,^2\Bigr\} + 2 Re \Bigl\{ 2 M q^2 k\cdot q A_2 A_3^* +\nn\\
&&+[p\cdot q (2 k\cdot q - q^2)\,A_2 + 2 M q^2\,A_3 -q^2 (q^2-k\cdot q)\,A_5]\,A_{14}^* + \nn
\\
&&+q^2[(k\cdot q - q^2) A_4 -2 M k\cdot q A_2 - p\cdot q A_3]\,A_6^* + q^2 [- k\cdot q (q-k)^2\,A_2 +
\nn\\
&&
+2 M (k\cdot q - q^2)\,A_4]\,A_5^* +[2 M p\cdot q (2 k\cdot q -q^2) A_2 +q^2 (q^2 +p^2 -k\cdot q) A_3]\,A_4^*\Bigr\},\label{eq:30}
\\
H_5&=&-2q^2Im\Bigl\{ [2M\,A_3-q\cdot p\,A_2-(q^2-k\cdot q)\,A_5]A_{14}^*+
[-2 M k\cdot q A_2+q\cdot p\,A_6- 
\nn\\&&
-(p^2+q^2-k\cdot q)A_4]A_3^*+ [k\cdot q (q-k)^2 A_2-2 M(k\cdot q-q^2) A_4]A_5^*+
\nn\\&&
+[(q^2-k\cdot q)A_4+2 M k\cdot q A_2]A_6^*- 2M q\cdot p\,A_2\,A_4^*\Bigr \}.\label{eq:31}
\ea

The relations between invariant amplitudes and hadronic structure functions in Eqs.~(\ref{eq:7}) and (\ref{eq:8}), which depend on the nucleon polarization states, are more complicated and are given them in Appendix~B.

Using the relations (\ref{eq:26}), we obtain very simple expressions for the contribution of the non-resonant mechanism in terms of the electromagnetic form factors
(further we will omit the upper index of the form factors keeping in mind that they are different for $p\, \bar p$ and $n\, \bar n$)
$$H_1=8M^2G^2(k\cdot q)^2(m^2q^2-d_1d_2)|G_M(q^2)|^2, $$
\begin{equation}\label{eq:32}
H_2=-2\frac{G^2}{(1-\tau)^2}
\{4M^2q^2d_1d_2(1-\tau)^2|G_M(q^2)|^2-(p\cdot q)^2[d_1d_2|G_M(q^2)-G_E(q^2)|^2+
\end{equation}
$$+4m^2M^2(1-\tau)(|G_E(q^2)|^2-
\tau |G_M(q^2)|^2)]\}, $$
$$H_3=2(k\cdot q)^2\frac{G^2}{(1-\tau)^2}
[4m^2M^2(1-\tau)(|G_E(q^2)|^2-\tau |G_M(q^2)|^2)+d_1d_2|G_M(q^2)-G_E(q^2)|^2], $$
$$H_4=-\frac{p\cdot q}{k\cdot q}H_3, \ \ H_5=0, $$
where $G=g_{\pi^0 N N }/(Md_1d_2)$, $\tau=q^2/(4M^2)$.

The product of the leptonic and hadronic tensors, which define the matrix element squared (\ref{eq:3}), reads
$$L_{\mu\nu}\,H^{\mu\nu}(0)= -2s H_1+ [-sm^2 +(s_1-M^2+t_1-t_2)(s_2-M^2+t_2-t_1)]H_2 \,+$$
$$\big\{s[m^2-2(t_1+t_2)]-(s_1+M^2-t_1-t_2)(s_2+M^2-t_1-t_2)\big\}H_3 \,+$$
\begin{equation}\label{eq:unpol}
2 [(t_1-t_2)(s+t_1+t_2-M^2) +s_1(t_2-M^2)-s_2(t_1-M^2)] H_4 + 8(k_1 k_2 p_1 p_2) H_5,
\end{equation}
where we used the electron polarization 4-vector $\eta=k_1/m_e$ and went to the limit $m_e\to 0 $.

The double differential distributions over $(s_1,\,s_2), \,\,(s_1,\,s_{12})$ and $(t_1,\,t_2),\,\,(t_1,\,s_2)$ can be obtained analytically. The first two distributions have simple forms:
\ba\label{eq:dsigmas1s2}
\frac{d\,\sigma}{d\,s_1\,ds_2}&=&\frac{\alpha^2\,g^2_{\pi^0 p \bar{p}}}{24\pi P_0}\bigg[P_1\left|G_M\right|^2
 +\frac{P_2}{(s-4M^2)^2}\bigg(P_3\left|G_M - G_E\right|^2 + m^2\left|sG_M - 4M^2G_E\right|^2\bigg)\bigg],
\\
P_0&=&(s_1-M^2)^2(s_2-M^2)^2s^3,\nn\\
P_1&=&8M^8-16M^6(s_1+s_2)+M^4\big[8s_1 s_2 +10(s_1+s_2)^2-8m^2 s\big]\nn\\
&& -2M^2(s_1+s_2)\big[4s_1 s_2+(s_1+s_2)^2 -4m^2 s\big]-m^2s\big[3(s_1+s_2)^2-4s_1 s_2\big],\nn\\
P_2&=&-8M^6+4M^4(s+m^2+s_1+s_2)-4M^2\big[(s+m^2)(s_1+s_2)\nn\\
&&-2s_1 s_2\big]+s(s_1+s_2)^2-4s_1 s_2(s_1+s_2-m^2),\nn\\
P_3&=&4M^2\big[M^4-M^2(s_1+s_2)+s_1 s_2-m^2 s\big].\nn
\ea
The $(s_1,\,s_{12})-$distribution is obtained from Eq. (\ref{eq:dsigmas1s2}) with the replacement $s_2\rightarrow 2M^2 +m^2 +s -s_1-s_{12}.$
The analytical forms of the  $(t_2,\,t_1)-$ and $(t_1,\,s_2)-$distributions, are much more involved and not reported in this paper. Note that all double differential non-resonant cross sections are symmetrical under the substitution $s_1\leftrightarrow s_2,\, t_1 \leftrightarrow t_2.$
%%%%%%%%%%%%%%%%%%%%%
\subsection{Choice of the form factors}
%%%%%%%%%%%%%%%%%%%%%%
It is obvious that a key moment in our calculations is the choice of the electromagnetic form factors in the time-like region, and the corresponding data used to fit different theoretical models of the form factors. Our numerical results are obtained for two different parameterizations of the two-component model based on the vector dominance (VDM) at low and intermediate energies and predictions of the perturbative QCD at the large ones. Recently, precise data  where obtained by direct beam scan  \cite{CMD-3:2018kql,BESIII:2019hdp,BESIII:2021tbq} or  radiative return measurements of the $e^++e^- \to N +\bar{N}$ cross section \cite{BESIII:2021rqk,Lees:2013uta,Lees:2013xe}   from the threshold up to $\sqrt{s}=6.5$ GeV . A general parametrization, including these data is not yet available, but a comparison with the parametrization used here was done in Ref. \cite{Tomasi-Gustafsson:2020vae}, showing that they give a description of the new data on the individual proton form factors,  even without refitting, that is sufficient for the present purposes. However,  it is not evident that these simple parametrizations based on few parameters will be successful in describing simultaneously the new precise sets of data, on proton, neutron, electric and magnetic form factors in both space- and time-like regions.

To account for the VDM properties,  the Dirac ($F_1$) and Pauli ($F_2$) form factors are divided by the isotopic vector ($F_{1,2}^V$) and scalar
($F_{1,2}^S$) parts which are normalized  in such a way that
$$F_{1,2}^p =\frac{1}{2}\big(F_{1,2}^S + F_{1,2}^V\big), \ \ F_{1,2}^n =\frac{1}{2}\big(F_{1,2}^S - F_{1,2}^V\big).$$
In the parameterizations used here, the vector part is fulfilled by the $\rho$ meson and the scalar part by $\omega,\,\phi$ meson contributions.

The first parametrization is taken from the papers \cite{Iachello:1972nu,Pacetti:2015iqa,Iachello:2004aq} and is labeled as the "old" one. The second parametrization, labeled as "new", is taken from
the paper \cite{Bijker:2004yu}. In Fig.~3  the real and imaginary parts of the electric and magnetic form factors of a proton and neutron for both parametrizations:  old (upper row)
and new (bottom row),  are plotted.

%%%%%%%%%%%%%%%%%%%%%%%%%%%%%%%%%%%%%%%%%%%%%%%%%%%%%%%%%%%%%%%%%%%%%%%%%%%%%%%%%%%%%%%%%%
\begin{figure}
\centering
\includegraphics[width=0.224\textwidth]{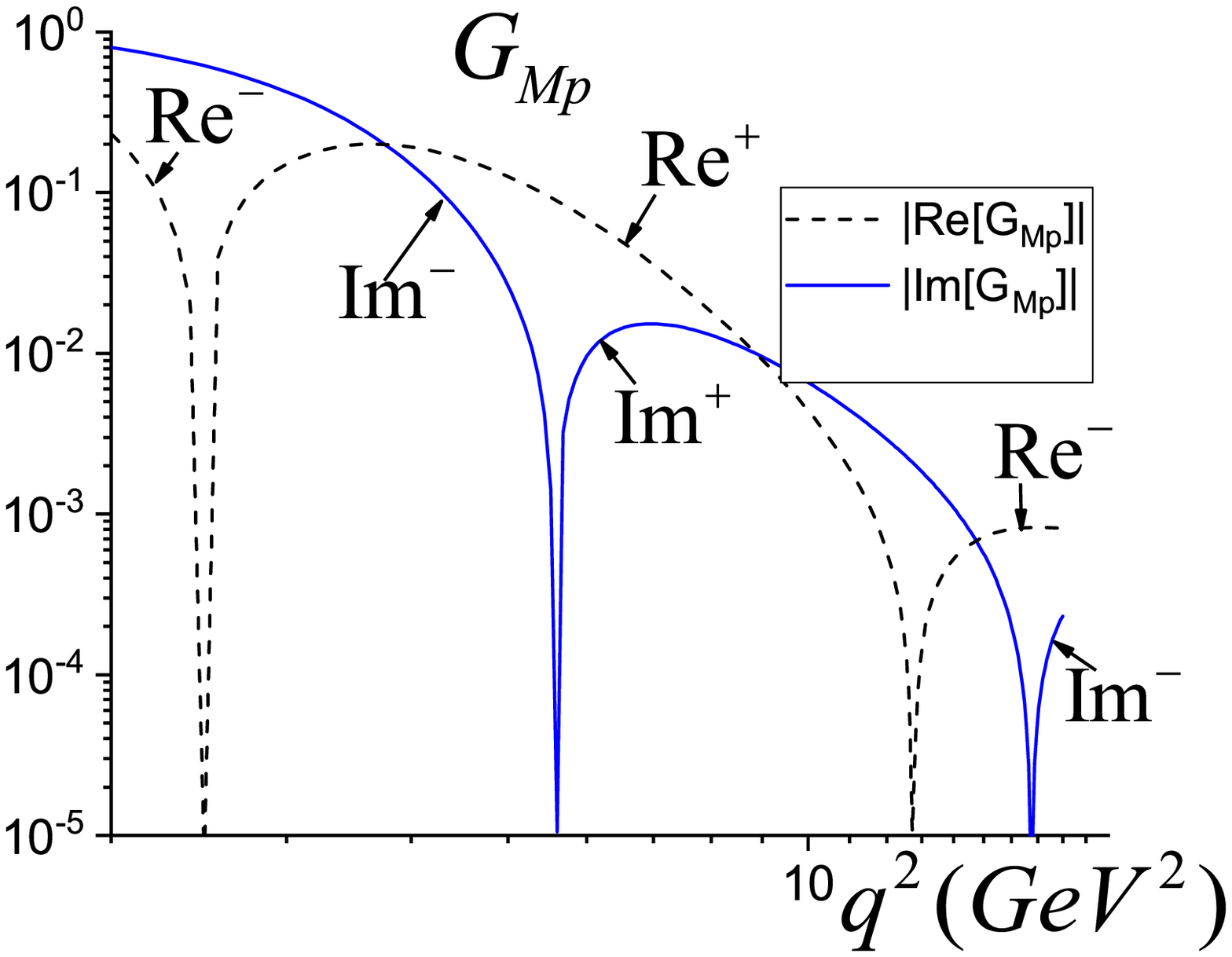}
\hspace{0.1cm}
\includegraphics[width=0.224\textwidth]{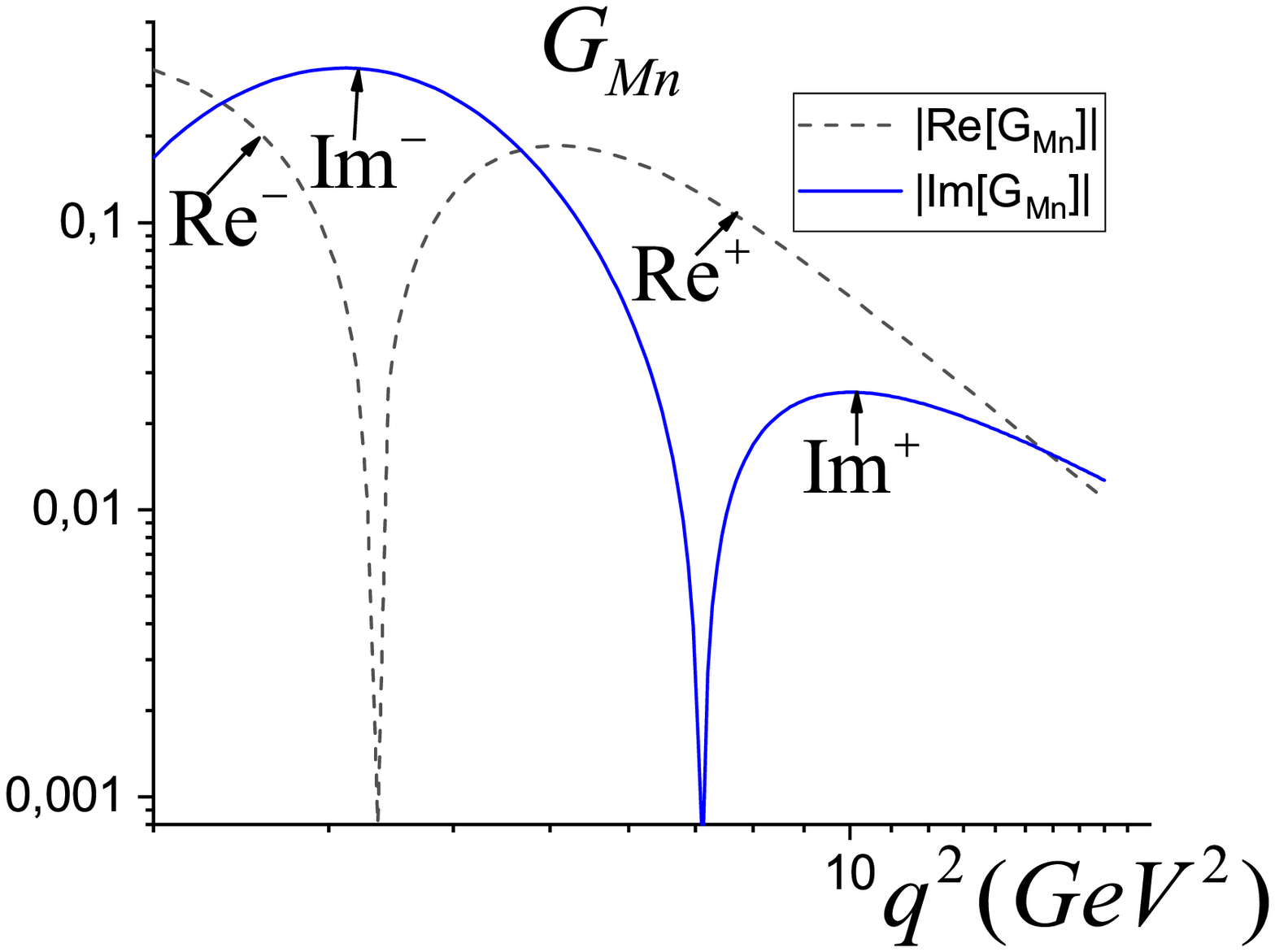}
\hspace{0.1cm}
\includegraphics[width=0.224\textwidth]{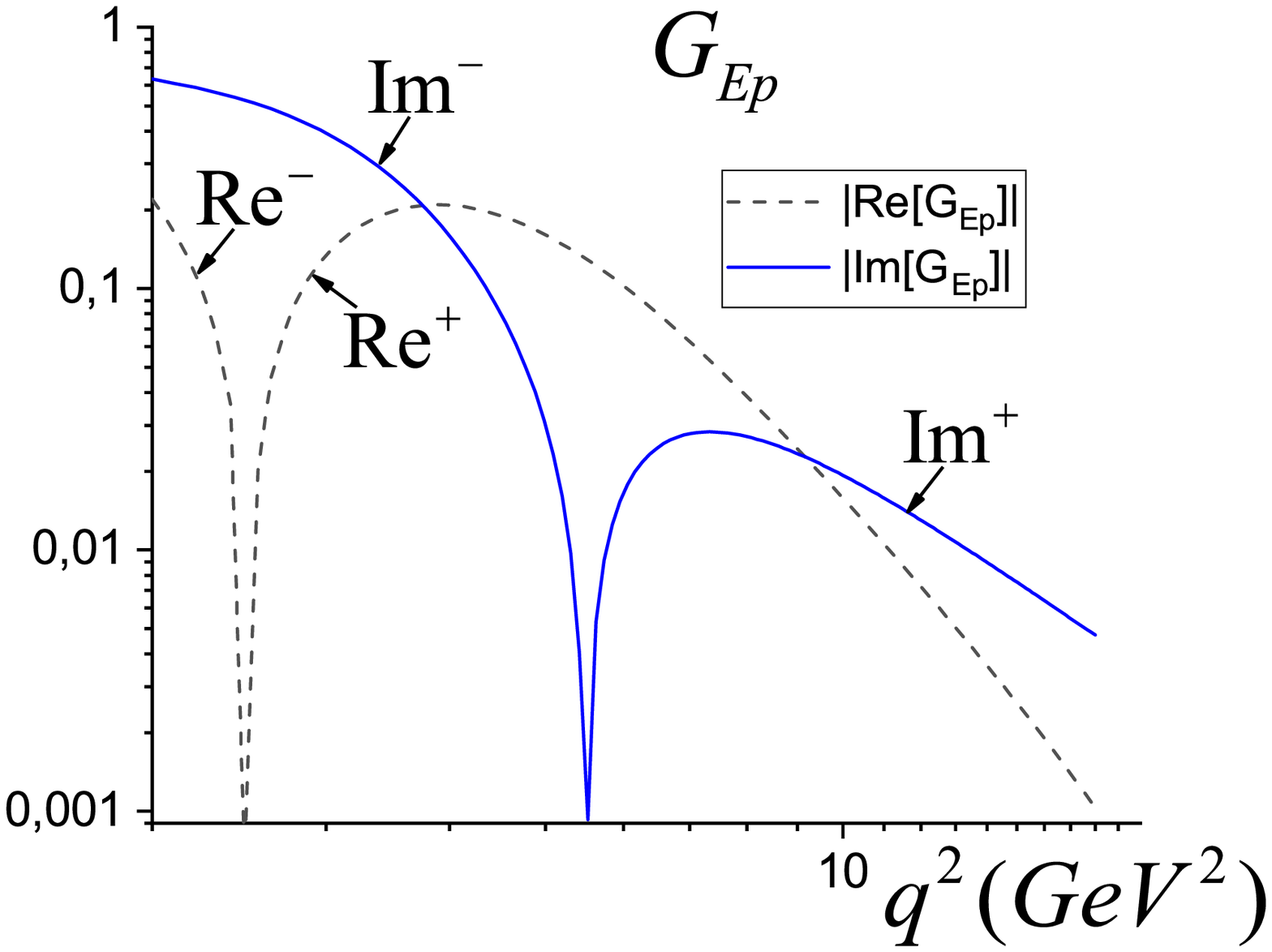}
\hspace{0.1cm}
\includegraphics[width=0.224\textwidth]{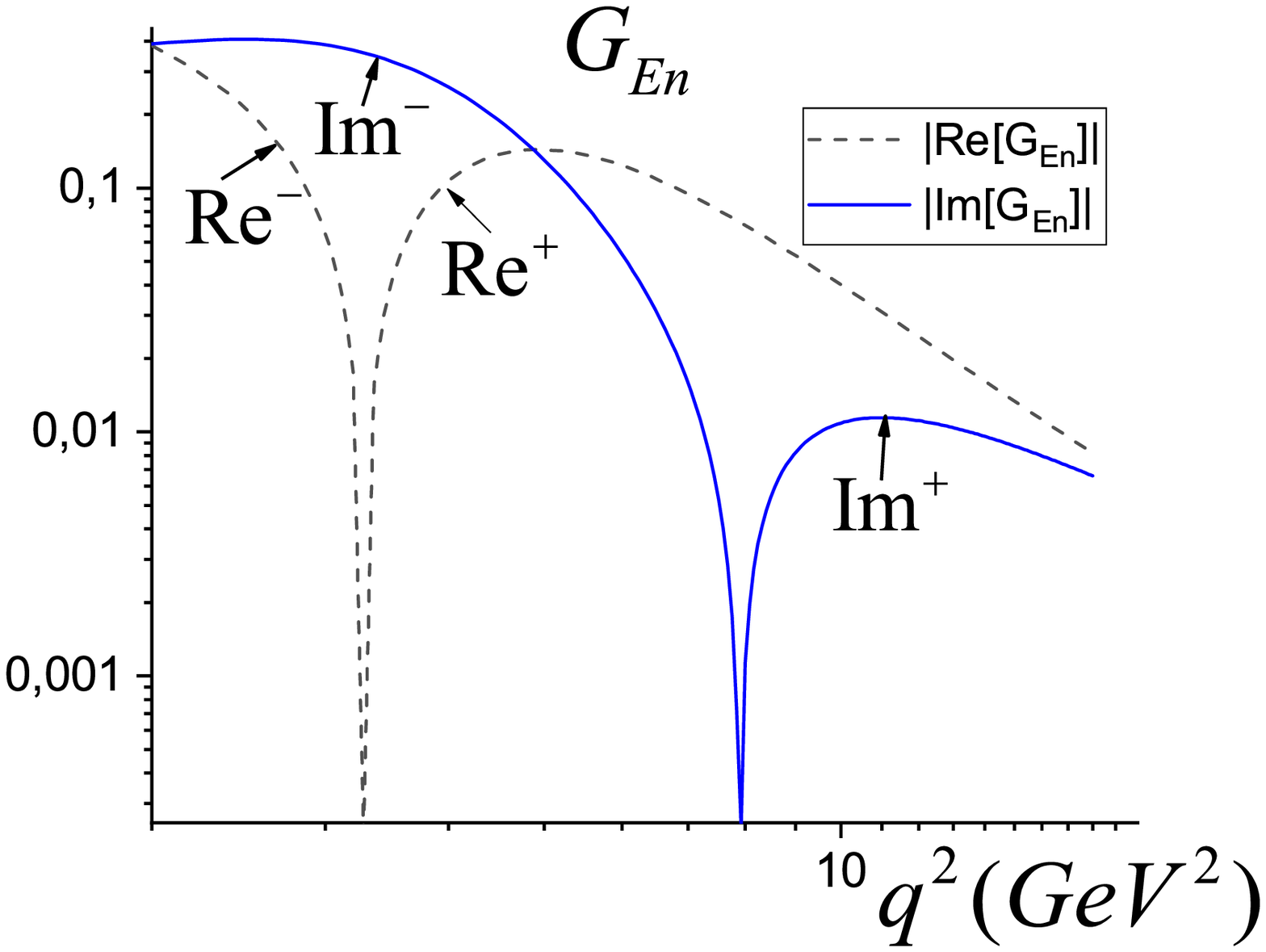}
\vspace{0.2cm}
\includegraphics[width=0.224\textwidth]{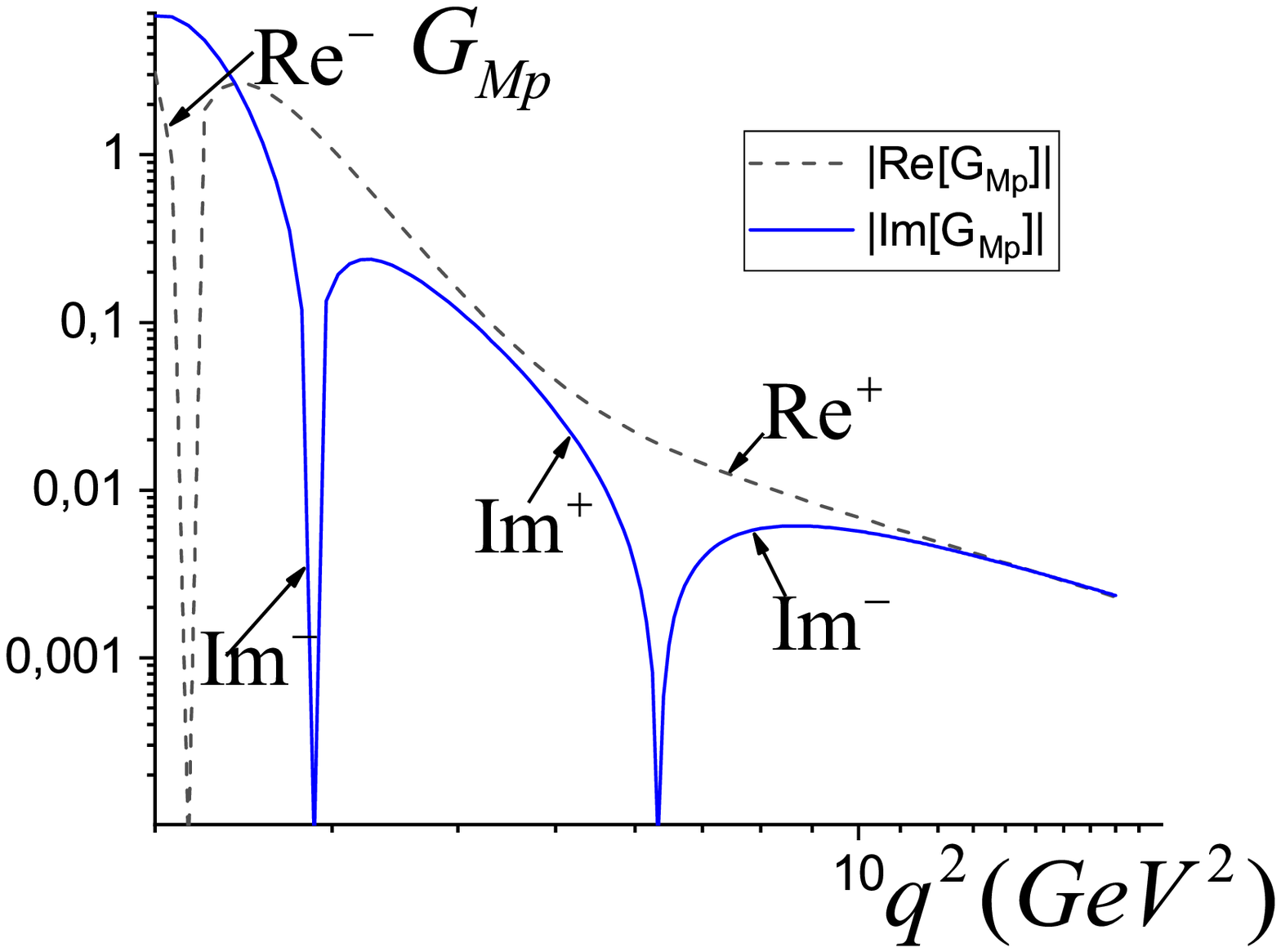}
\hspace{0.1cm}
\includegraphics[width=0.224\textwidth]{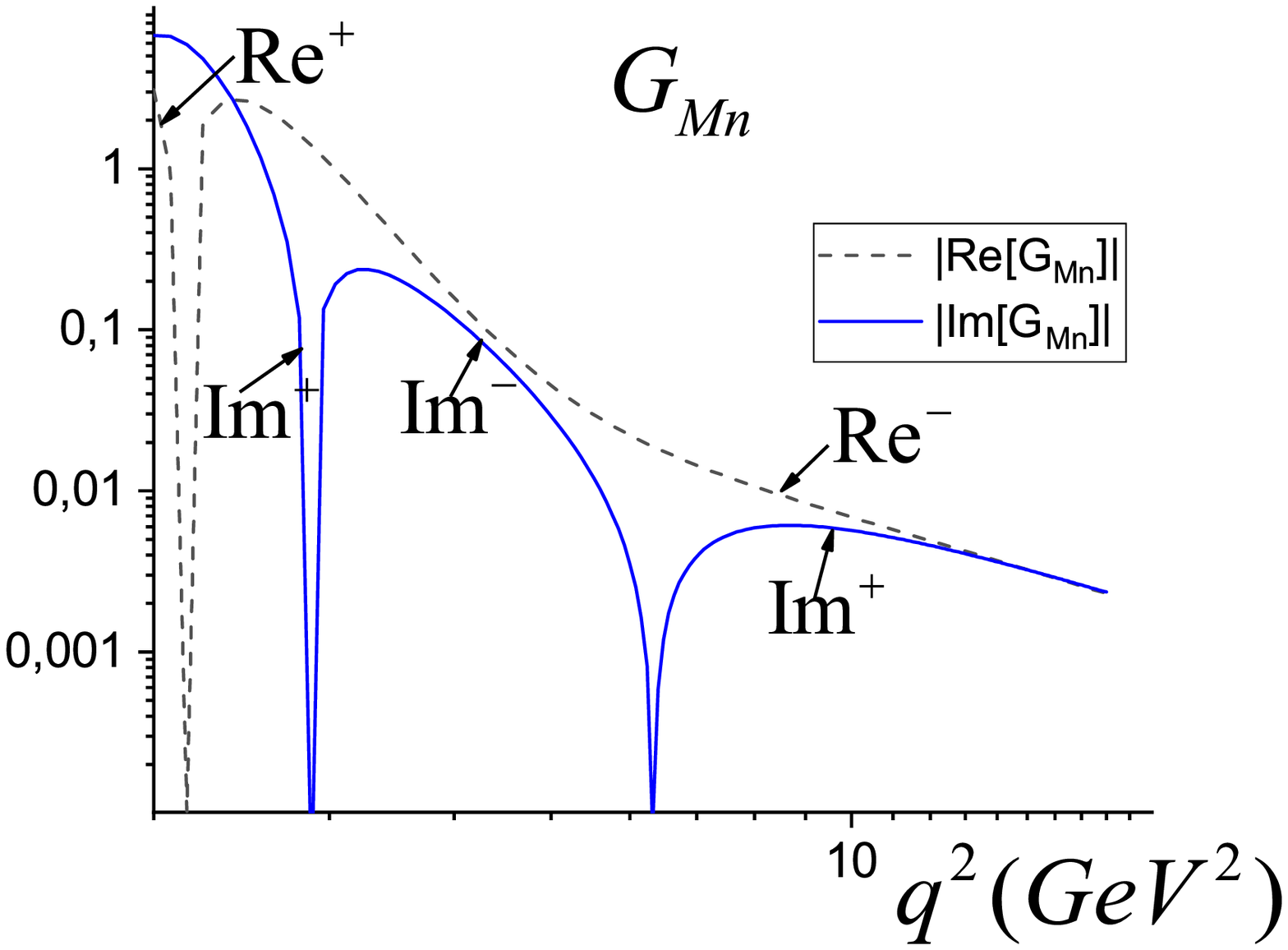}
\hspace{0.1cm}
\includegraphics[width=0.224\textwidth]{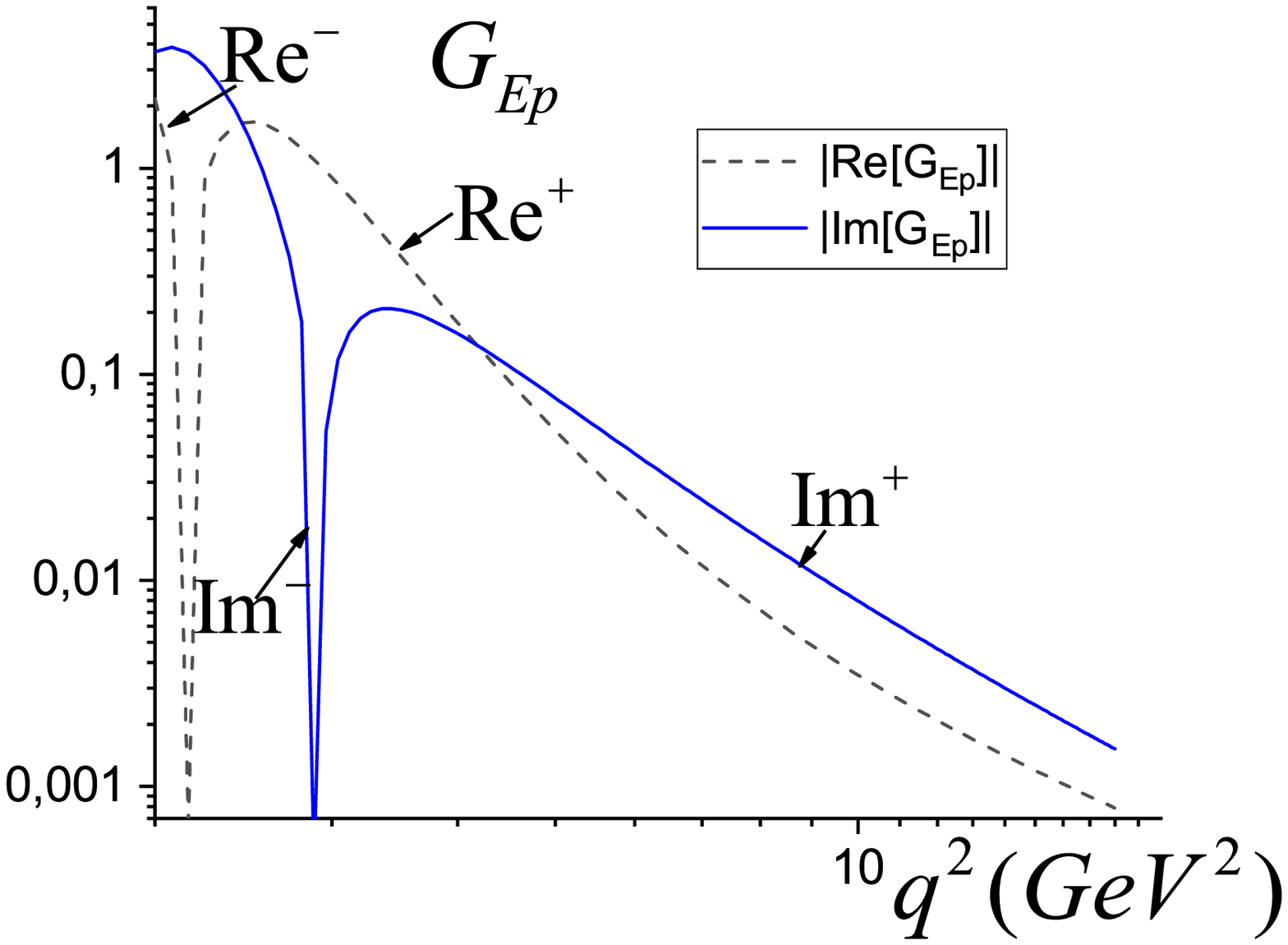}
\hspace{0.1cm}
\includegraphics[width=0.224\textwidth]{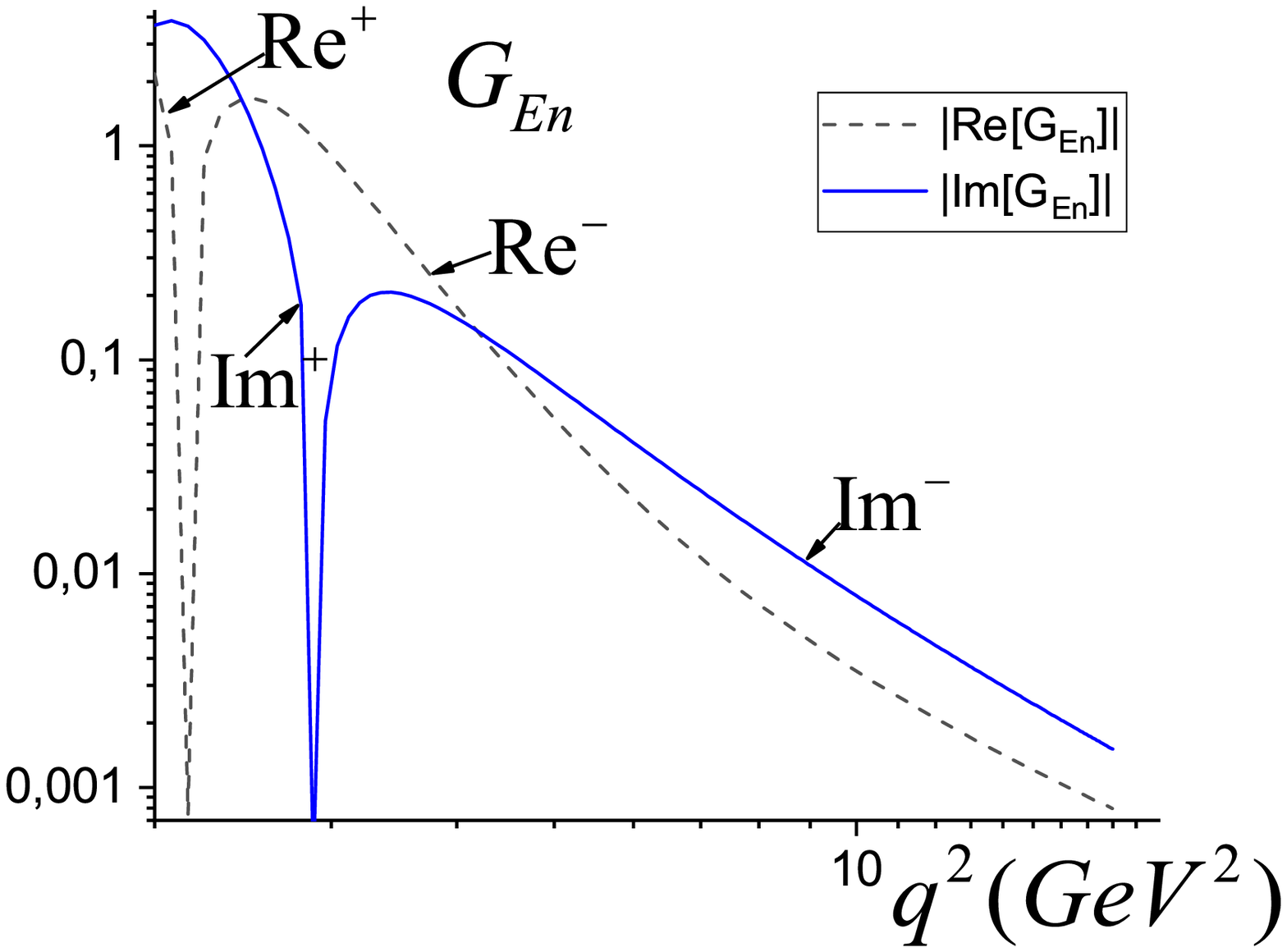}
\vspace{0.2cm}
\parbox[t]{0.9\textwidth}{\caption{The moduli of the real and imaginary parts of the proton and neutron electromagnetic form factors as calculated from Refs  \cite{Iachello:1972nu,Pacetti:2015iqa,Iachello:2004aq} ("old" version, upper row) and  \cite{Bijker:2004yu} ("new" version,  lower row). The notation $(Re,Im)^+\, (Re, Im)^-$ means that the real or imaginary part is positive (negative). }\label{fig.3}}
\end{figure}

%%%%%%%%%%%%%%%%%%%%%%%%%%%%%%%%%%%%%%%%%%%%%%%%%%%%%%%%%%%%%%%%%%%%%%%%%%%%%%%%%%%%%%%%%%%

In Fig.~4 the moduli are shown for comparison with the original papers. As one can see, above the threshold of the process (\ref{eq:1}), the  moduli of all form factors, except $G_{Mp},$ are larger for the old parametrization than for the new one. This characteristic 
affects directly the corresponding values of the differential (see Figs.~7,\,8) and total cross sections.

\begin{figure}
\centering
\includegraphics[width=0.4\textwidth]{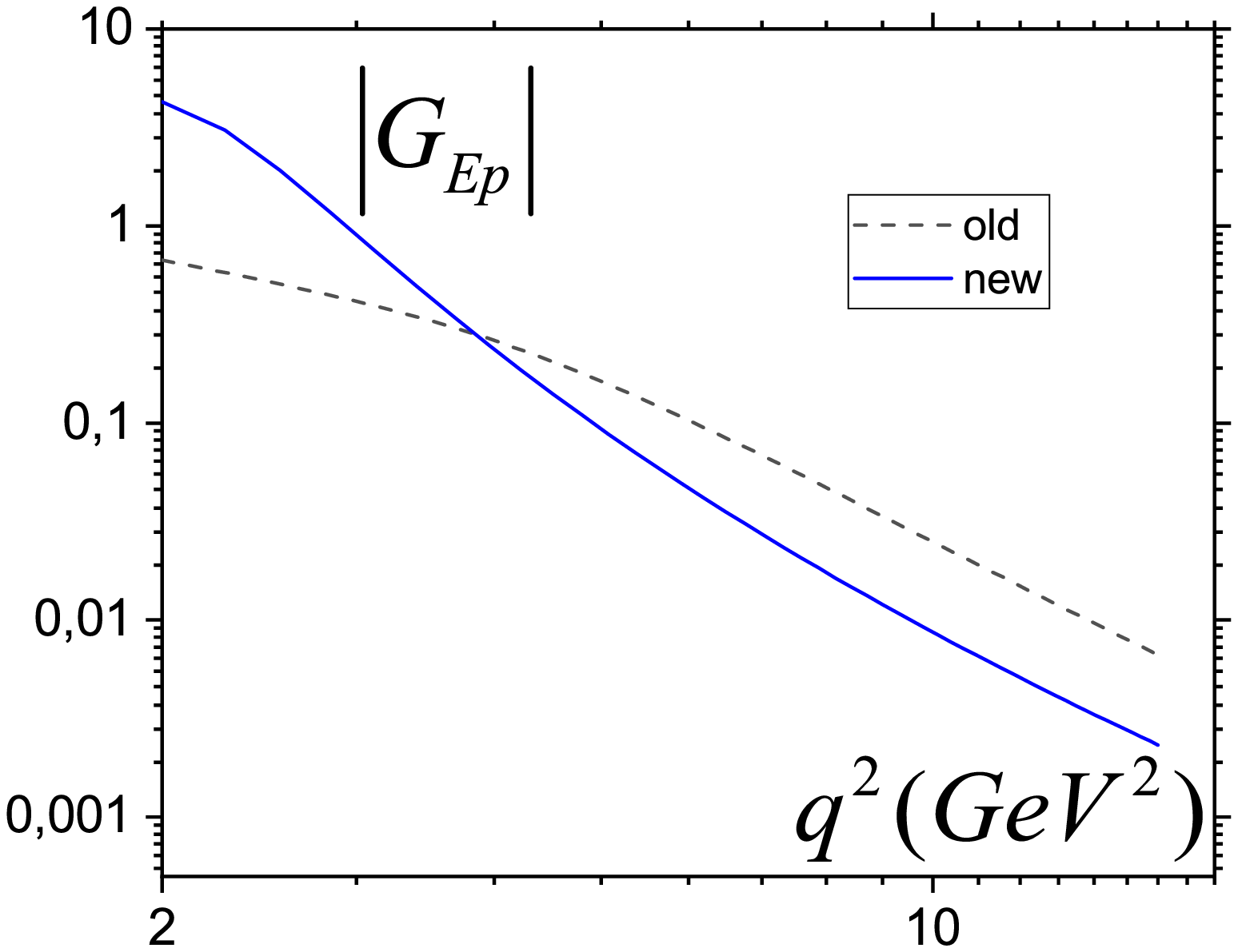}
\hspace{0.1cm}
\includegraphics[width=0.4\textwidth]{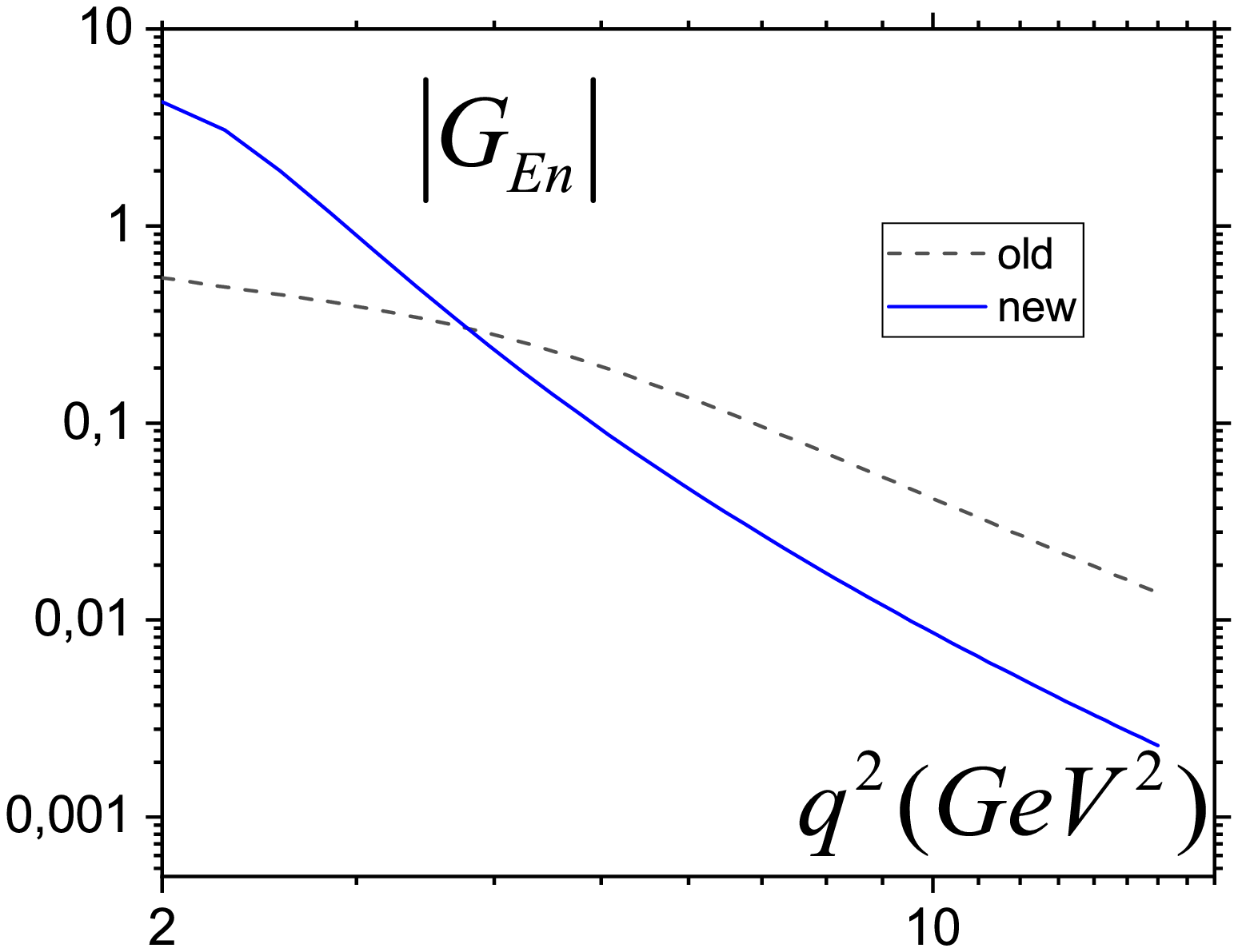}
\hspace{0.1cm}
\includegraphics[width=0.4\textwidth]{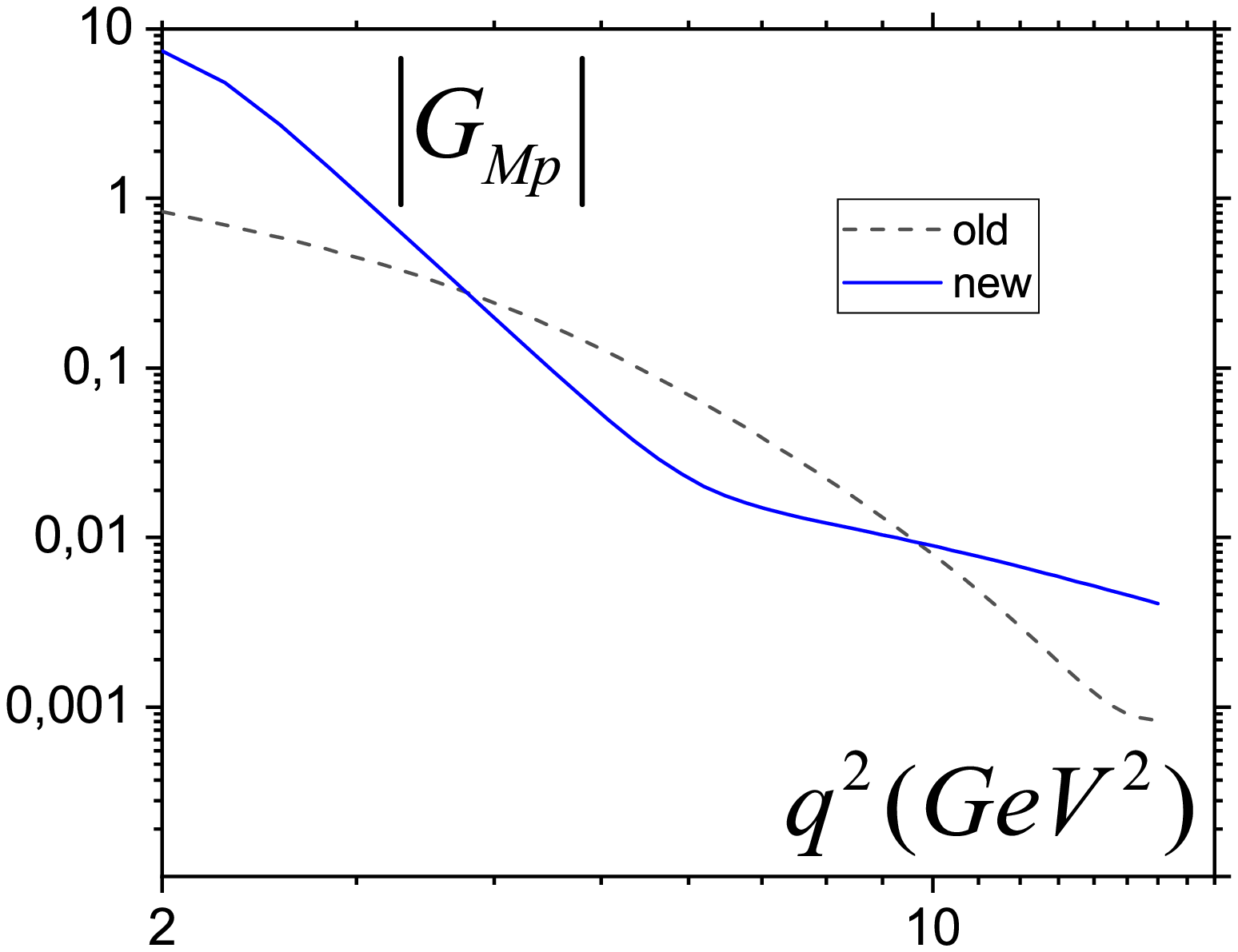}
\hspace{0.1cm}
\includegraphics[width=0.4\textwidth]{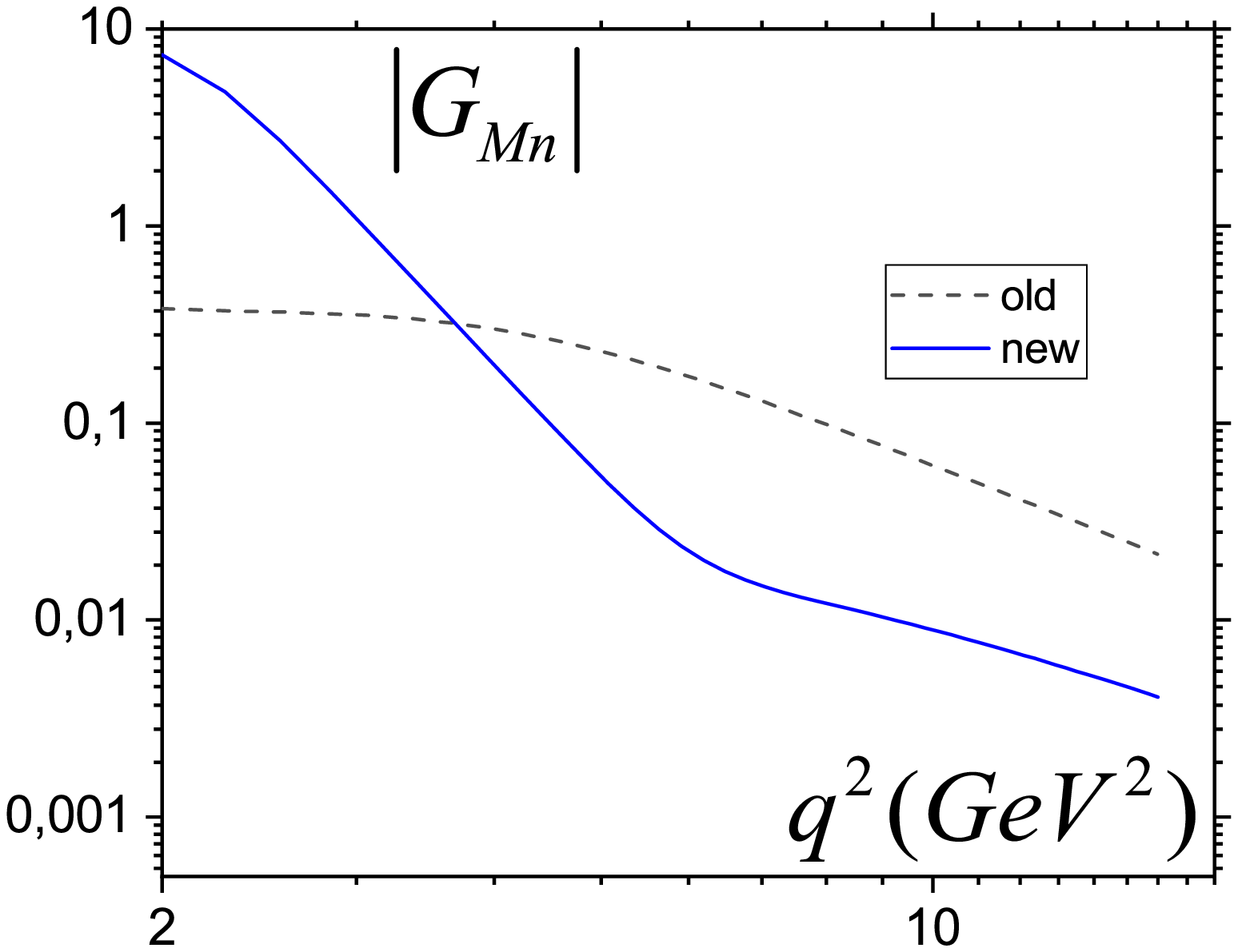}
\vspace{0.2cm}
\parbox[t]{0.9\textwidth}{\caption{Moduli of the electromagnetic form factors as given in the cited papers  \cite{Iachello:1972nu,Pacetti:2015iqa,Iachello:2004aq} ("old" version)
and  \cite{Bijker:2004yu} ("new" version).}\label{fig.4}}
\end{figure}

%%%%%%%%%%%%%%%%%%%%%%%%%%%%%%%%%%%%%%%%%%%%%%%%%%%%%%%%%%%%%%%%%%%%%%%%%%%%%%%%%%%%%%%%%%%%%%%%%%%%%%%%%

\begin{figure}
\centering
\includegraphics[width=0.21\textwidth]{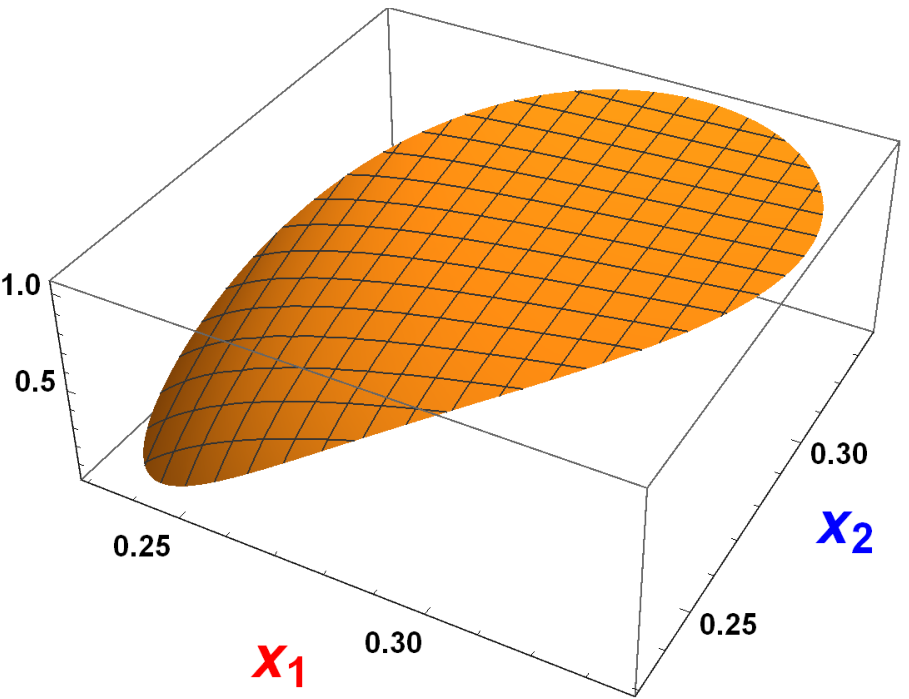}
\hspace{0.3cm}
\includegraphics[width=0.21\textwidth]{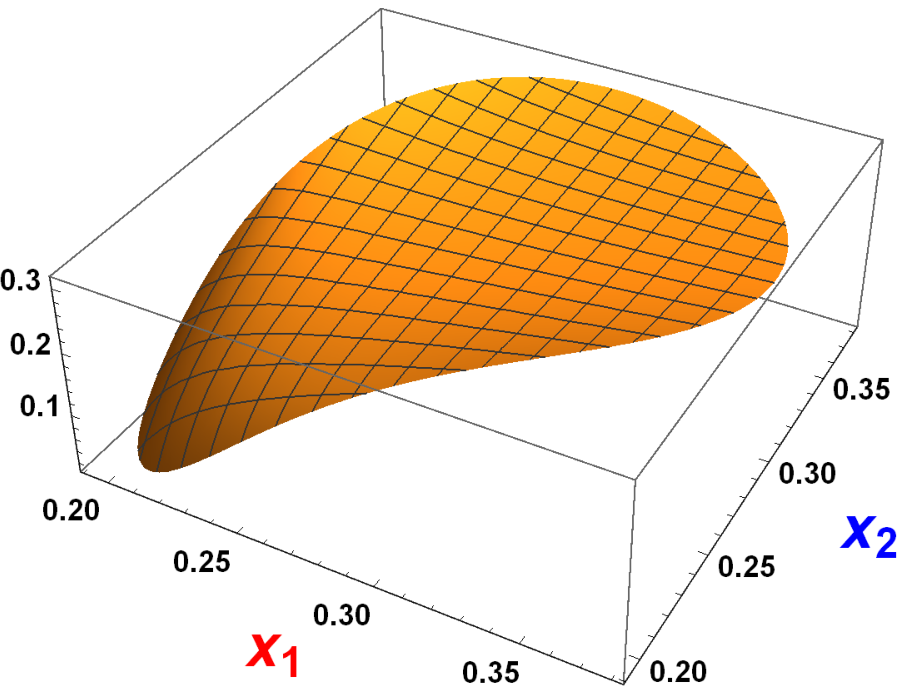}
\hspace{0.3cm}
\includegraphics[width=0.21\textwidth]{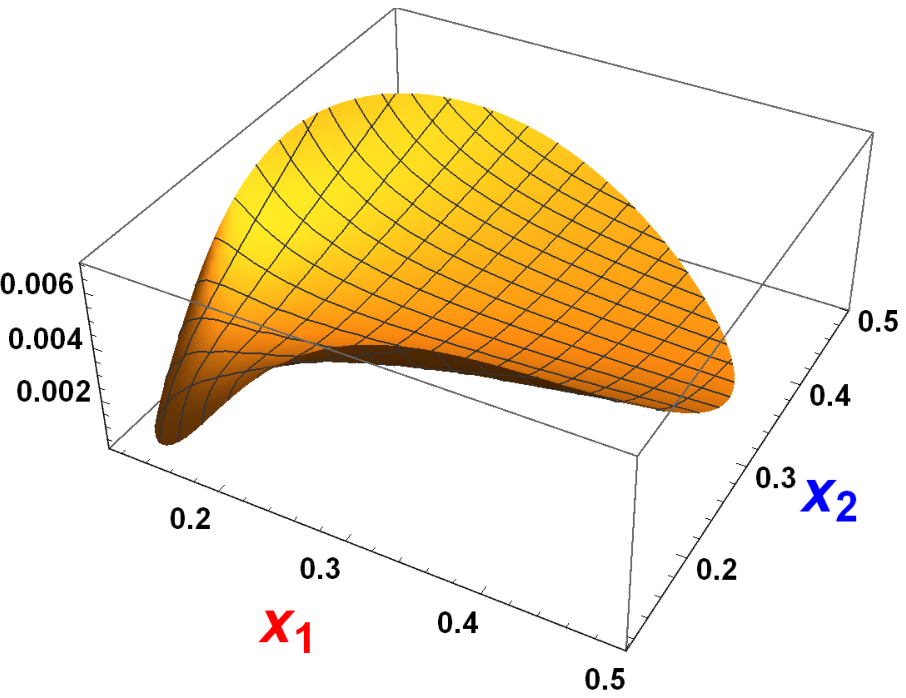}
\hspace{0.3cm}
\includegraphics[width=0.21\textwidth]{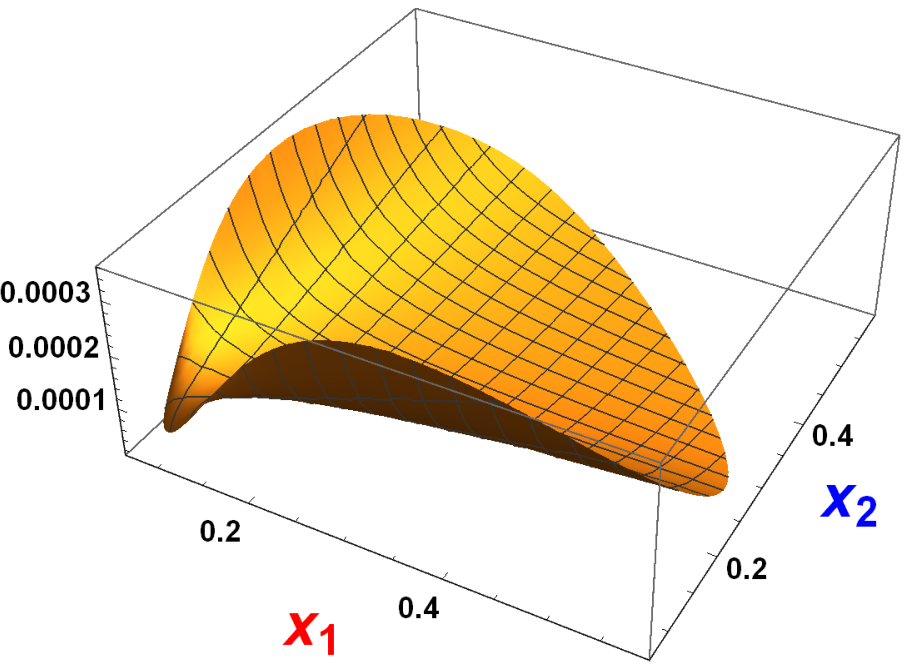}
\vspace{0.5cm}
\includegraphics[width=0.21\textwidth]{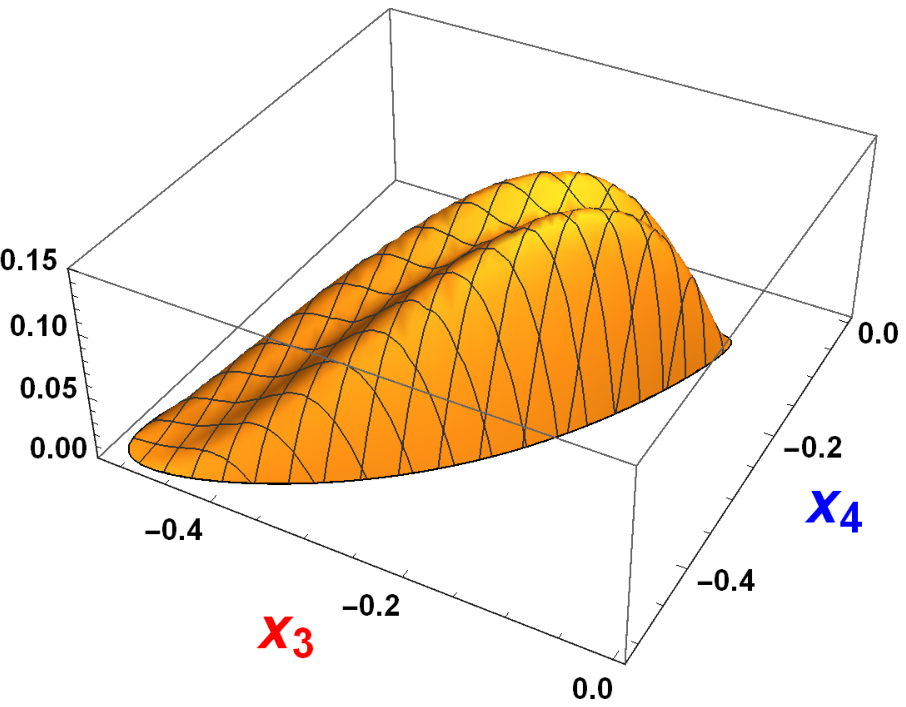}
\hspace{0.3cm}
\includegraphics[width=0.21\textwidth]{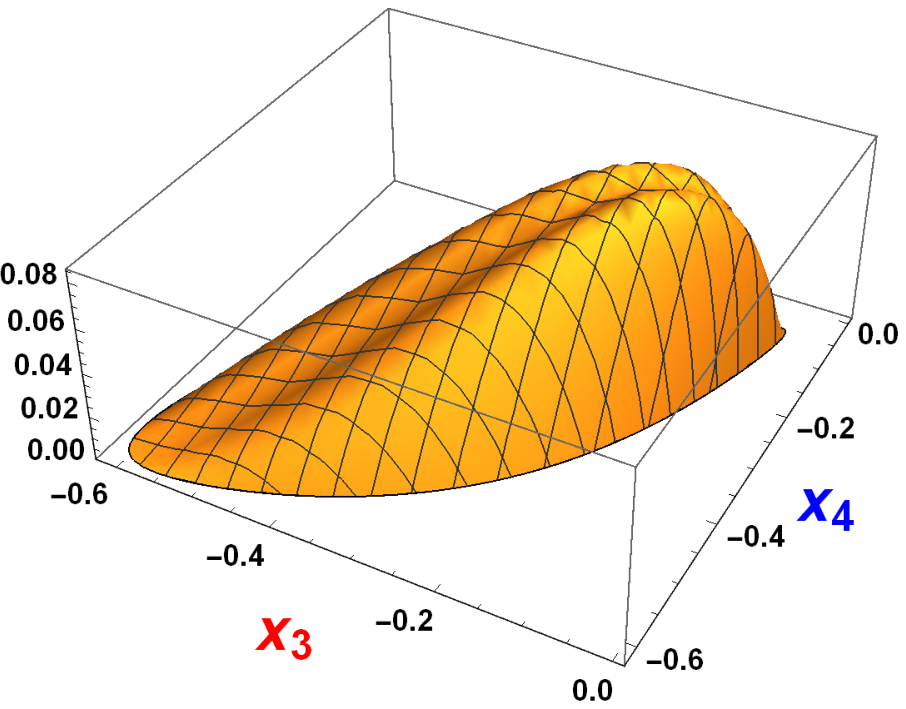}
\hspace{0.3cm}
\includegraphics[width=0.21\textwidth]{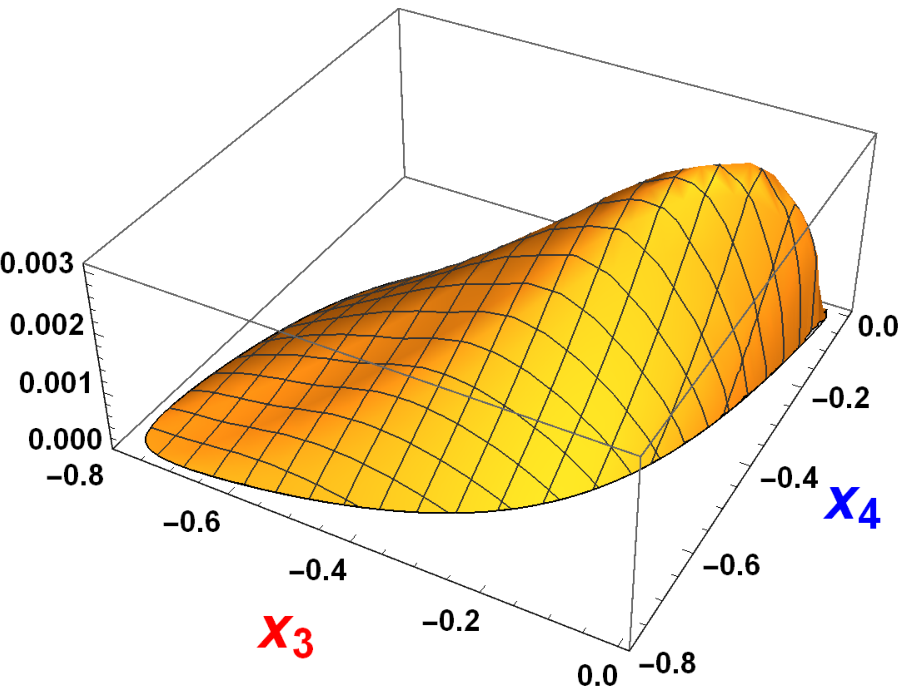}
\hspace{0.3cm}
\includegraphics[width=0.21\textwidth]{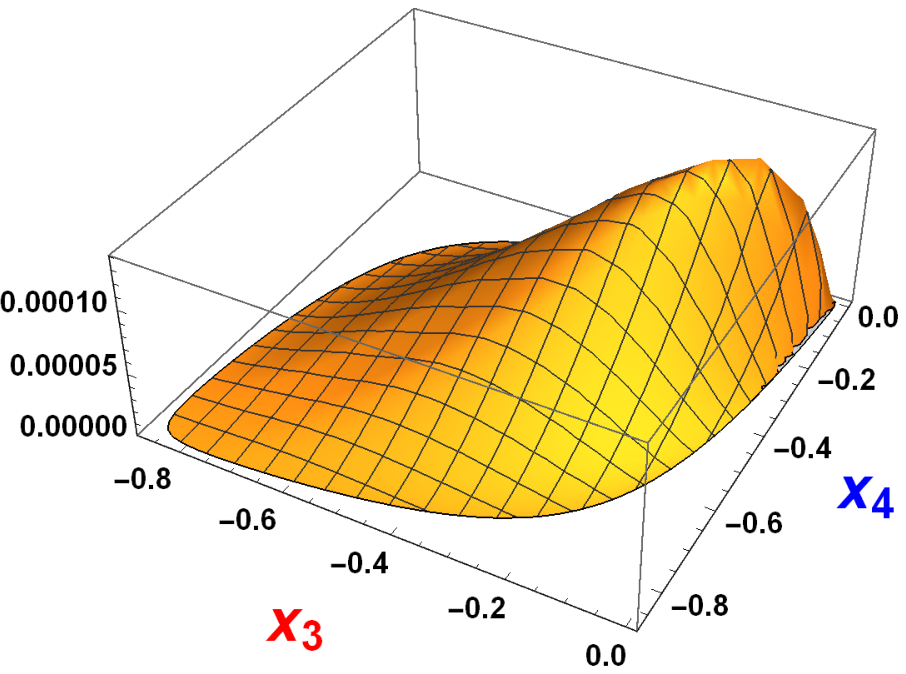}
\vspace{0.5cm}
\includegraphics[width=0.21\textwidth]{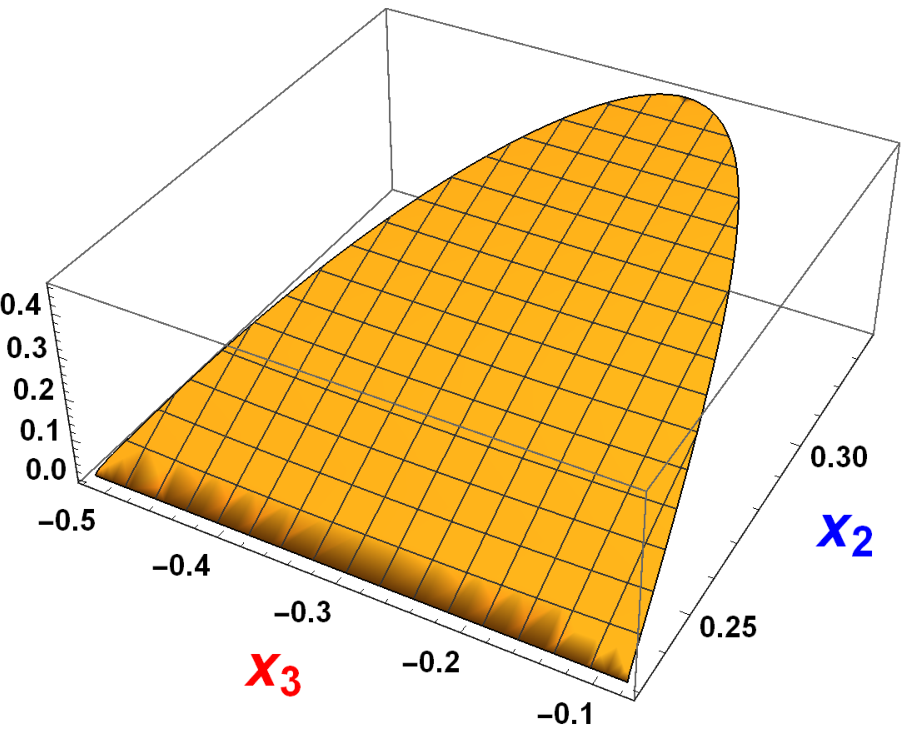}
\hspace{0.3cm}
\includegraphics[width=0.21\textwidth]{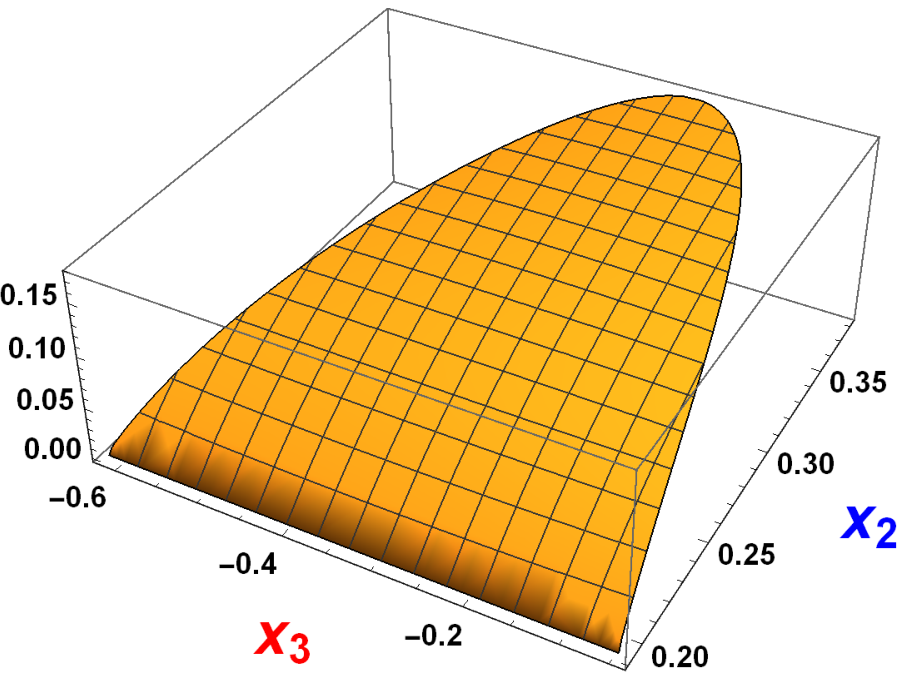}
\hspace{0.3cm}
\includegraphics[width=0.21\textwidth]{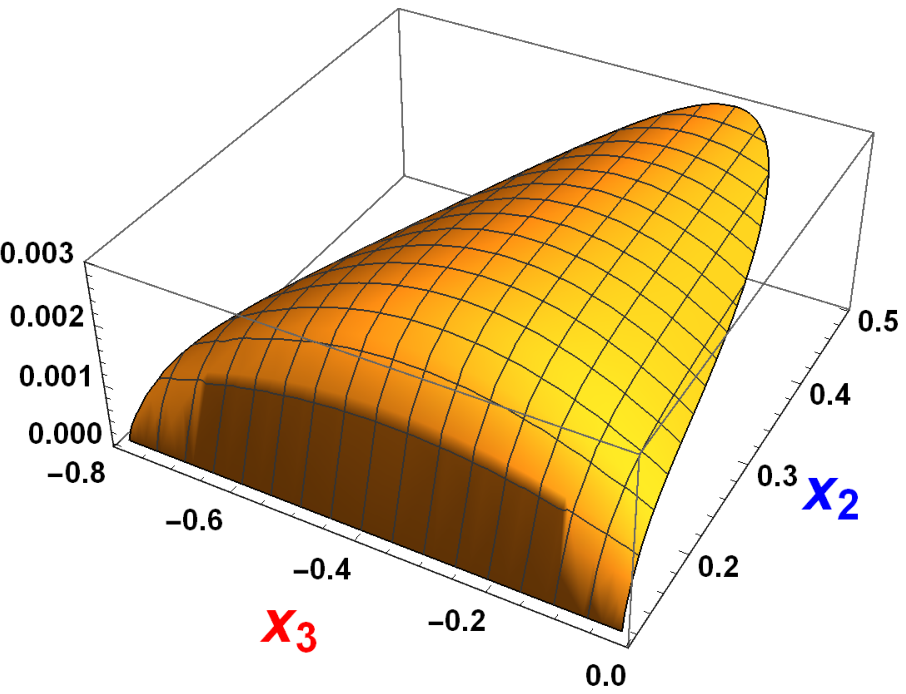}
\hspace{0.3cm}
\includegraphics[width=0.21\textwidth]{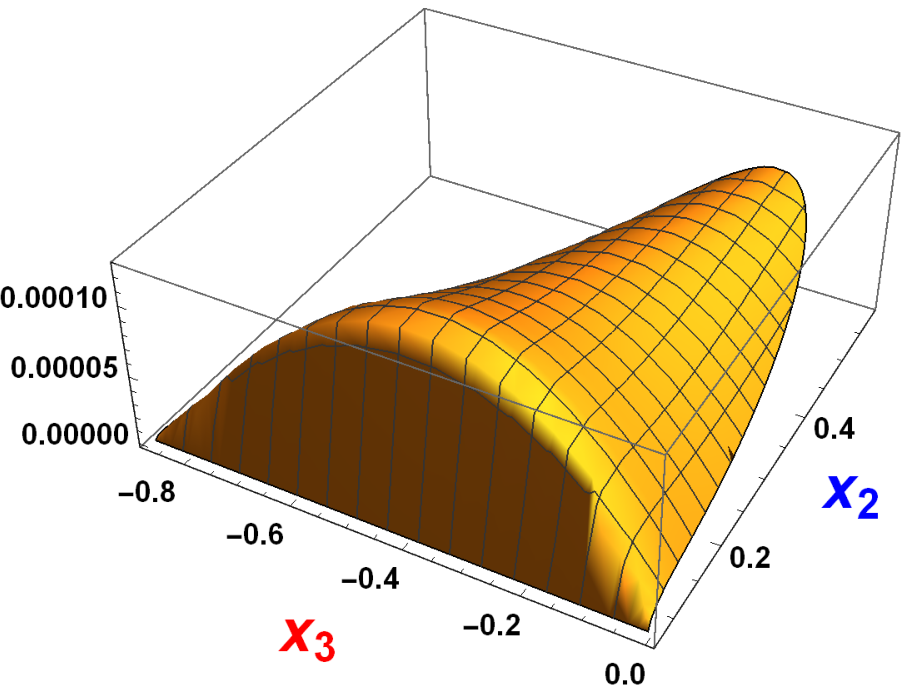}
\vspace{0.3cm}
\includegraphics[width=0.21\textwidth]{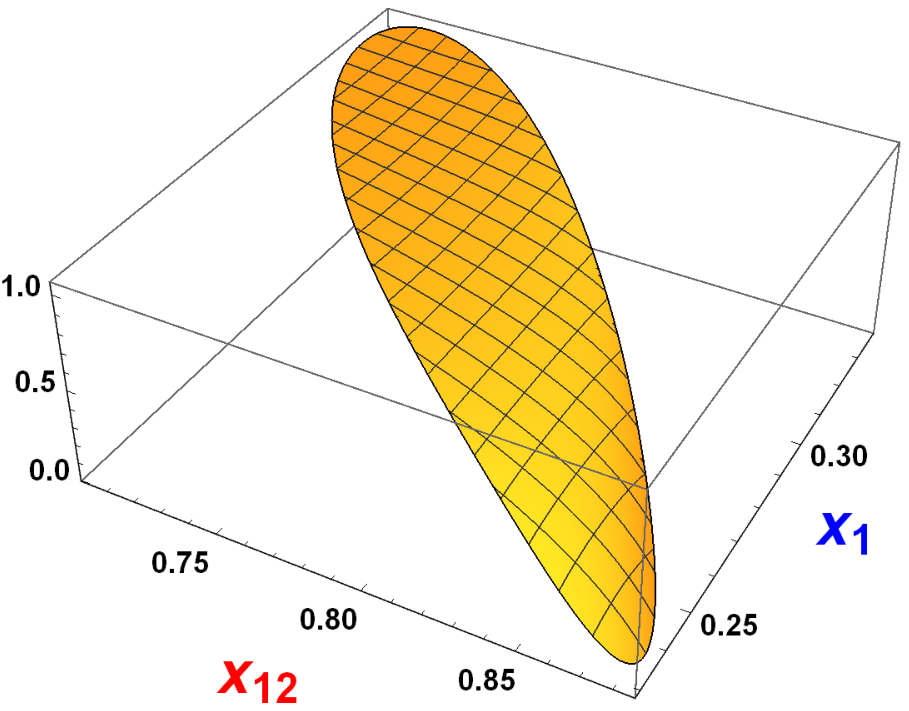}
\hspace{0.3cm}
\includegraphics[width=0.21\textwidth]{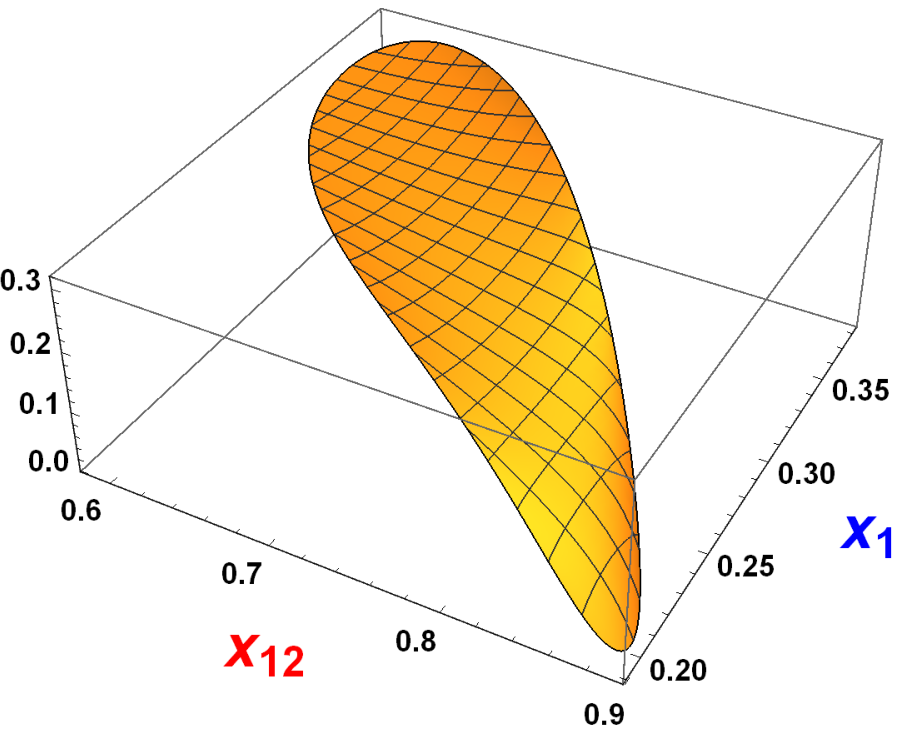}
\hspace{0.3cm}
\includegraphics[width=0.21\textwidth]{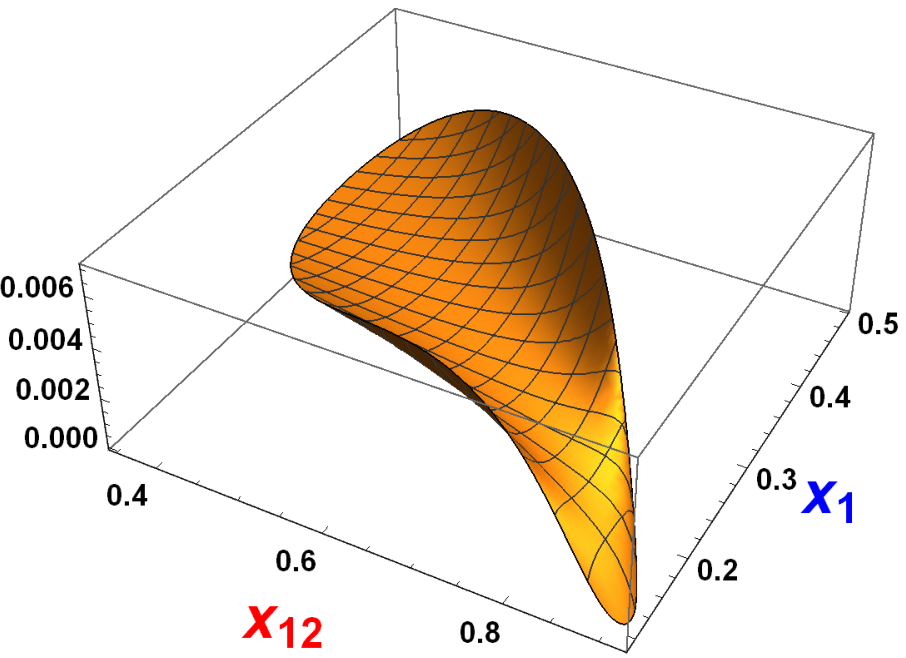}
\hspace{0.3cm}
\includegraphics[width=0.21\textwidth]{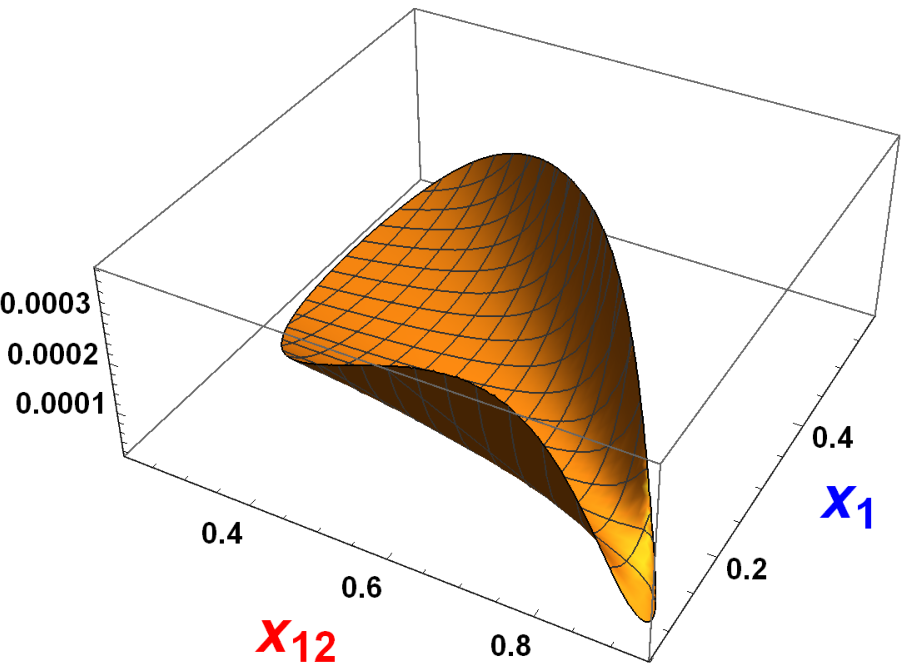}
 \parbox[t]{0.9\textwidth}{\caption{Double differential distributions of the process (\ref{eq:1}),  for the $\pi^0 p \bar p$-channel and the 'old version' of the form factor
 parametrization  over the dimensionless invariant variables:  $(x_1,\,x_2),\,\,x_{1,2}=s_{1,2}/s$ (first row), $(x_3,\,x_4),\,\,x_{3,4}=t_{1,2}/s$ (second row), 
 $(x_3,\,x_2)$ (third row), and $(x_1,\,x_{12}),\,\,x_{12}=s_{12}/s $( forth row),  and for different values of $s$:  $s=5$ GeV$^2$ (first column), $s=6$ GeV$^2$ (second column), $s=10$ GeV$^2$ (third column) and
 $s=16$ GeV$^2$  (fourth column). }\label{fig.5}}
\end{figure}

In Figs.~5,\,6 we plot the double differential distributions for the $\pi^0 p \bar p-$ and $\pi^0 n \bar n-$channel at different energies of the colliding electron and positron beams. At chosen parametrization of the form factors (here we use the old version), the differential
cross section of the $\pi^0 n \bar n-$channel is systematically larger than the $\pi^0 p \bar p-$channel (the same is valid also for the new version). In our numerical calculations we chose the value of the neutral pion-nucleon constant interaction as $g^2_{\pi^0 N N }/(4\pi) = 13.5$ \cite{Machleidt:2000ge}.

The integration of the double differential cross section (\ref{eq:dsigmas1s2}) with respect to the variable $s_2$ at fixed value of $s_1$ or $s_{12}$,
in the limits (\ref{eq:s2pm}) or (\ref{eq:s1s12pm}), gives the single differential cross section over $s_1$ or $s_{12}$, respectively.
\ba
\frac{d\,\sigma}{d\,s_1} &=& \frac{\alpha^2\,g^2_{\pi p p}}{24\pi\,N}\bigg\{\left|G_M^p-G_E^p\right|^2\bigg[2M^2D\bigg(\frac{K_1}{s_1^2}-
\frac{2m^2s(s-4M^2)(s_1-M^2)^2}{K}\bigg)+4M^2(s_1-M^2)K_L\,L\bigg]+\nn\\
&&
\hspace{-1truecm}+\left|4M^2G_E^p-sG_M^p\right|^2\bigg[-\frac{D K_2}{s_1 K} + 2(s_1-M^2)(s-2s_1-2M^2+2m^2)L\bigg]+\nn\\
&&
\hspace{-1truecm}+\left|G_M^p\right|^2(s-4M^2)^2\bigg[D\bigg(\frac{K_3}{s_1^2}-\frac{3m^2s(s_1-M^2)^2}{K}\bigg) + 2(s_1-M^2)[(s_1-M^2)^2-m^2s]L\bigg]\bigg\}\nn\\
N&=&s^3(s-4M^2)^2(s_1-M^2)^2, \ K=M^6 -2M^4s_1+M^2(s_1^2-3m^2s) +m^2s(s-s_1+m^2),\nn\\
D&=&\sqrt{M^4-2M^2(s+s_1)+(s-s_1)^2}\cdot\sqrt{M^4-2M^2(s_1+m^2)+(s_1-m^2)^2},\nn\\
L&=&\ln\bigg(\frac{C+D}{C-D}\bigg), \ \ C=M^4-M^2(s+m^2)+s_1(s-s_1)+m^2(s+s_1),\nn\\
K_1&=&-M^6(s+4s_1)+M^4[s^2+(s_1+m^2)(s+4s_1)]+M^2[s_1(4s_1^2-2s^2+s s_1)-\nn\\
&&-m^2(s^2-2s s_1+8s_1^2)]+s_1[s_1(s^2-s s_1-4s_1^2)+m^2(-s^2+5s s_1+4s_1^2)],\nn\\
K_2&=&-M^6(s-8s_1)+M^4s_1(s-16s_1) +M^2[s_1^2(s+8s_1)+3m^2s(s-4s_1)]\nn\\
&&
-s[m^4(s-4s_1)+m^2(s^2-5s s_1+4s_1^2) + s_1^3],\nn\\
K_3&=&-M^6+M^4(s+5s_1+m^2)-M^2[s_1(2s+7s_1)+m^2(s+2s_1)]+\nn\\
&& s_1[s_1(s+3s_1)+m^2(s_1-2s)],\nn\\
K_L&=&-4M^6+M^4(s+8s_1)+M^2[4m^2s-2s_1(s+2s_1)] +s[s_1^2-2m^2(s-2s_1+2m^2)].\label{eq:eqs1}
\ea

The distribution, over the invariant mass squared of the proton-antiproton system, can be written in the following  form
\ba
\frac{d\,\sigma}{d\,s_{12}}&=&\frac{\alpha^2\,g^2_{\pi^0 p p }}{24\,\pi s^3(s-4M^2)^2}\bigg\{8M^2\left|G_M^p-G_E^p \right|^2\bigg(\frac{D_1\,Q_1}{s_{12}Q}
-\frac{Q_L}{s+m^2-s_{12}}L_1 \bigg)+\nn\\
&&+2m^2\left|4M^2G_E^p-sG_M^p\right|^2\bigg[\frac{(s-4M^2)D_1}{Q}-\frac{2(s+4M^2-2s_{12})}{s+m^2-s_{12}}L_1\bigg]+\nn\\
&&
+\left|G_M^p\right|^2(s-4M^2)^2\bigg[-\frac{6 m^2 s D_1}{Q}+\frac{4[m^2(s-2s_{12})+m^4+(s-s_{12})^2]}{s+m^2-s_{12}}L_1\bigg]\bigg\}\nn\\
D_1&=&\sqrt{s_{12}(s_{12}-4M^2)l_0}, \ \ l_0= (s+s_{12}-m^2)^2-4s s_{12}, \ \ Q=l_0 M^2+m^2s s_{12},\nn\\
Q_1&=&-m^2s s_{12}(3s-2s_{12})+2M^2[2m^2(s^2+s s_{12}-s_{12}^2)-m^4(s-s_{12})- (s-s_{12})^3],
\nn\\
Q_L&=&4M^2[m^4-2m^2 s_{12}+(s-s_{12})^2]-s[m^4 + m^2(4s-6s_{12})+(s-s_{12})^2], \nn\\
L_1&=&\ln\bigg(\frac{[D_1+s_{12}(s+m^2-s_{12})]^2}{4 s_{12} Q}\bigg).\label{eq:dss12}
\ea

\begin{figure}
\centering
\includegraphics[width=0.21\textwidth]{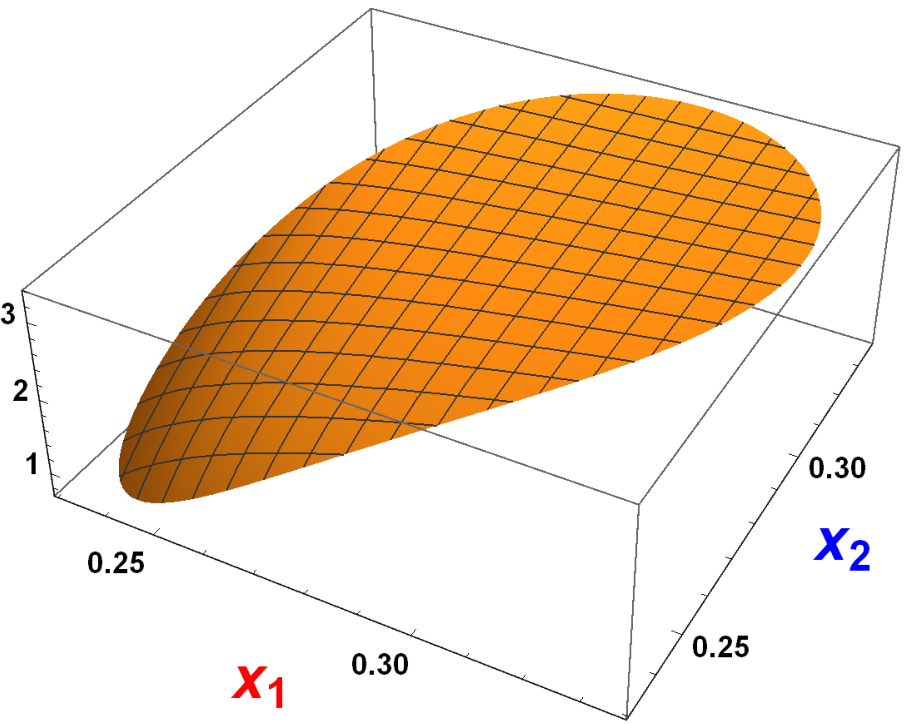}
\hspace{0.3cm}
\includegraphics[width=0.21\textwidth]{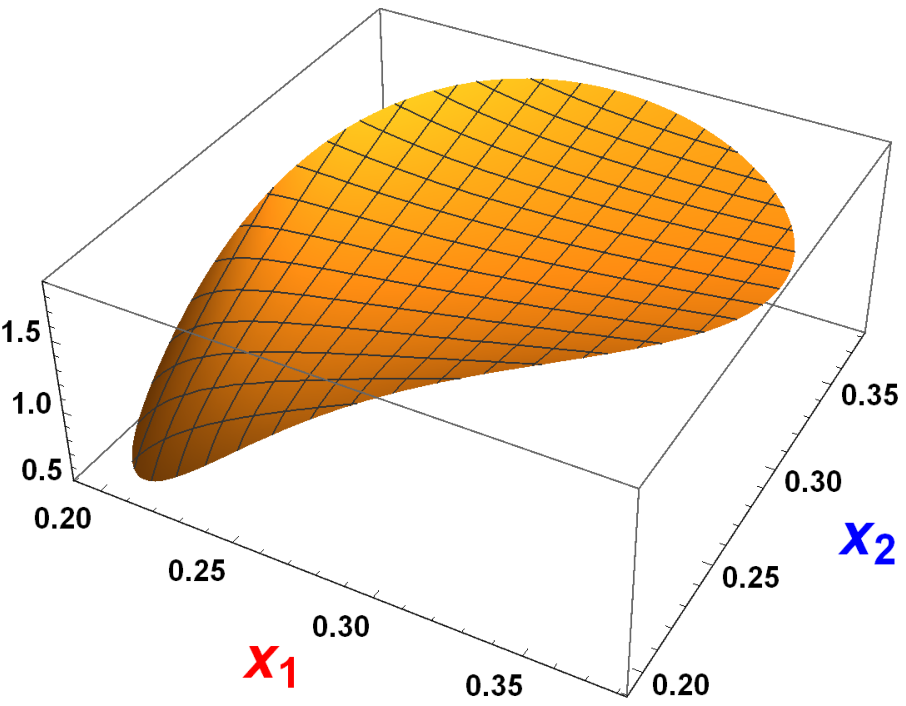}
\hspace{0.3cm}
\includegraphics[width=0.21\textwidth]{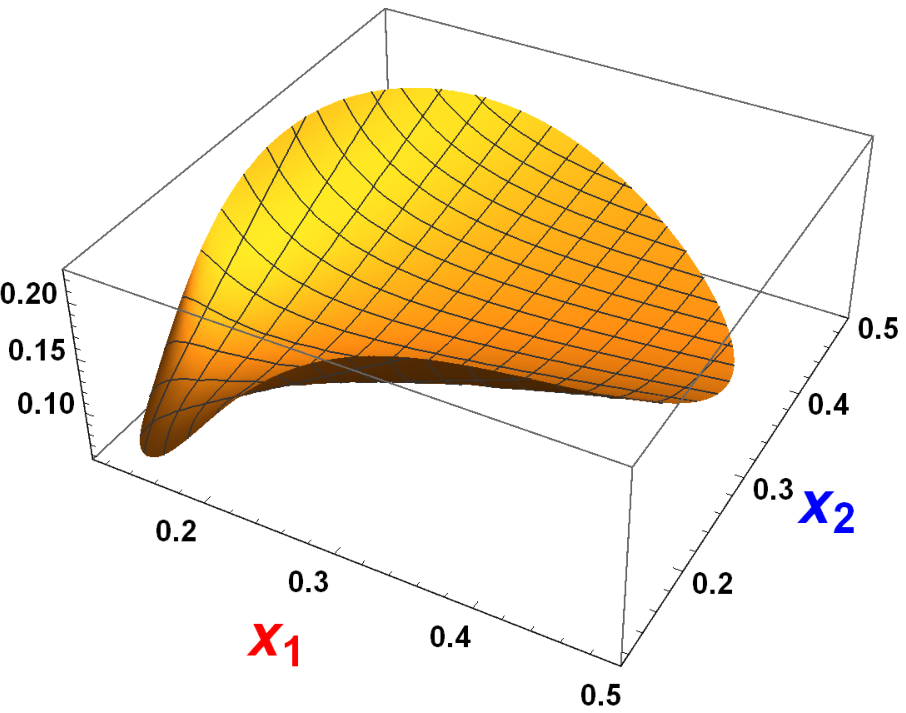}
\hspace{0.3cm}
\includegraphics[width=0.21\textwidth]{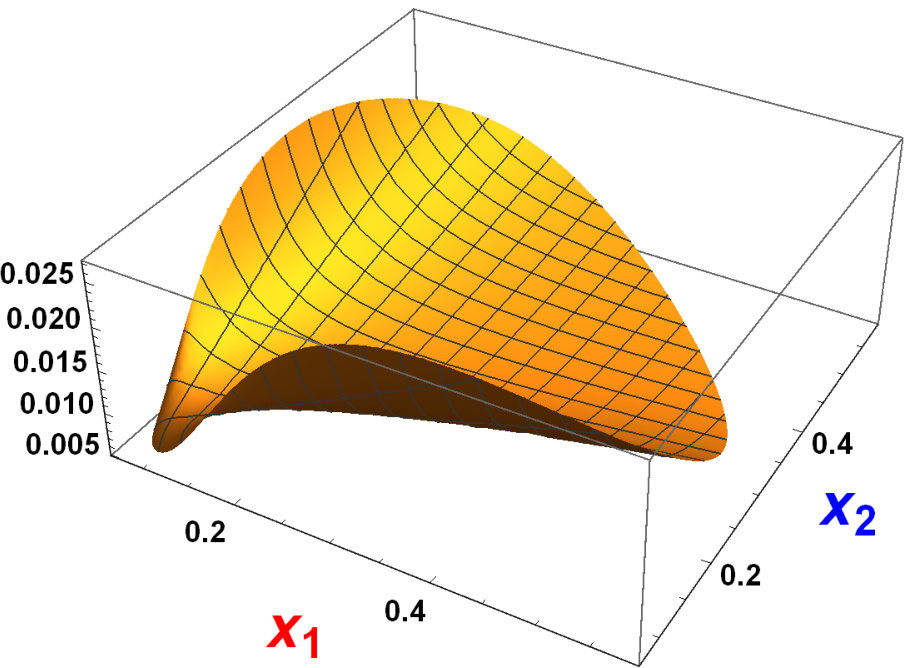}
\vspace{0.5cm}
\includegraphics[width=0.21\textwidth]{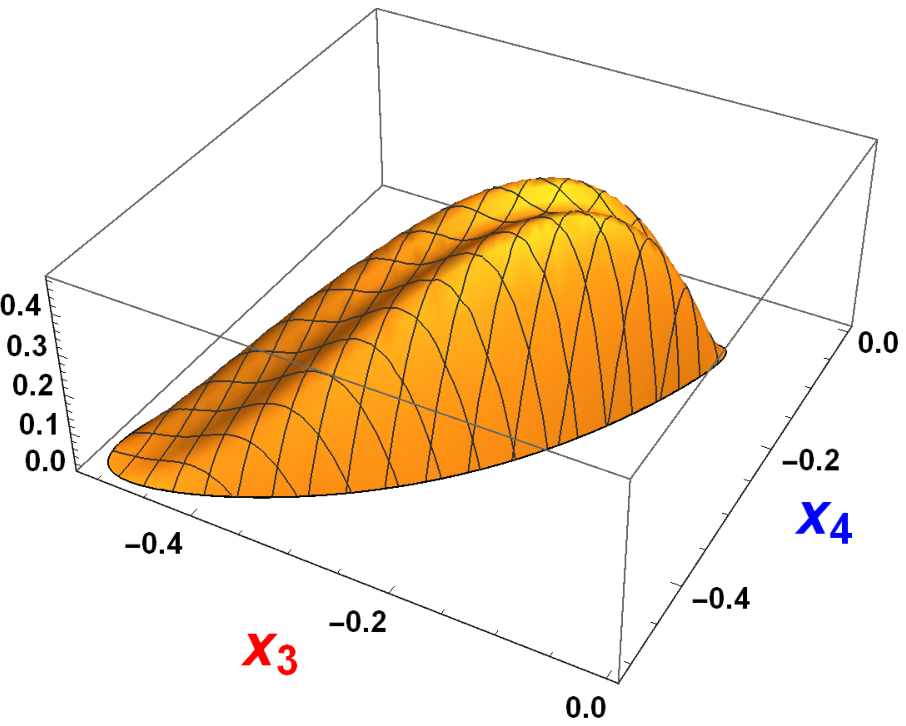}
\hspace{0.3cm}
\includegraphics[width=0.21\textwidth]{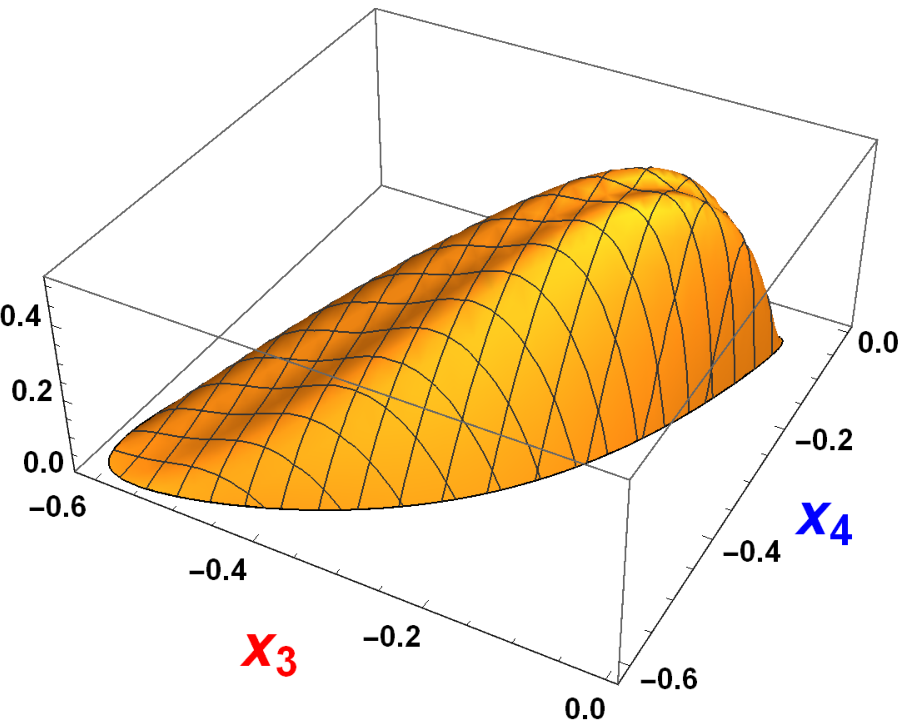}
\hspace{0.3cm}
\includegraphics[width=0.21\textwidth]{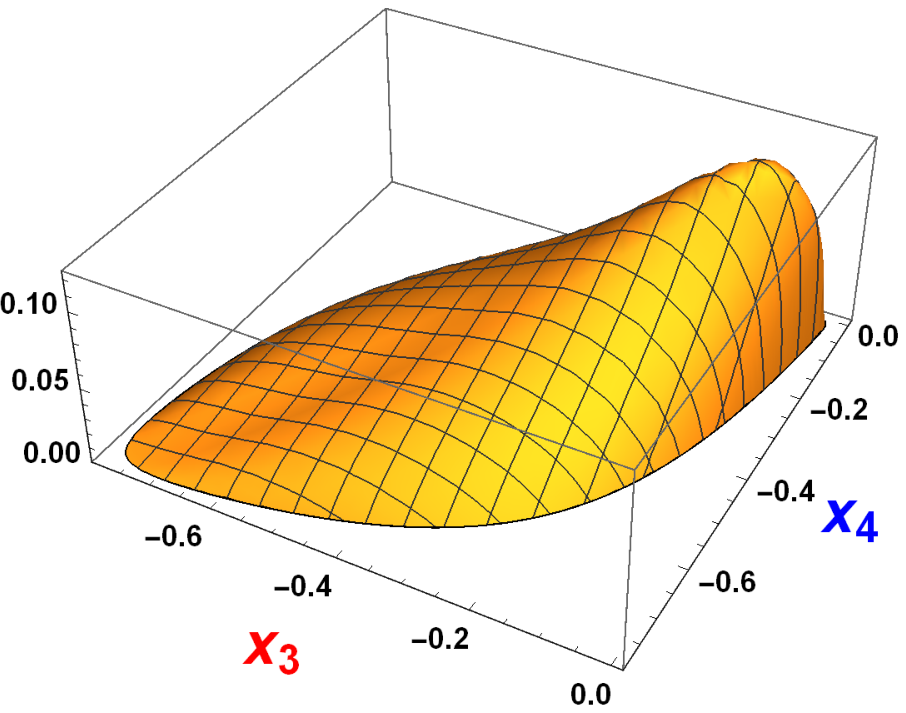}
\hspace{0.3cm}
\includegraphics[width=0.21\textwidth]{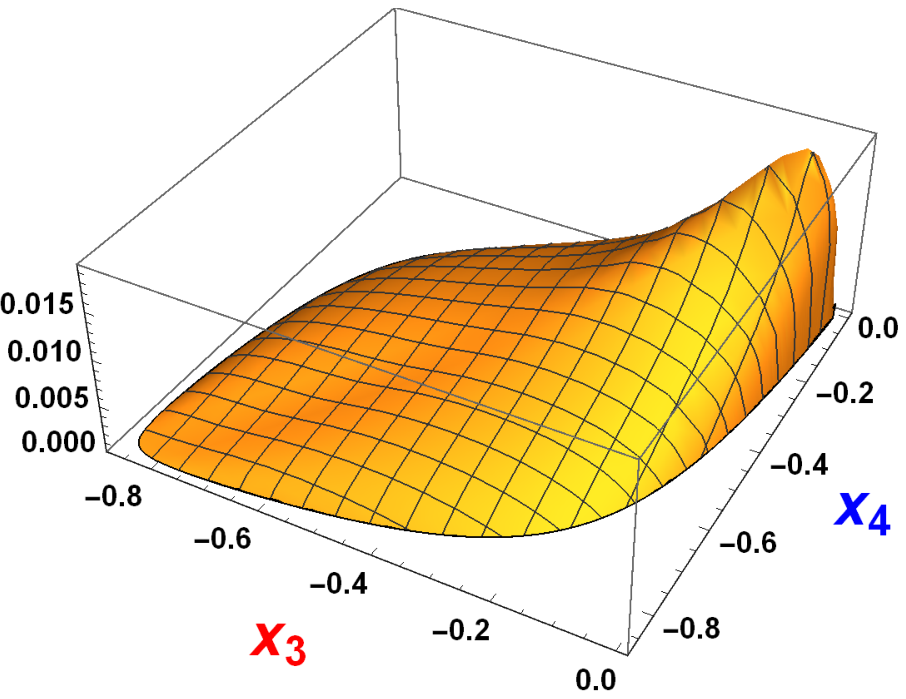}
\vspace{0.5cm}
\includegraphics[width=0.21\textwidth]{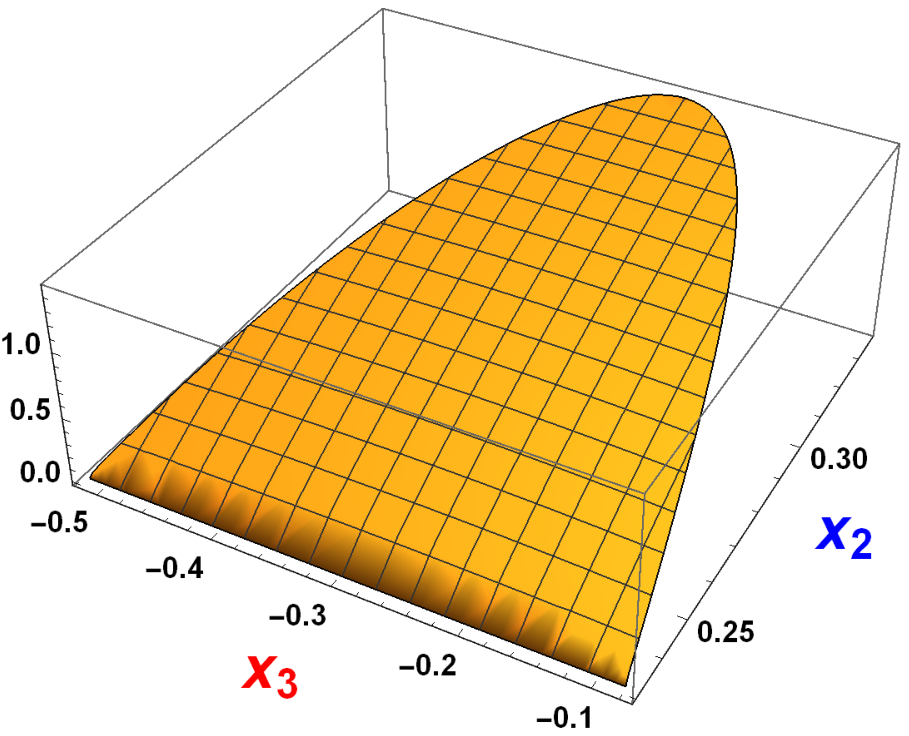}
\hspace{0.3cm}
\includegraphics[width=0.21\textwidth]{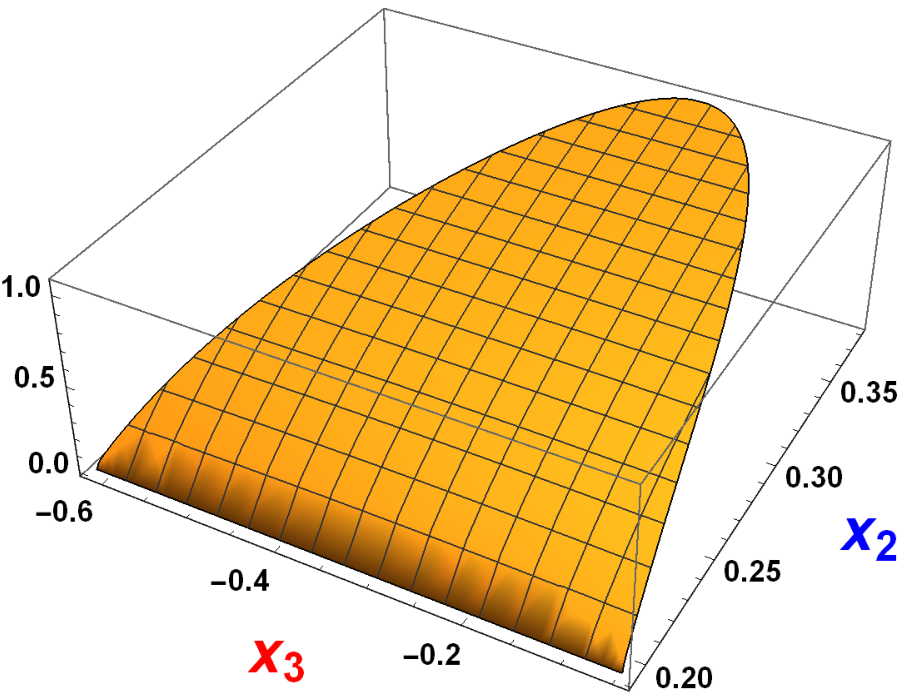}
\hspace{0.3cm}
\includegraphics[width=0.21\textwidth]{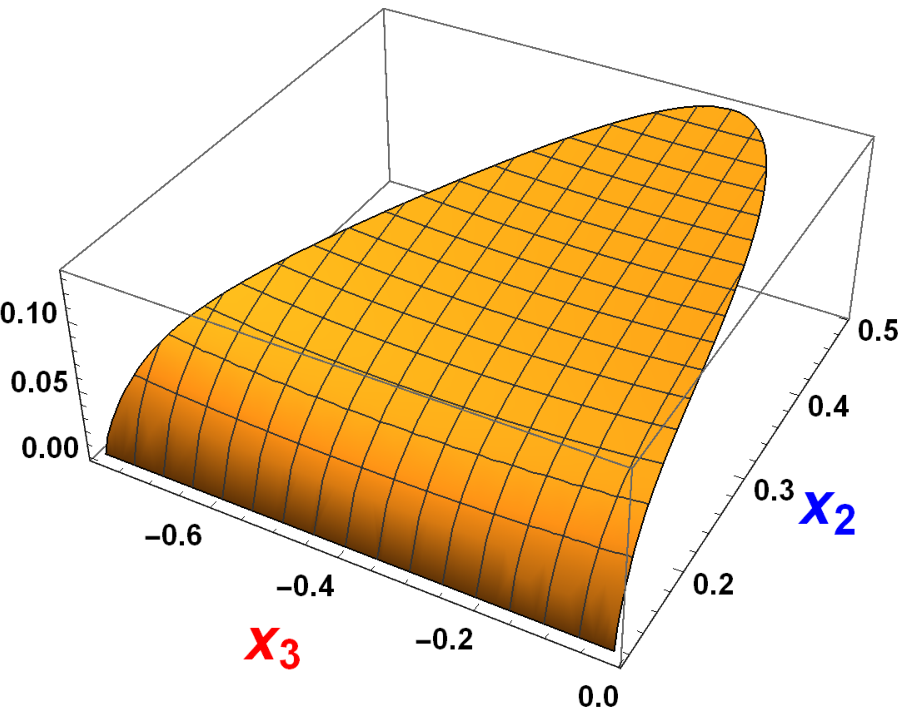}
\hspace{0.3cm}
\includegraphics[width=0.21\textwidth]{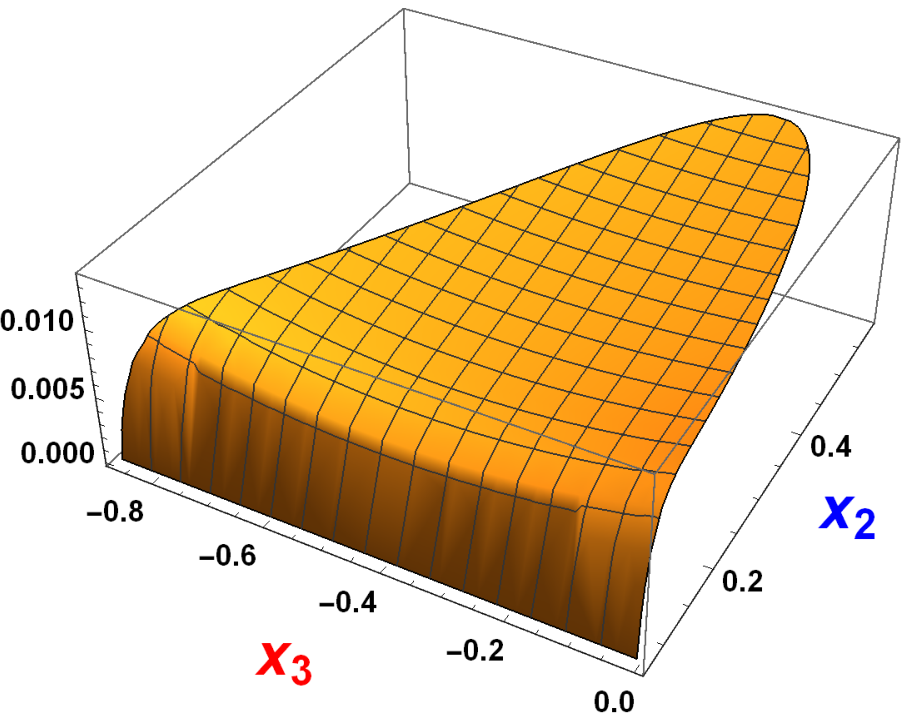}
\vspace{0.3cm}
\includegraphics[width=0.21\textwidth]{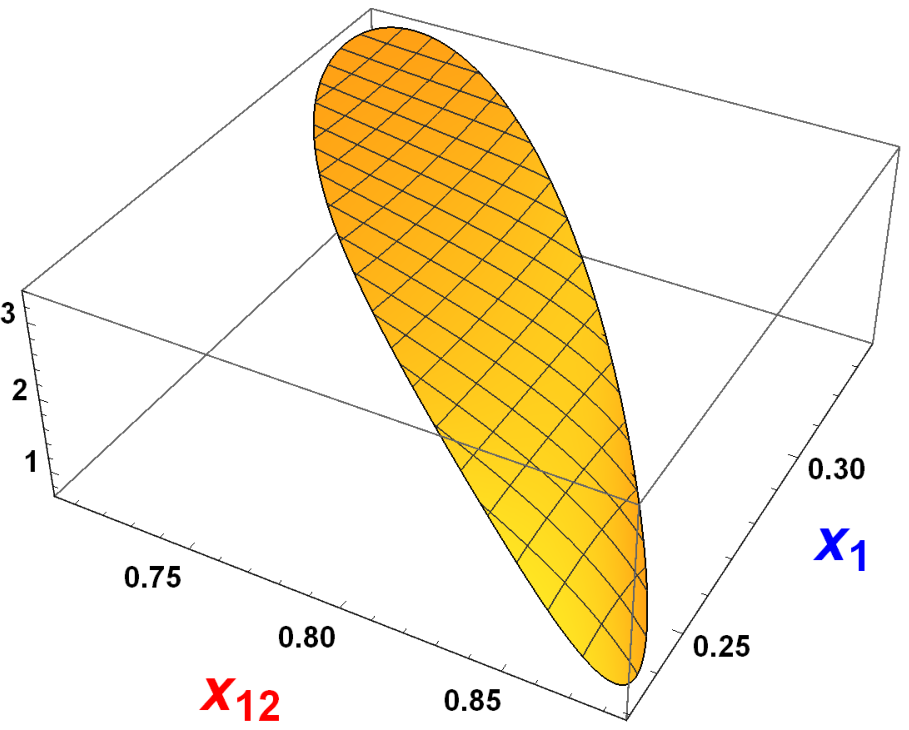}
\hspace{0.3cm}
\includegraphics[width=0.21\textwidth]{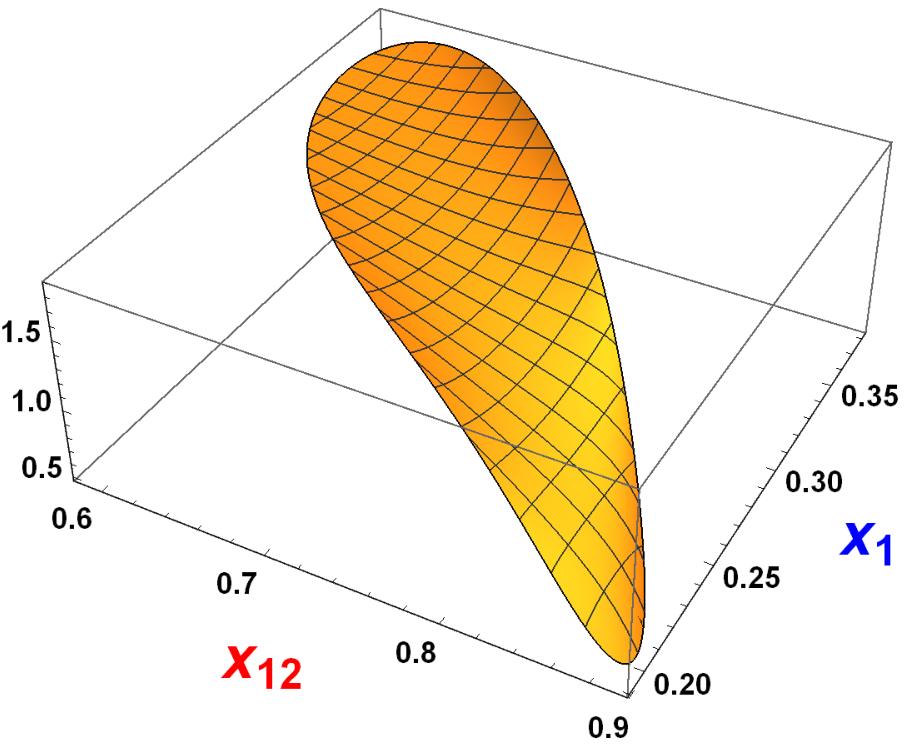}
\hspace{0.3cm}
\includegraphics[width=0.21\textwidth]{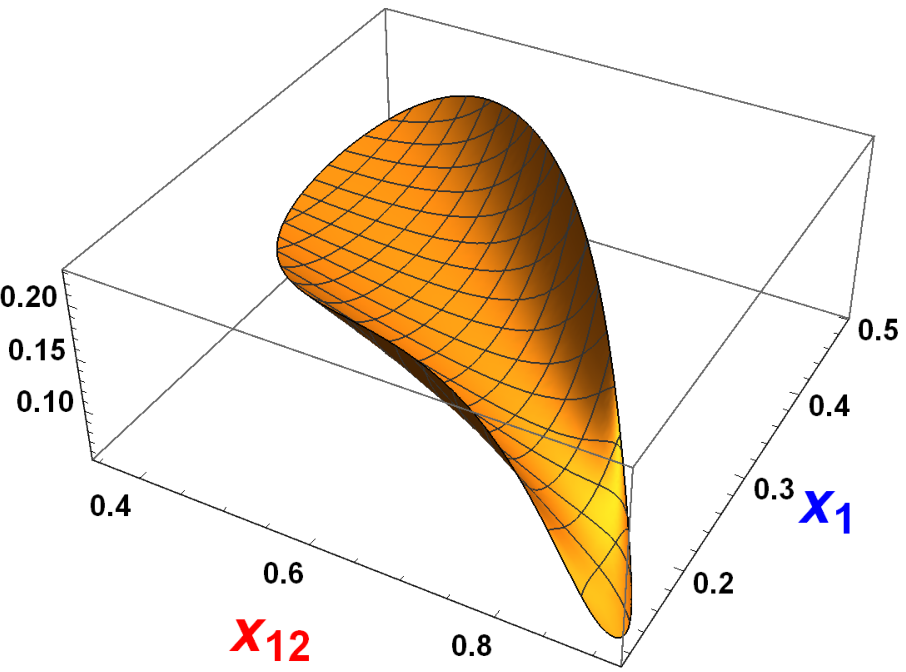}
\hspace{0.3cm}
\includegraphics[width=0.21\textwidth]{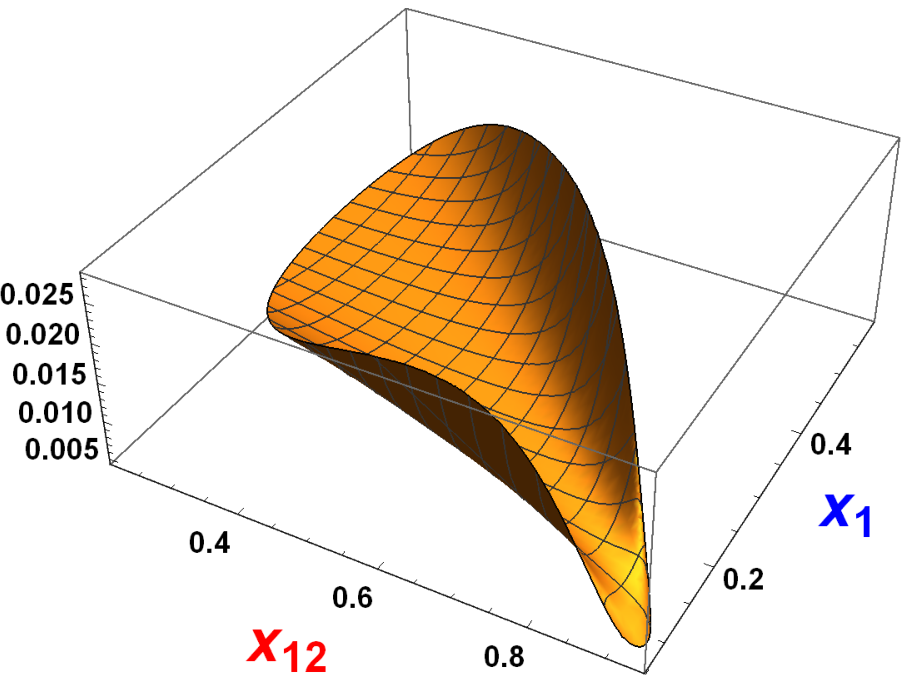}
 \parbox[t]{0.9\textwidth}{\caption{The same as in Fig.~5 but for the $\pi^0 n \bar n$-channel.}\label{fig.6}}
\end{figure}

%%%%%%%%%%%%%%%%%%%%%%%%%%%%%%%%%%%%%%%%%%%%%%%%%%%%%%%%%%%%%%%%%%%%%
%About form factors
%%%%%%%%%%%%%%%%%%%%%%%%%%%%%%%%%%%%%%%%%%%%%%%%%%%%%%%%%%%%%%%%%%%%%%%

As noted above, the $(t_1,\,t_2)$ double distribution is derived in analytical form but the single distribution over $t_1$ is obtained by numerical computation. In Figs.~7,8 the corresponding differential cross sections are plotted at different energies versus the dimensionless variables $x_1,\, x_3$ and $x_{12}$ for the 
$\pi^0 p \bar p-$ and $\pi^0 n \bar n-$channels, respectively.

\begin{figure}
\centering
\includegraphics[width=0.22\textwidth]{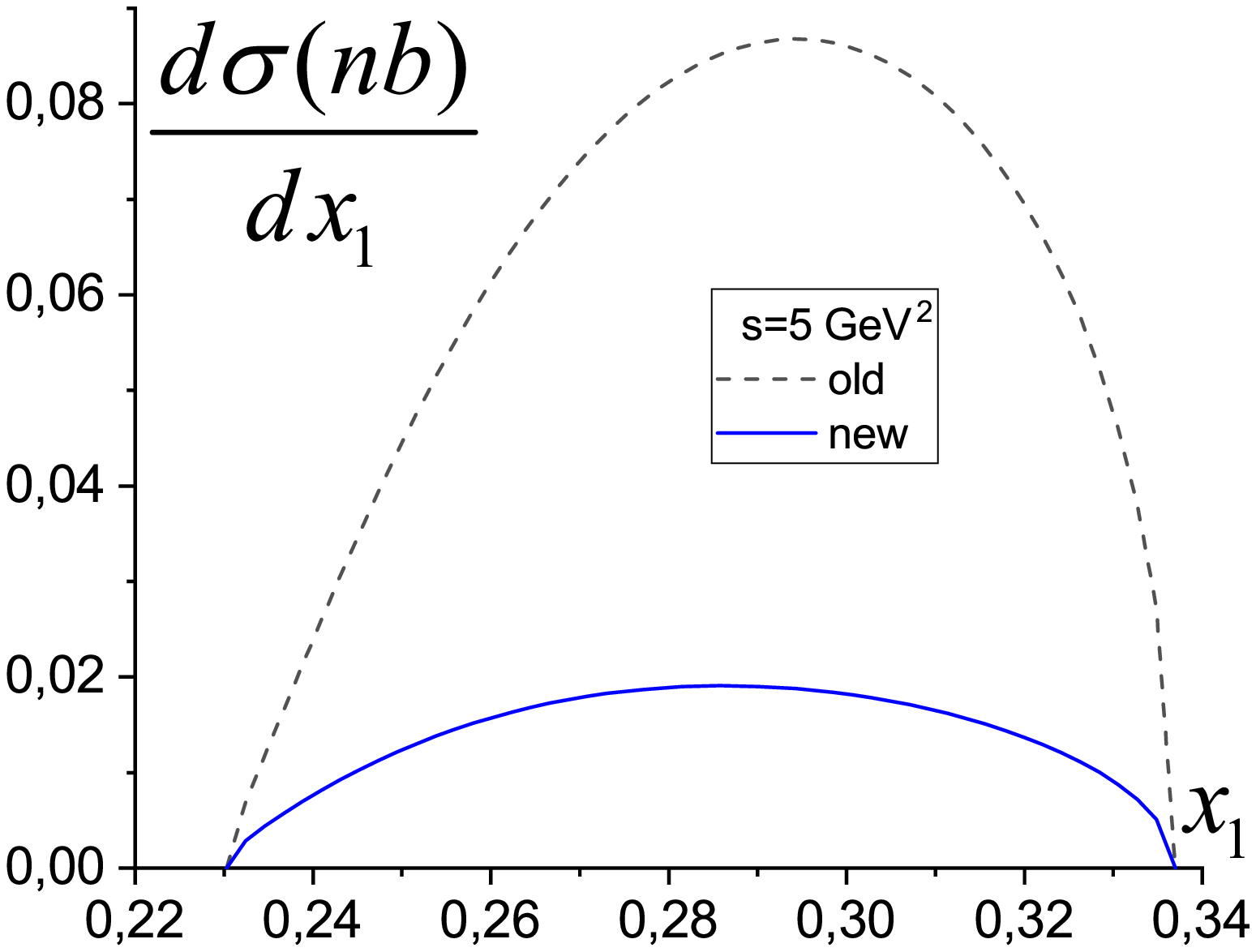}
\hspace{0.1cm}
\includegraphics[width=0.22\textwidth]{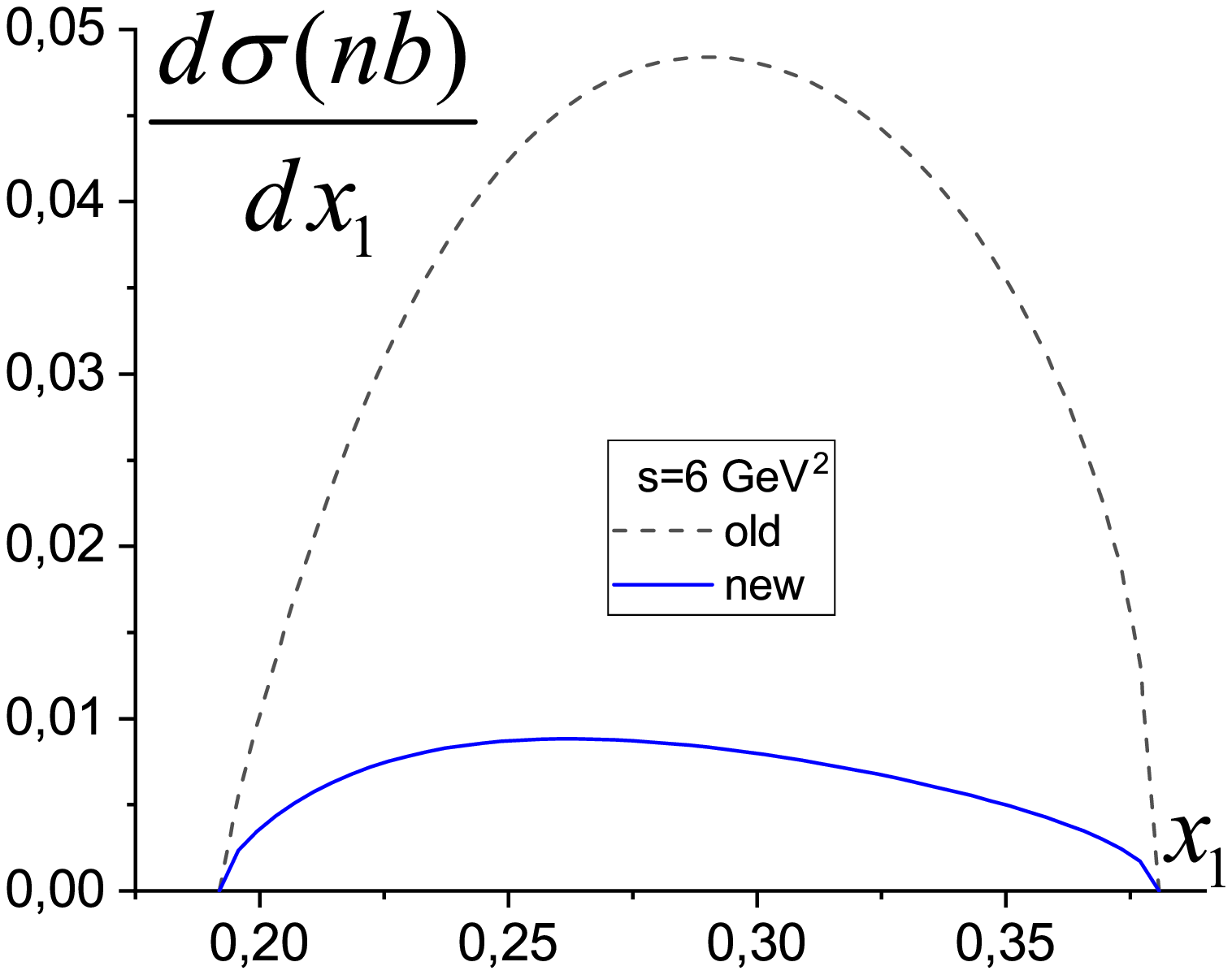}
\hspace{0.1cm}
\includegraphics[width=0.22\textwidth]{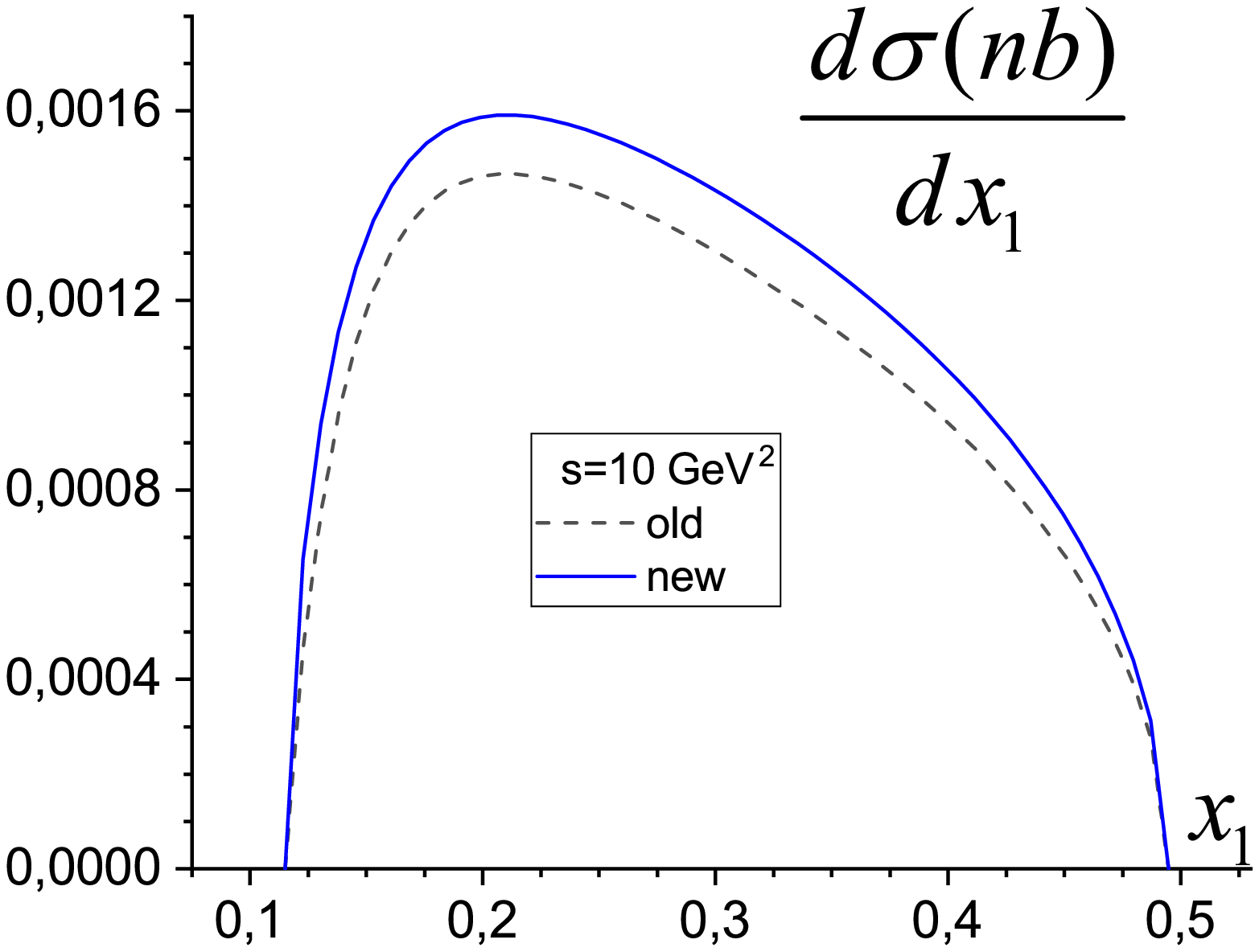}
\hspace{0.5cm}
\includegraphics[width=0.22\textwidth]{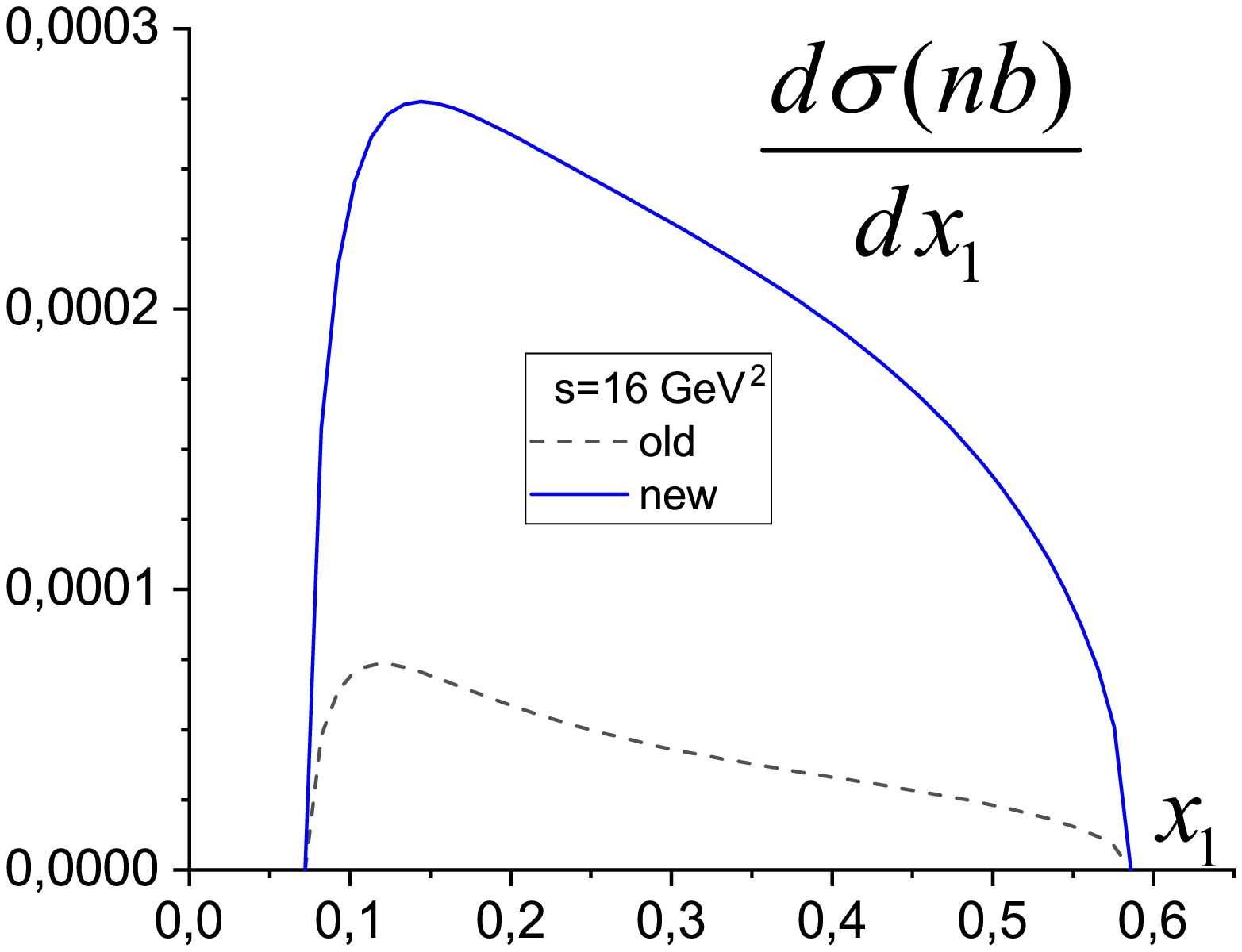}

\vspace{0.31cm}
\includegraphics[width=0.22\textwidth]{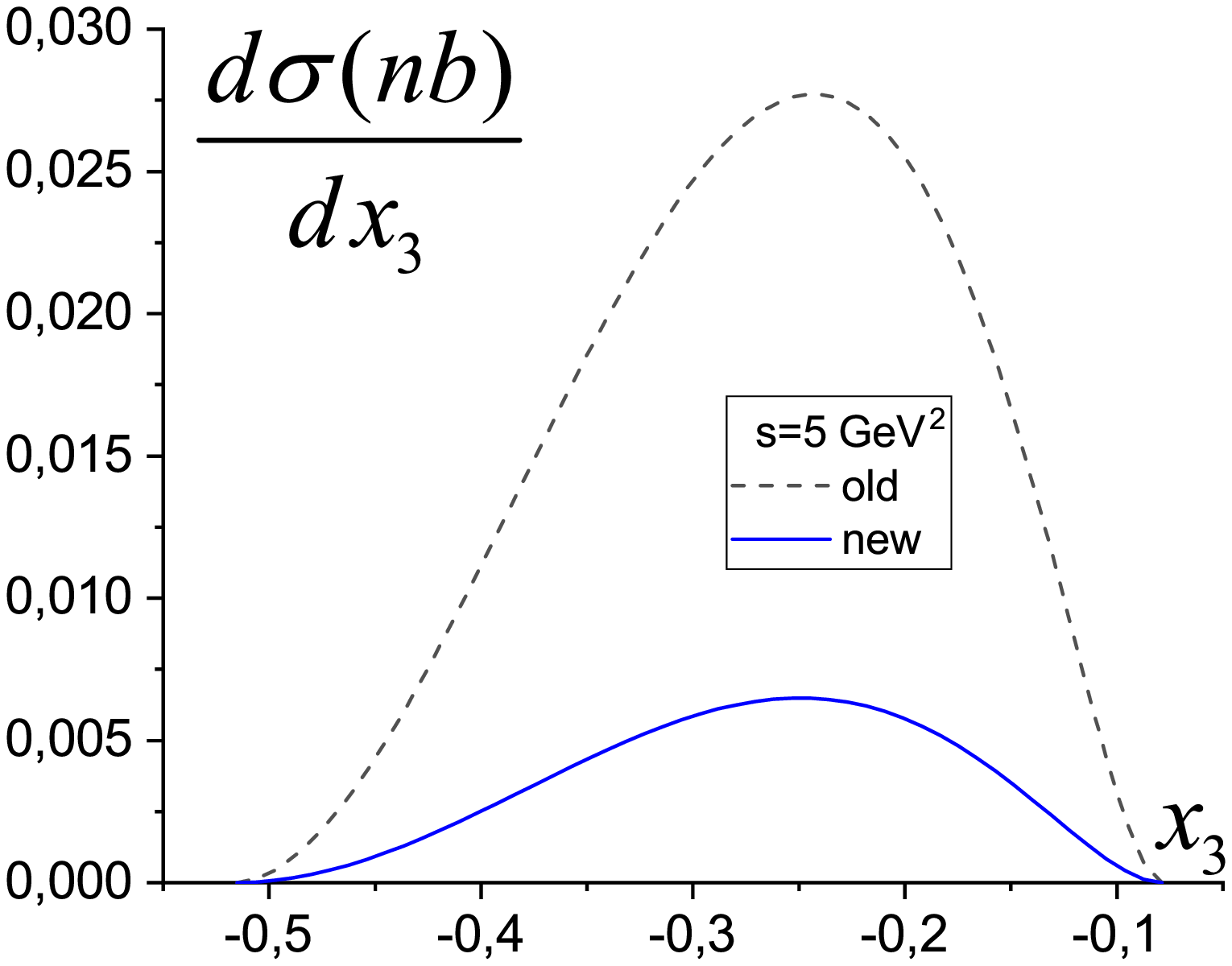}
\hspace{0.1cm}
\includegraphics[width=0.22\textwidth]{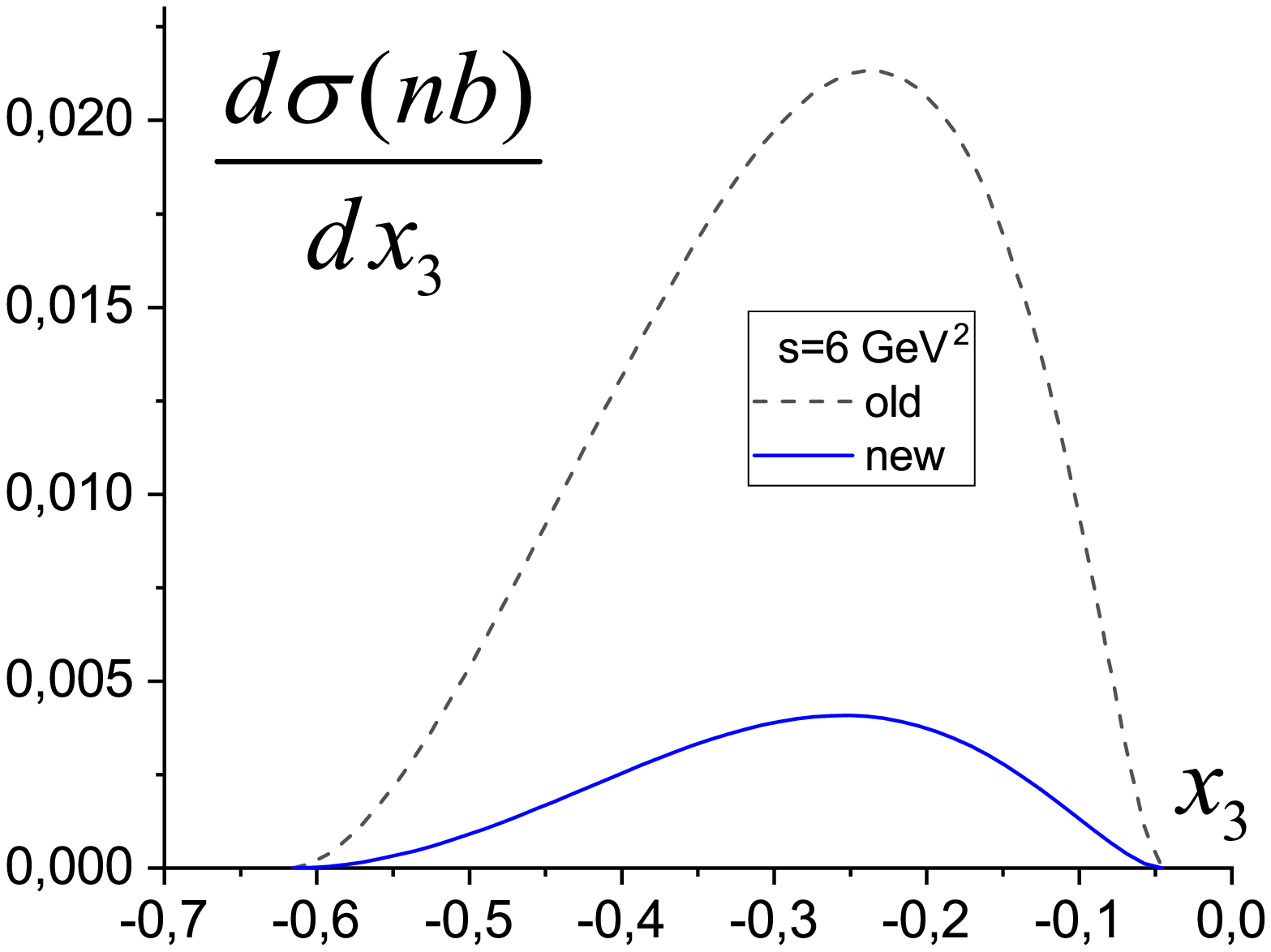}
\hspace{0.1cm}
\includegraphics[width=0.22\textwidth]{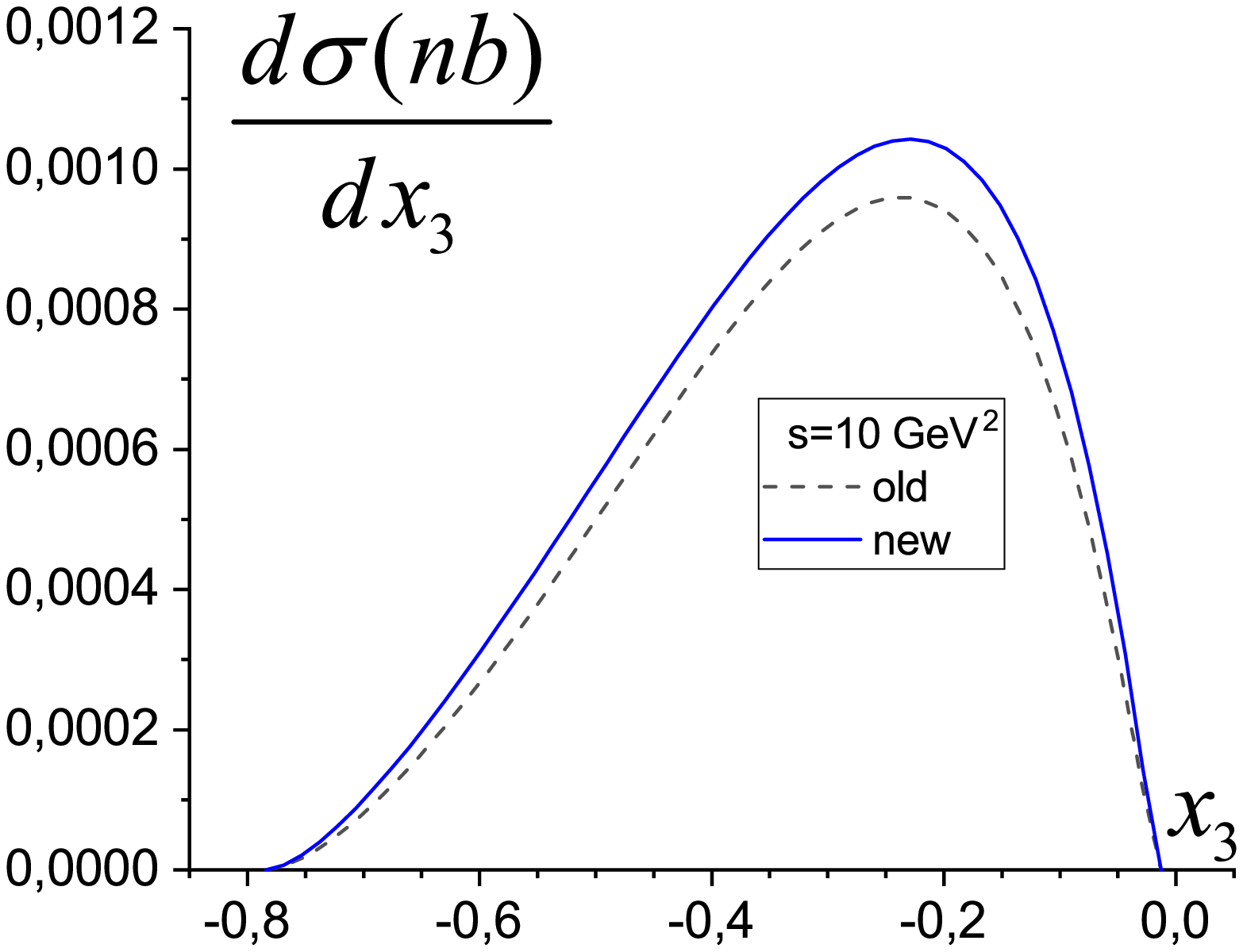}
\hspace{0.5cm}
\includegraphics[width=0.22\textwidth]{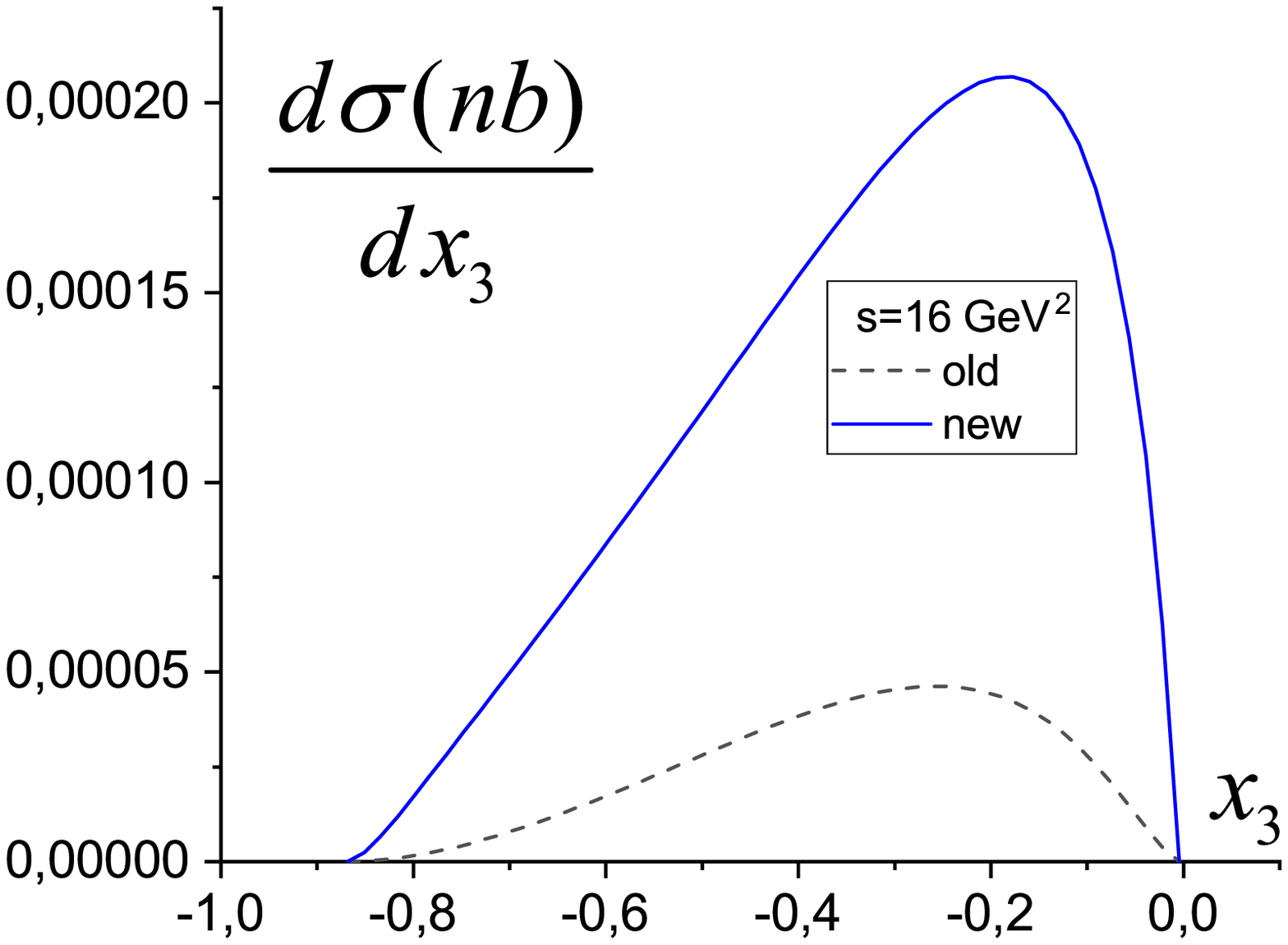}

\vspace{0.31cm}
\includegraphics[width=0.22\textwidth]{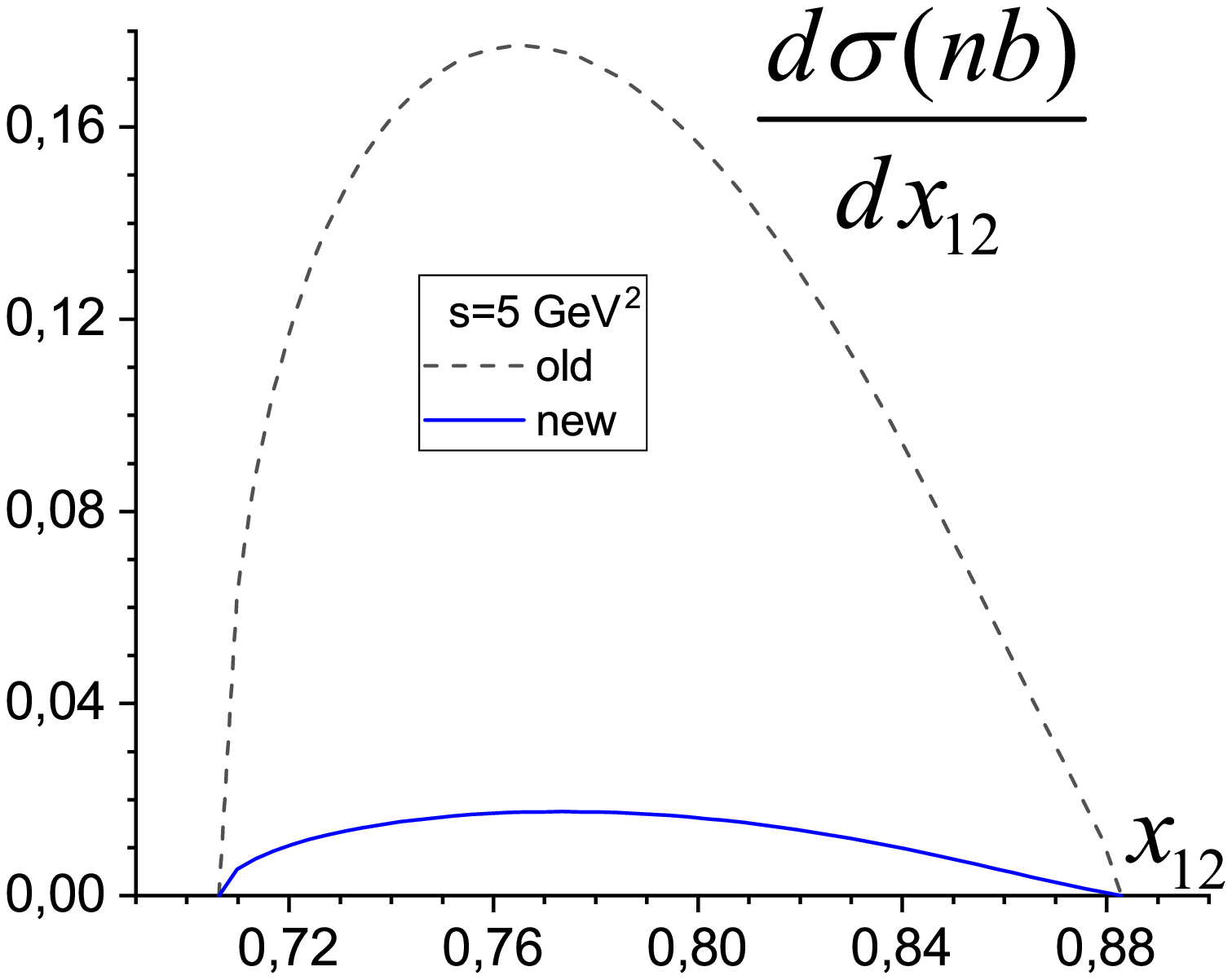}
\hspace{0.1cm}
\includegraphics[width=0.22\textwidth]{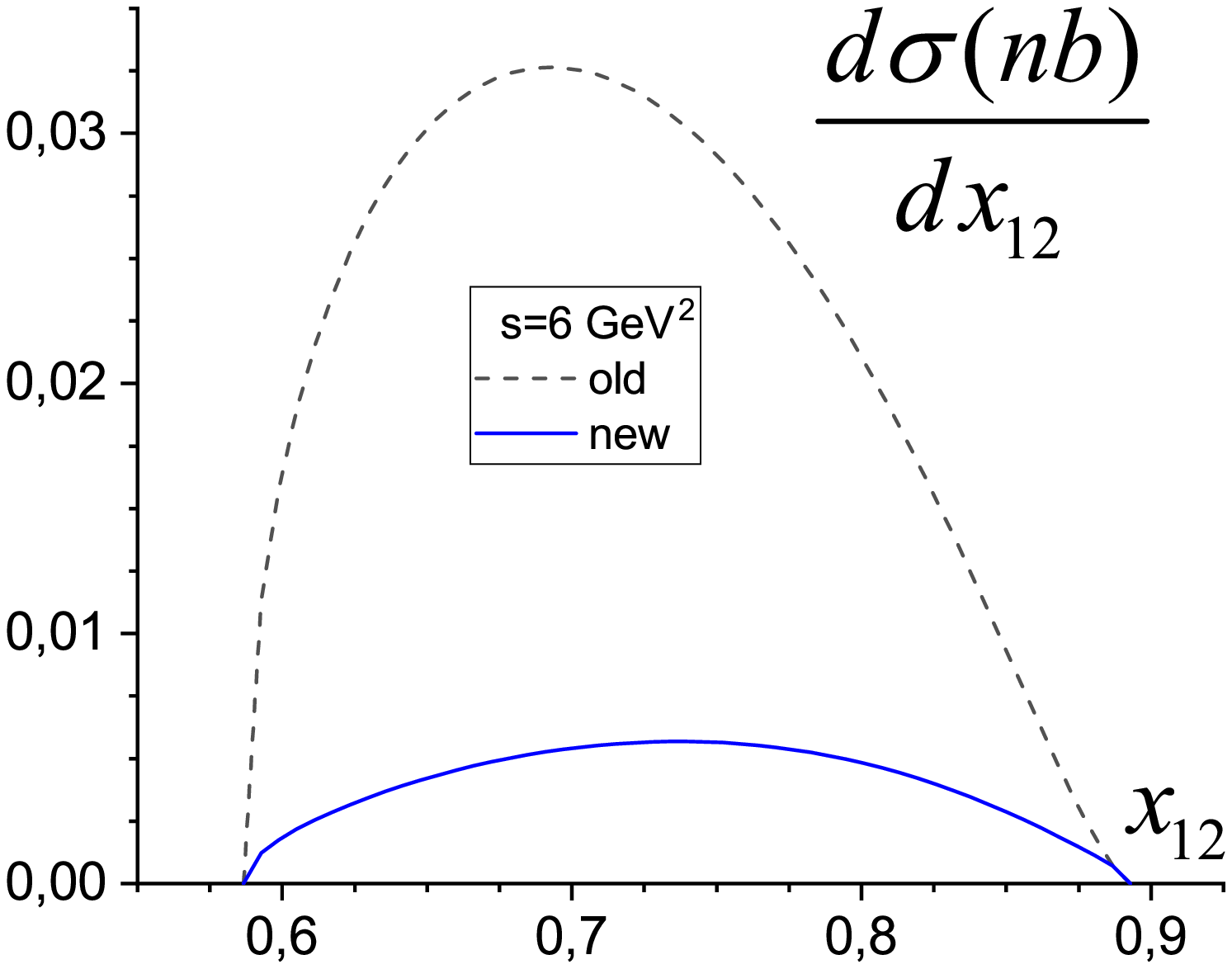}
\hspace{0.1cm}
\includegraphics[width=0.22\textwidth]{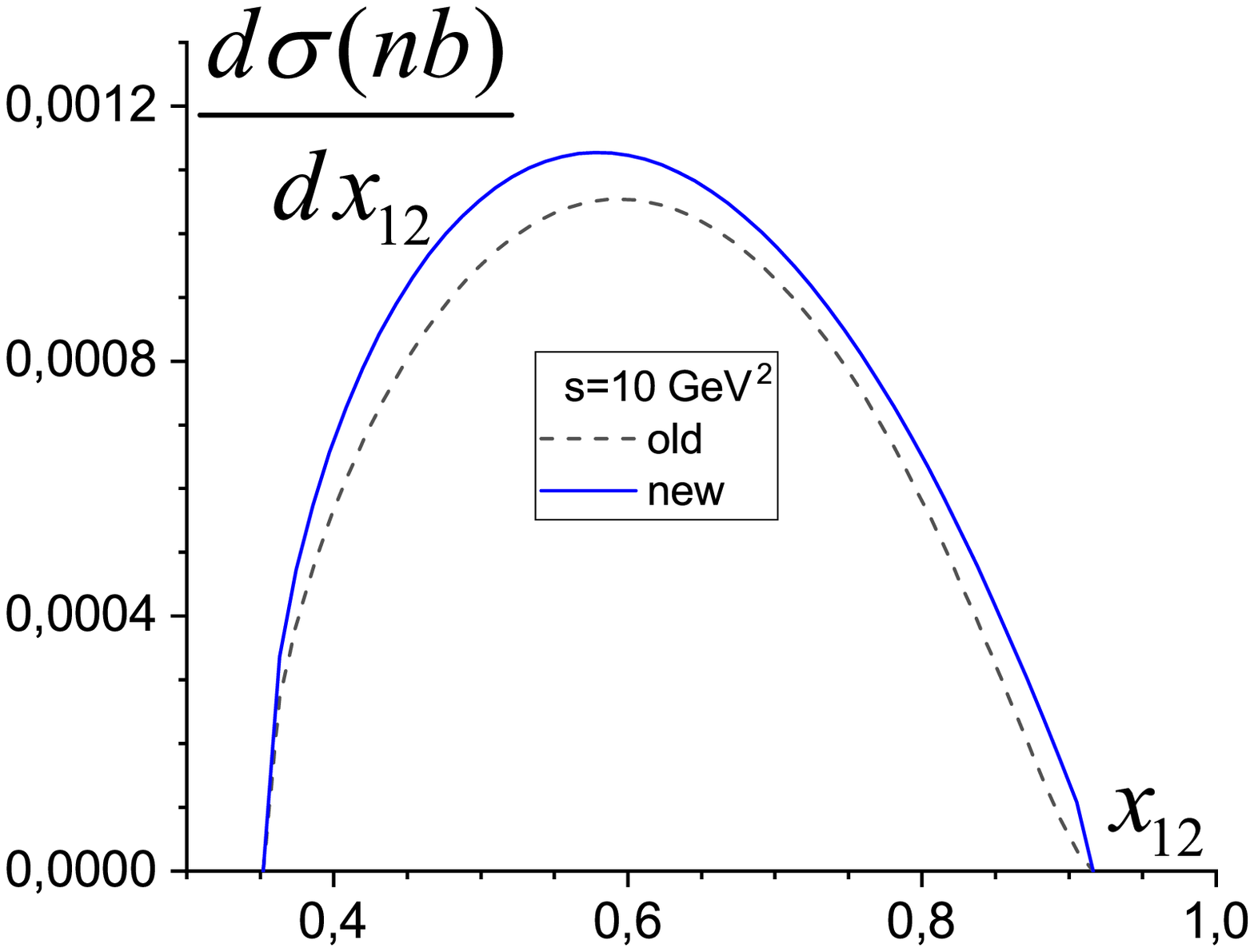}
\hspace{0.5cm}
\includegraphics[width=0.22\textwidth]{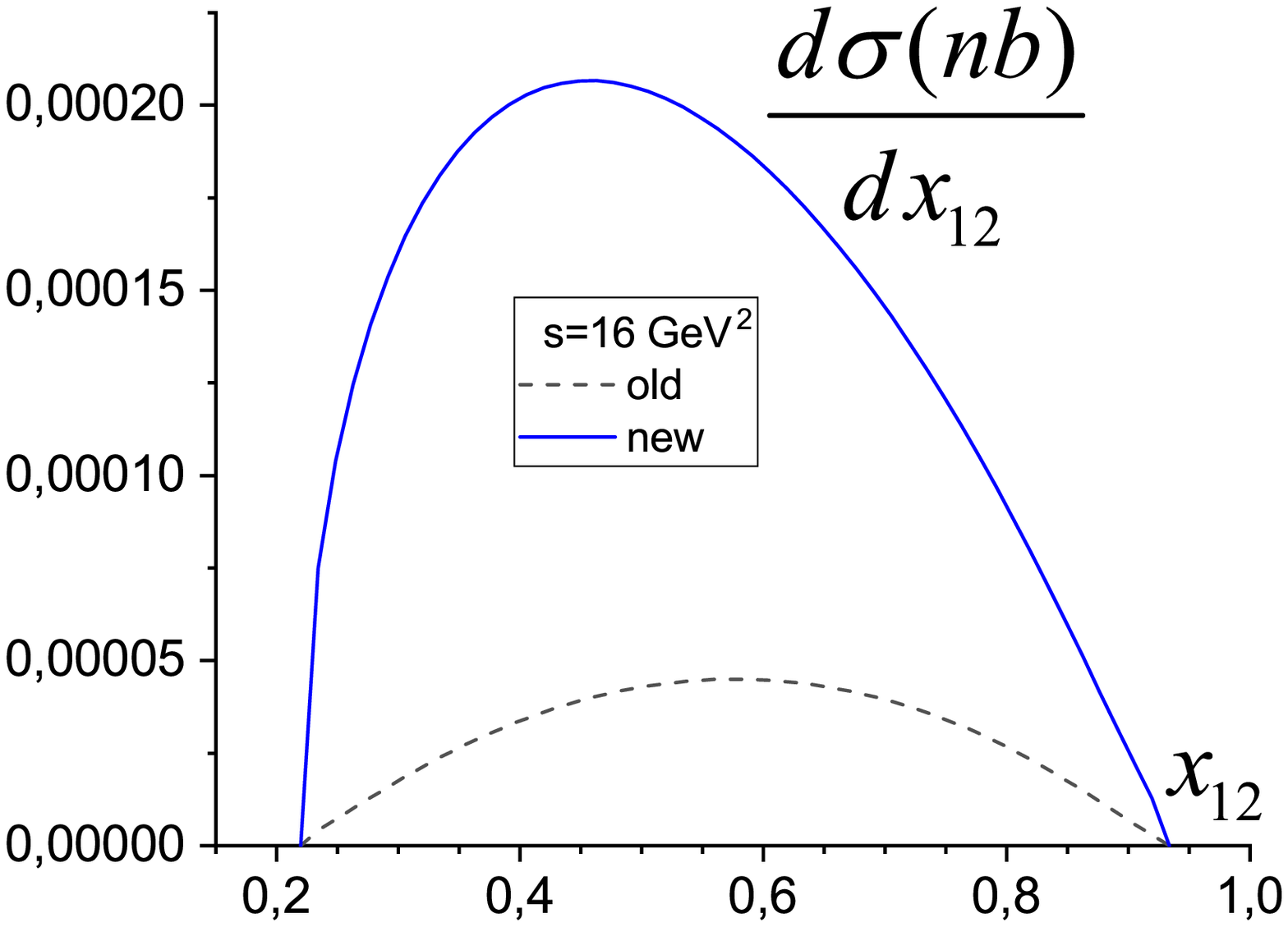}
\parbox[t]{0.9\textwidth}{\caption{Differential distributions for the $\pi^0 p \bar p$-channel 
over the dimensionless invariant variables : $x_1$ (upper row) $x_3$ (middle row) and over $x_{12}$ (lower row) and for different values of $s$:  $s=5$ GeV$^2$ (first column), $s=6$ GeV$^2$ (second column), $s=10$ GeV$^2$ (third column) and
 $s=16$ GeV$^2$  (fourth column). }\label{fig.7}}
\end{figure}
\begin{figure}
\centering
\includegraphics[width=0.22\textwidth]{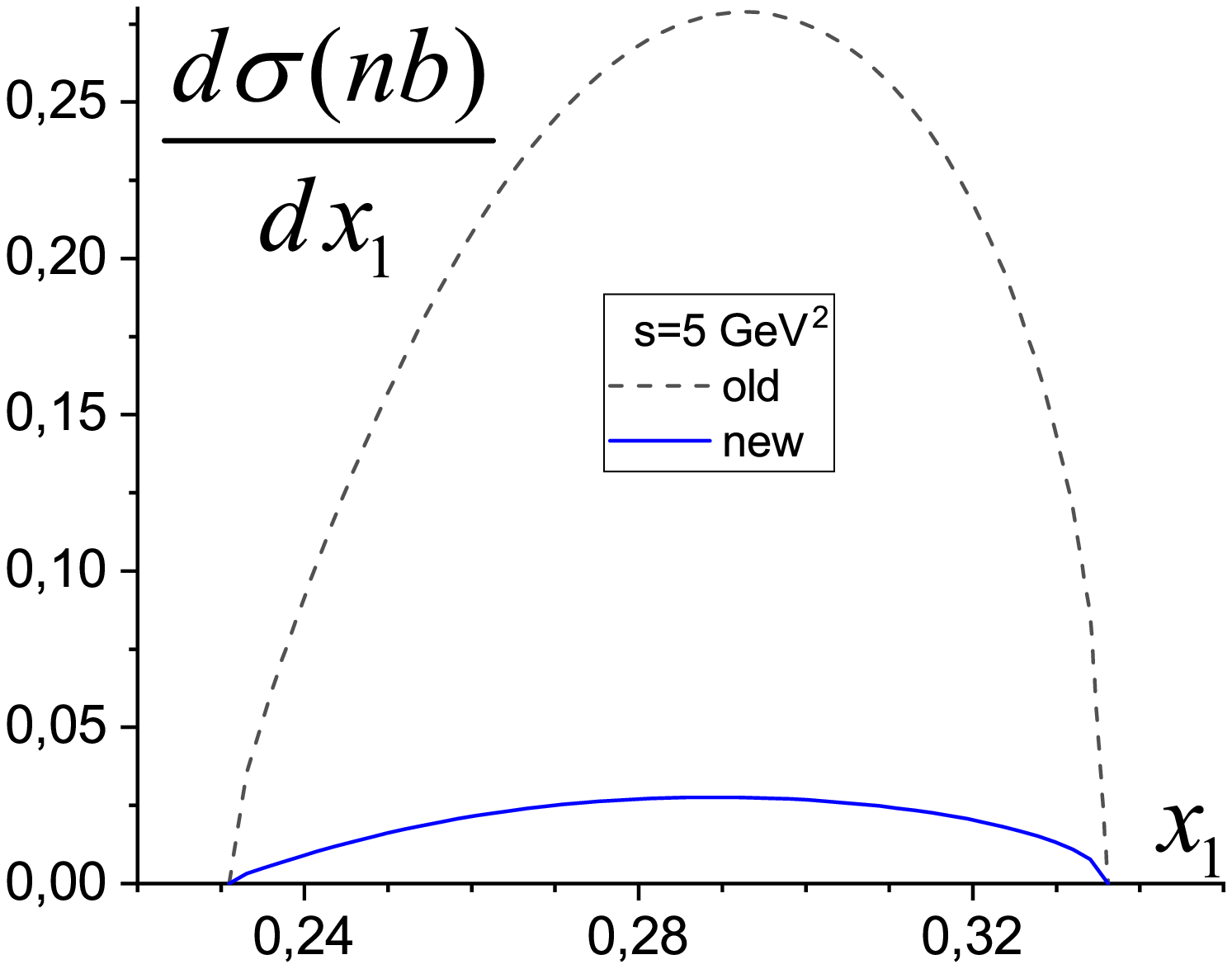}
\hspace{0.1cm}
\includegraphics[width=0.22\textwidth]{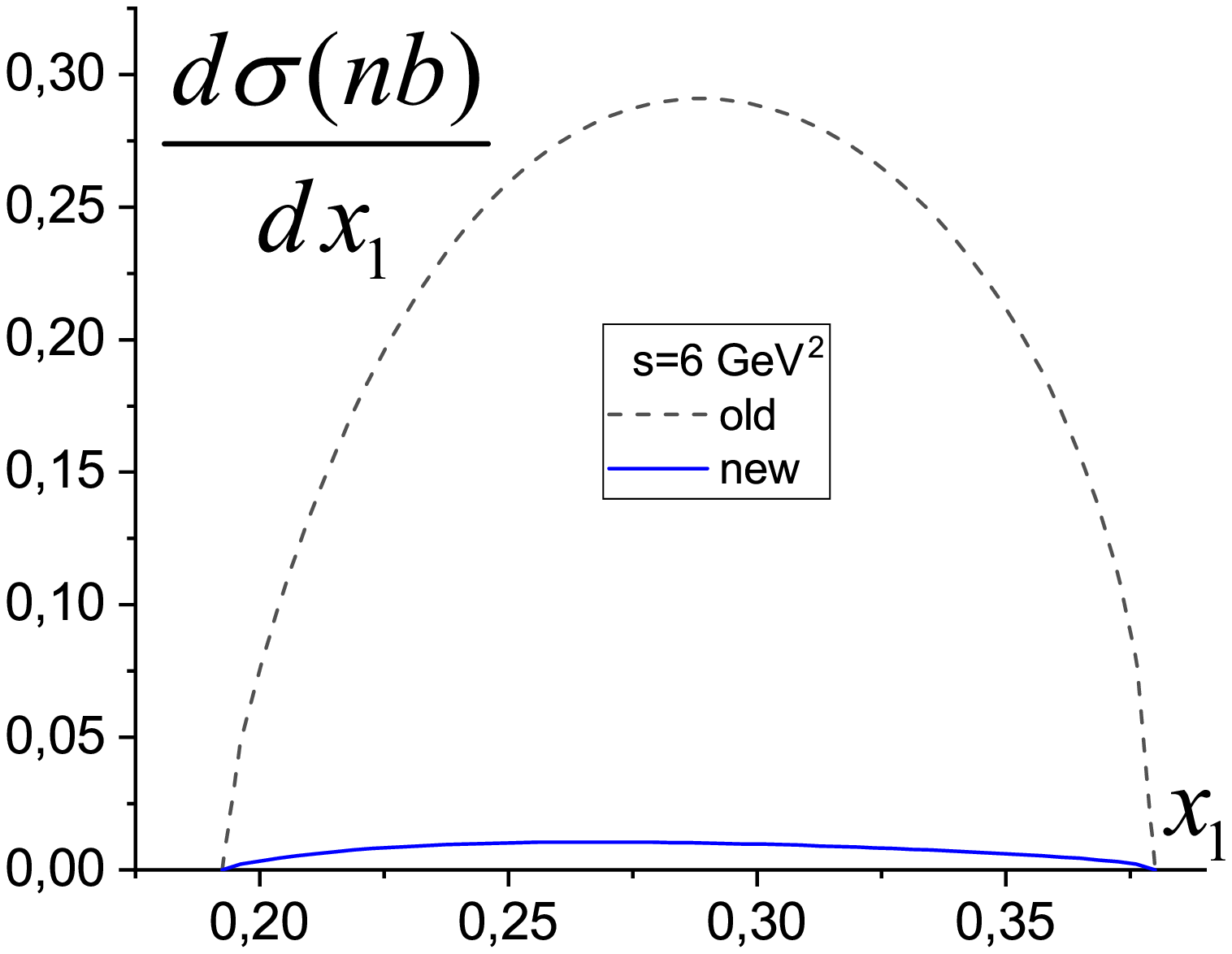}
\hspace{0.1cm}
\includegraphics[width=0.22\textwidth]{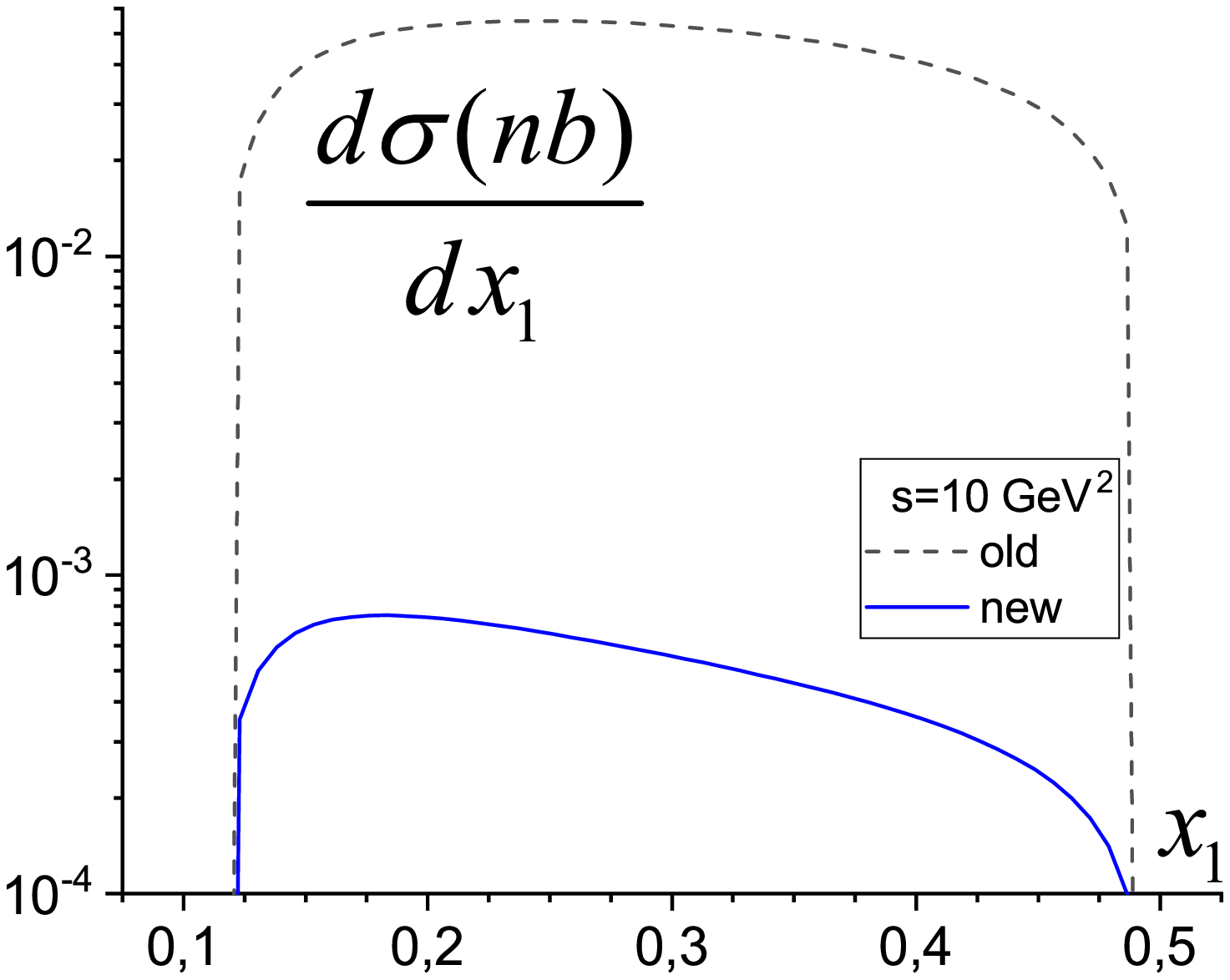}
\hspace{0.5cm}
\includegraphics[width=0.22\textwidth]{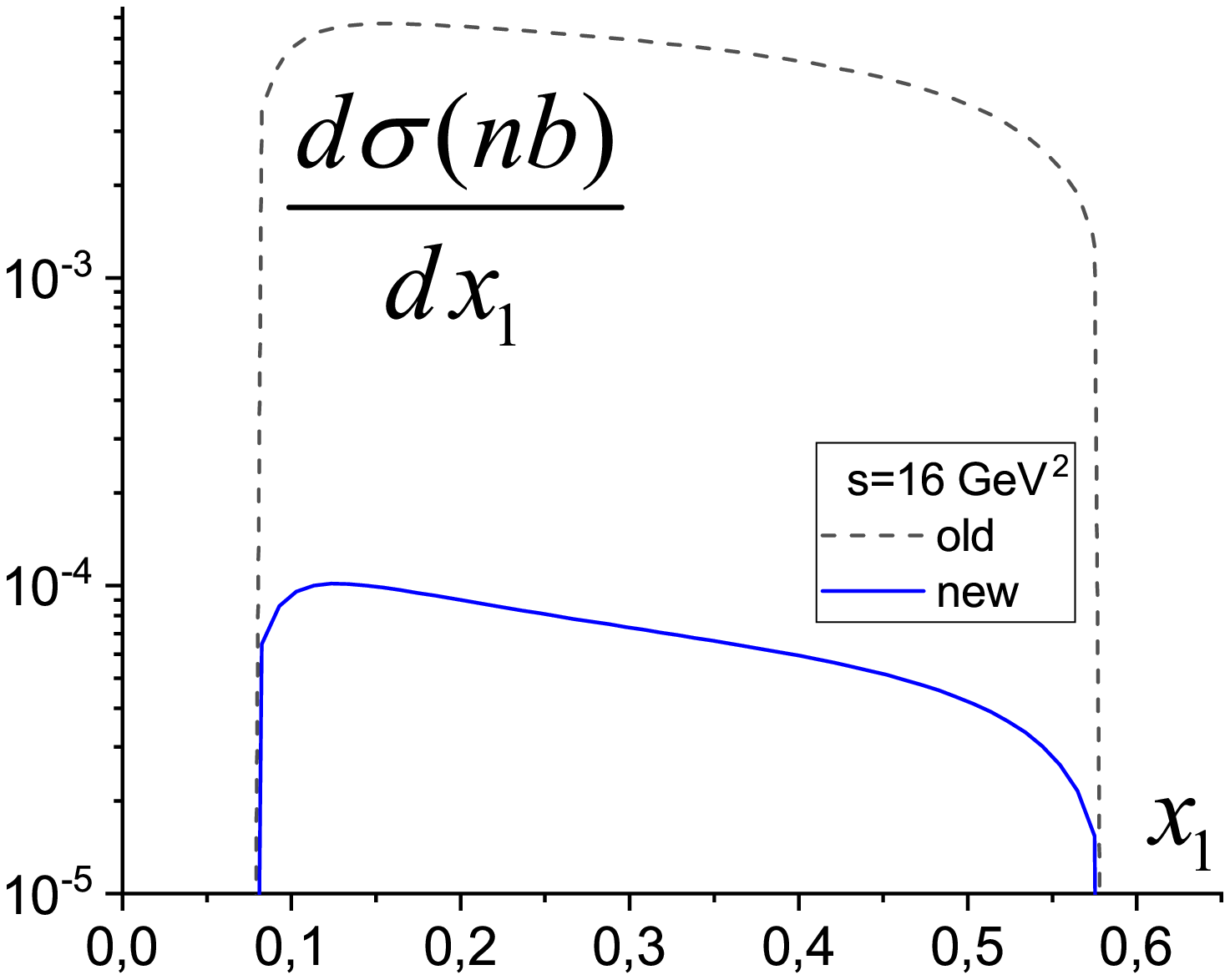}

\vspace{0.31cm}
\includegraphics[width=0.22\textwidth]{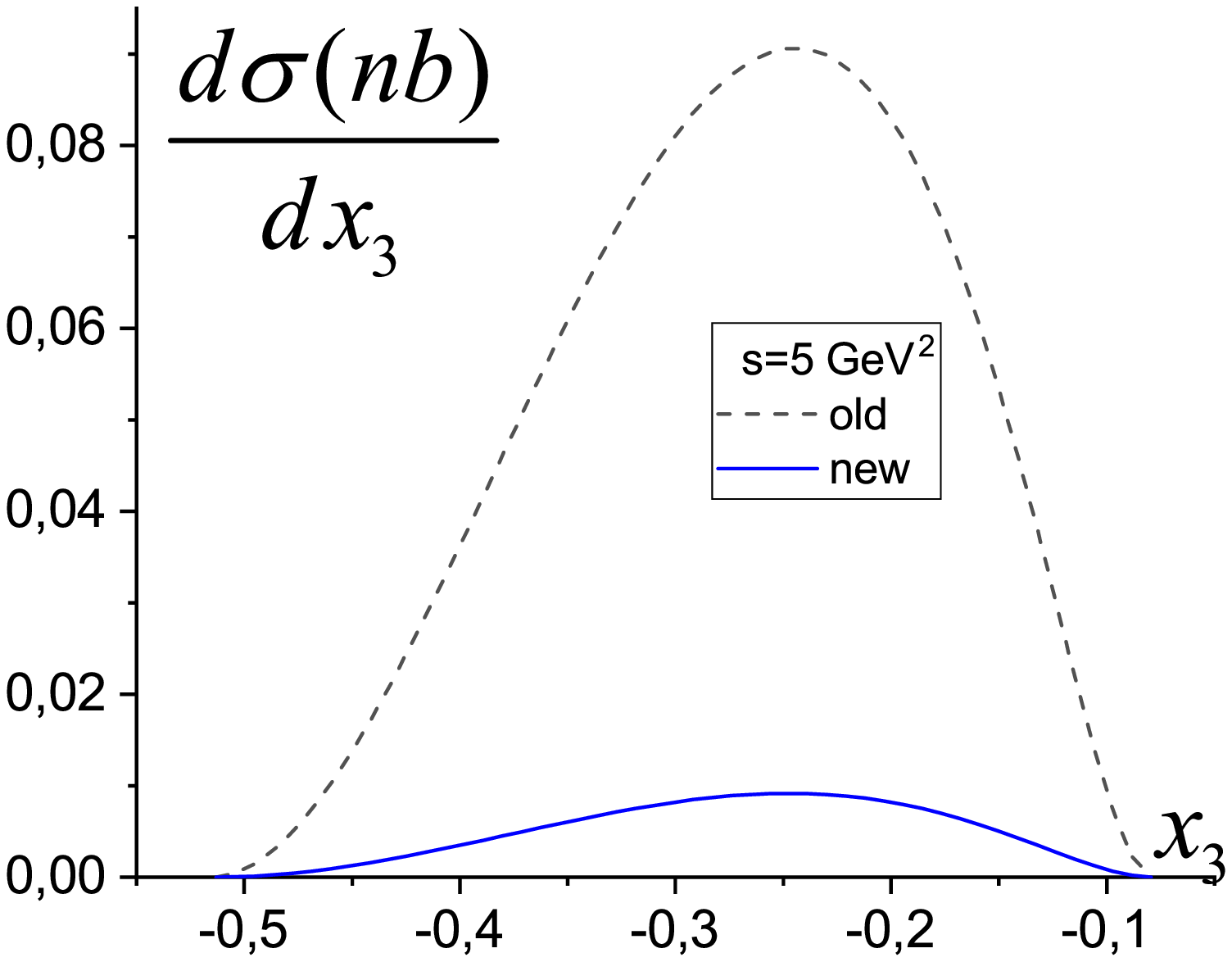}
\hspace{0.1cm}
\includegraphics[width=0.22\textwidth]{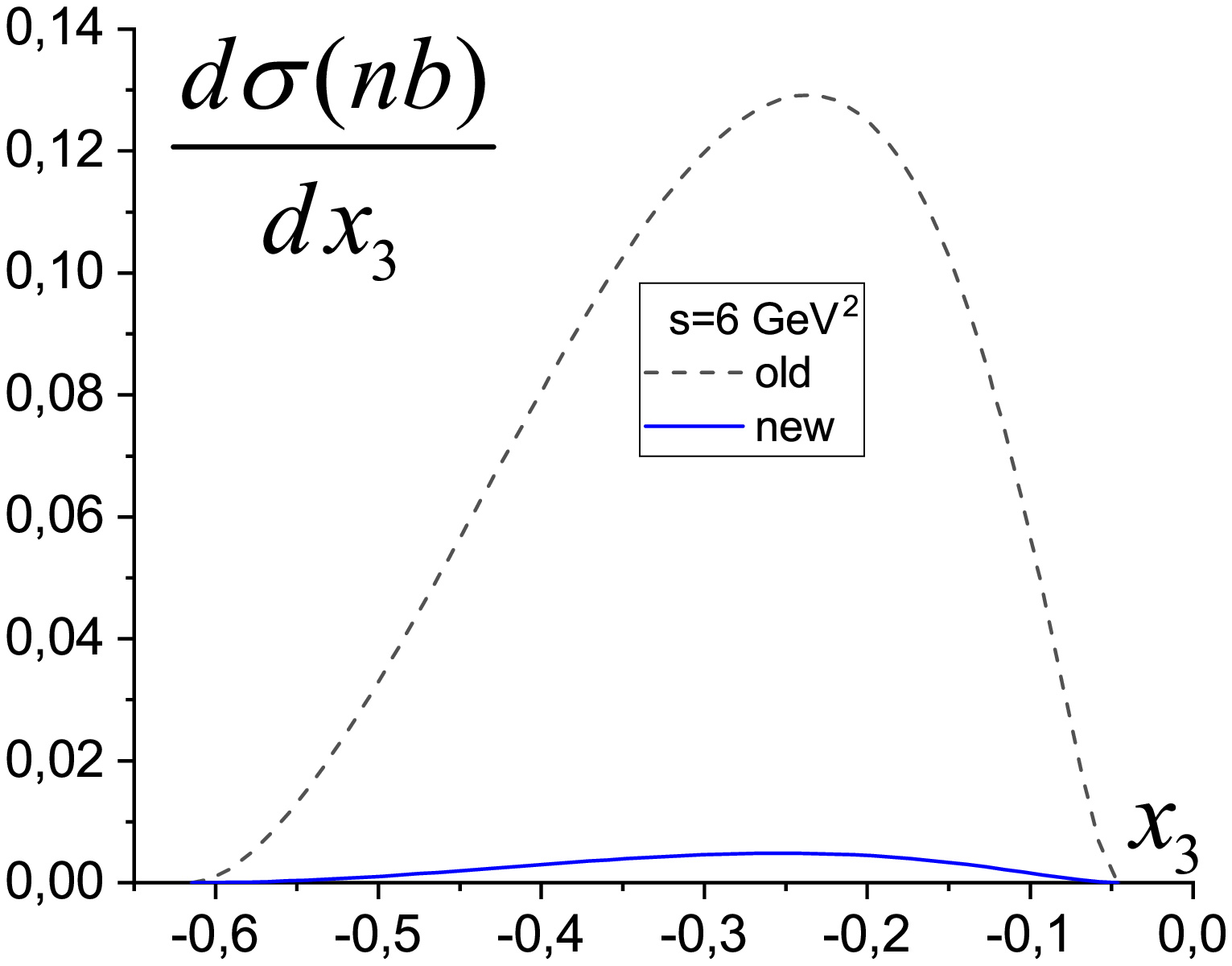}
\hspace{0.1cm}
\includegraphics[width=0.22\textwidth]{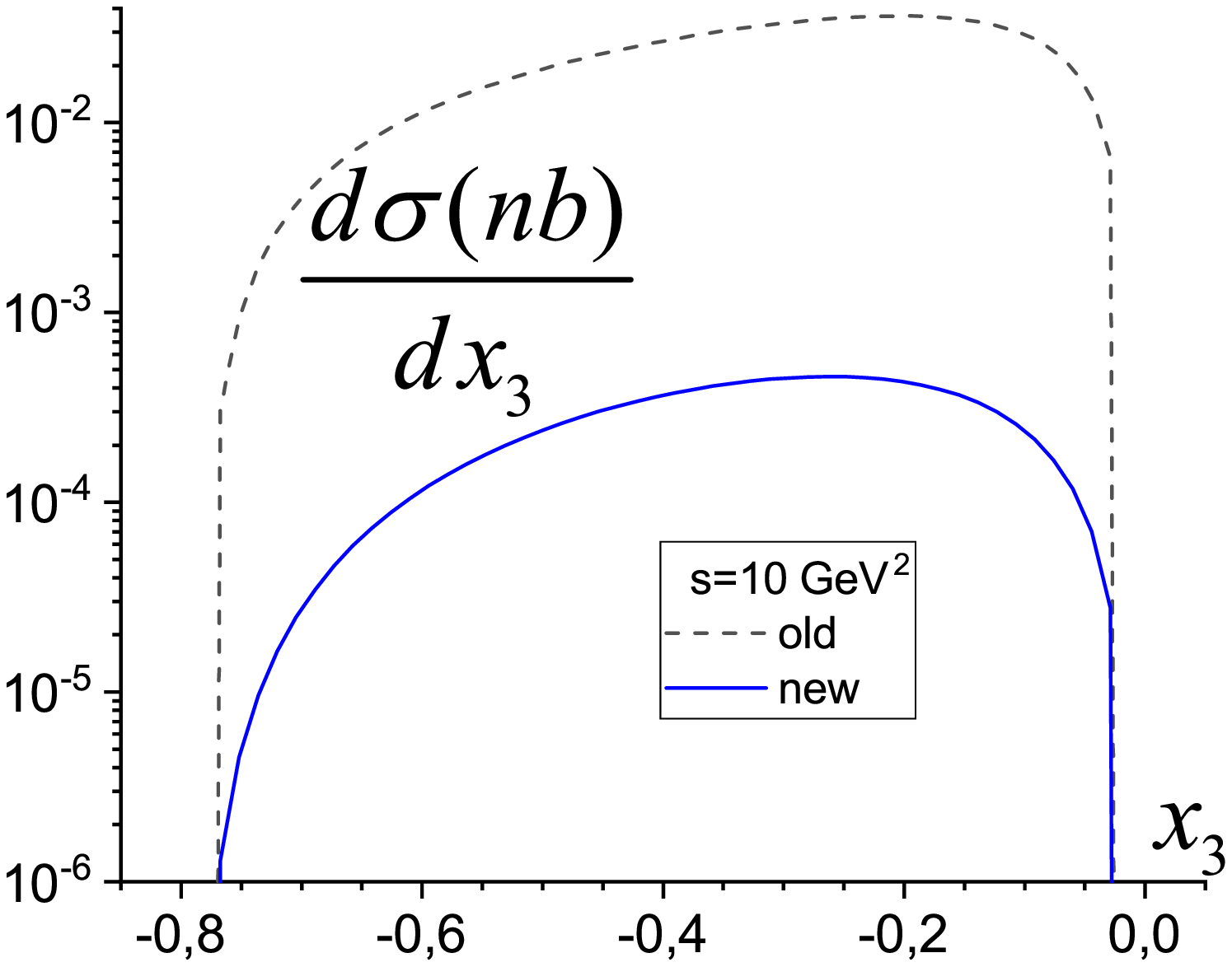}
\hspace{0.5cm}
\includegraphics[width=0.22\textwidth]{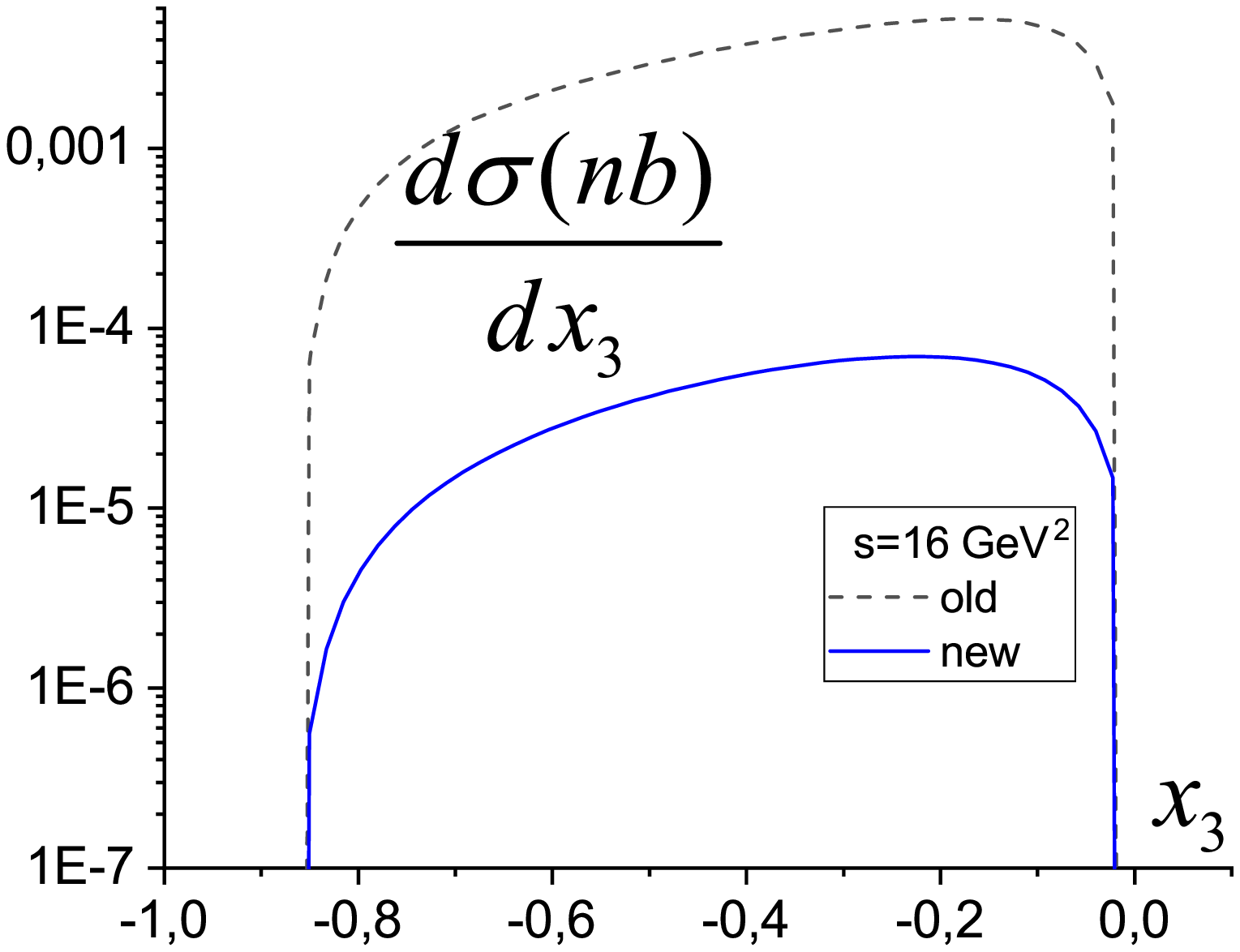}

\vspace{0.31cm}
\includegraphics[width=0.22\textwidth]{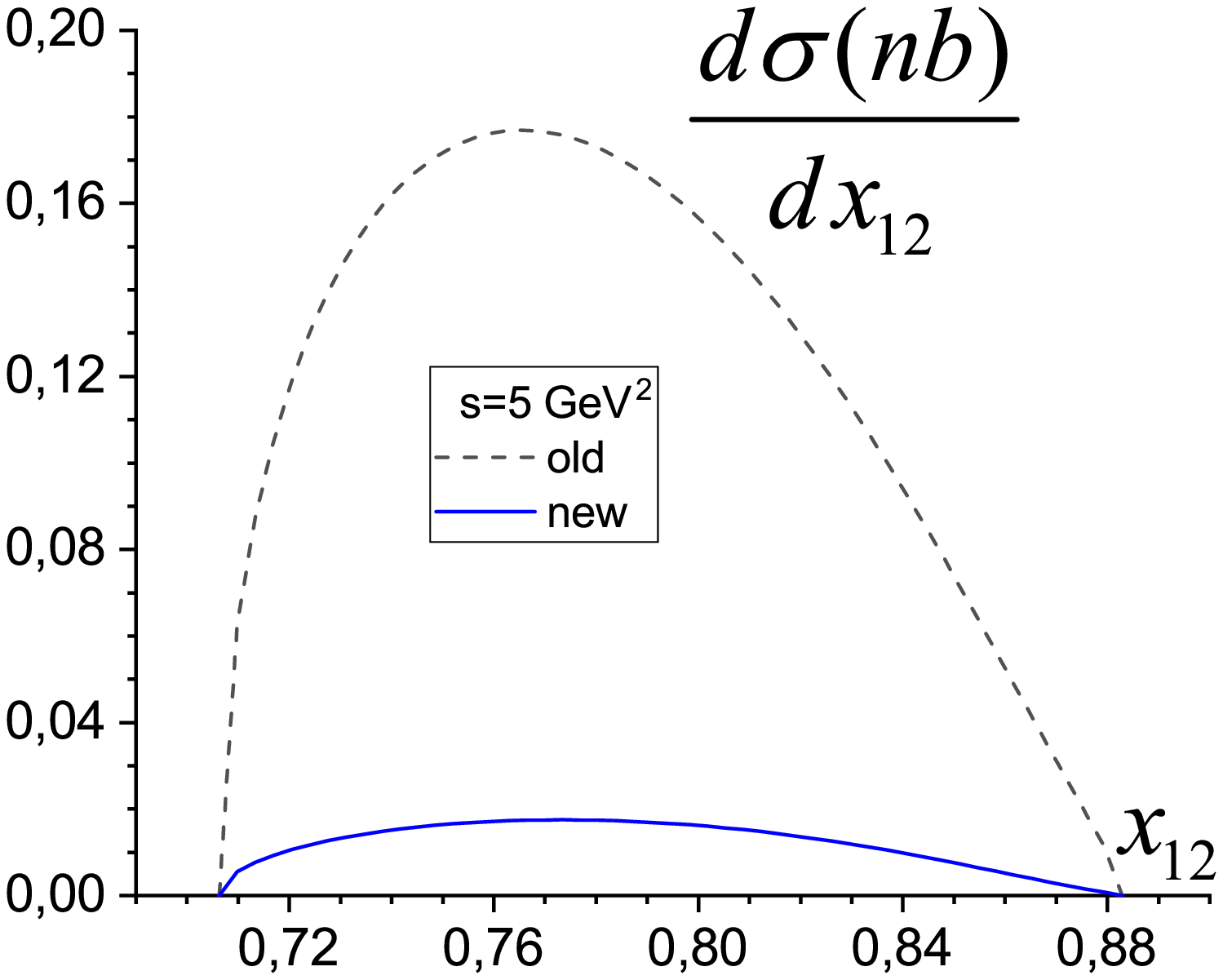}
\hspace{0.1cm}
\includegraphics[width=0.22\textwidth]{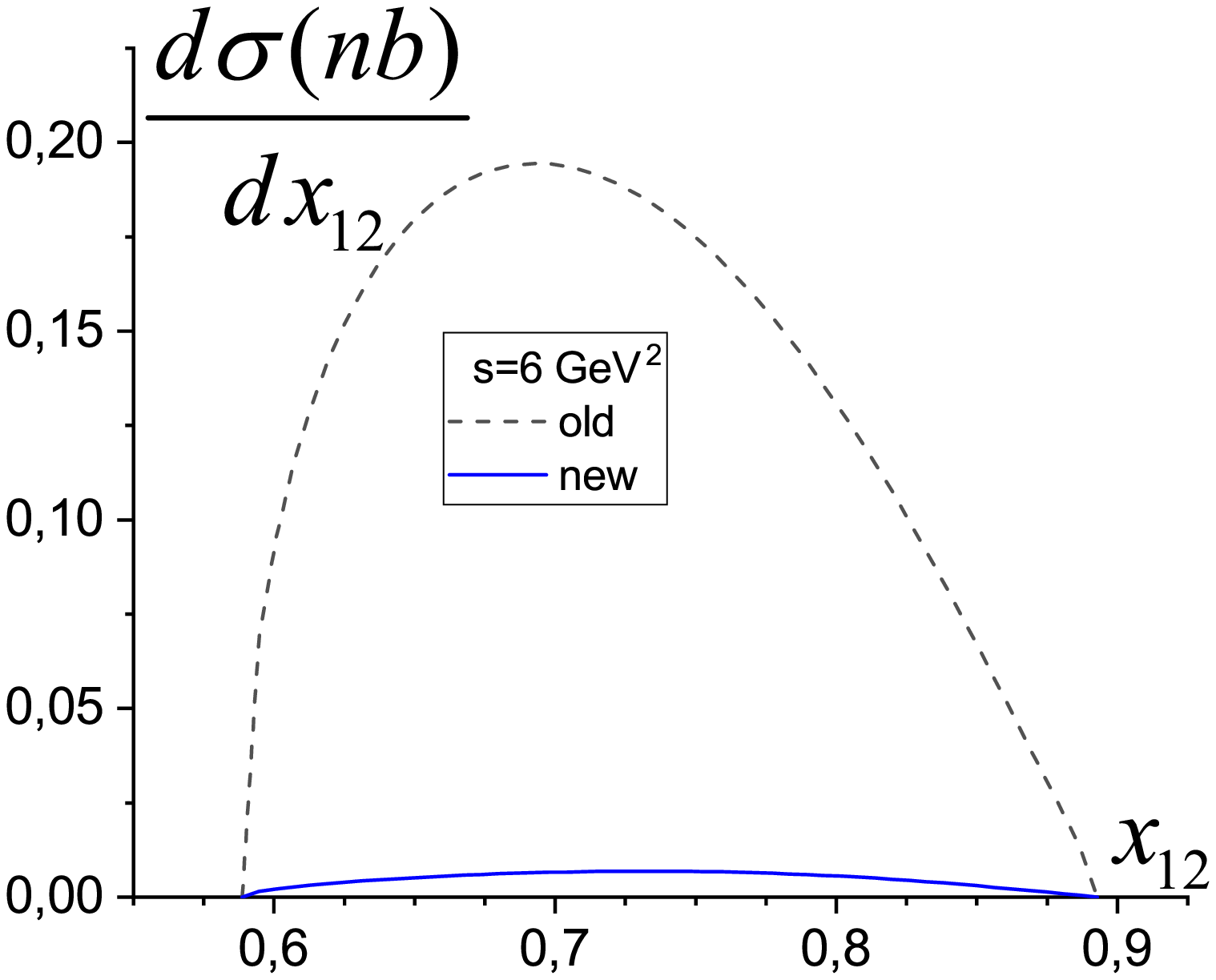}
\hspace{0.1cm}
\includegraphics[width=0.22\textwidth]{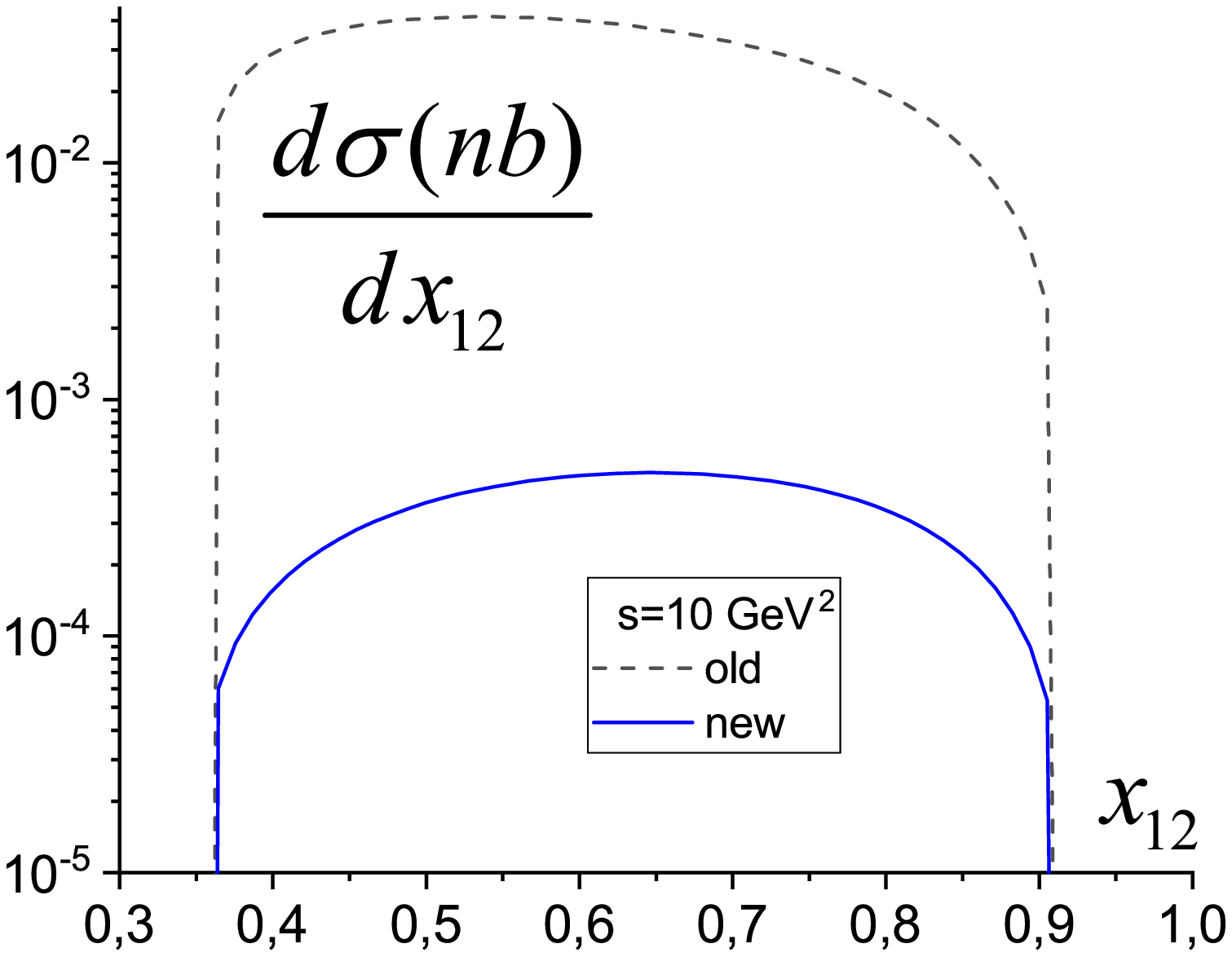}
\hspace{0.5cm}
\includegraphics[width=0.22\textwidth]{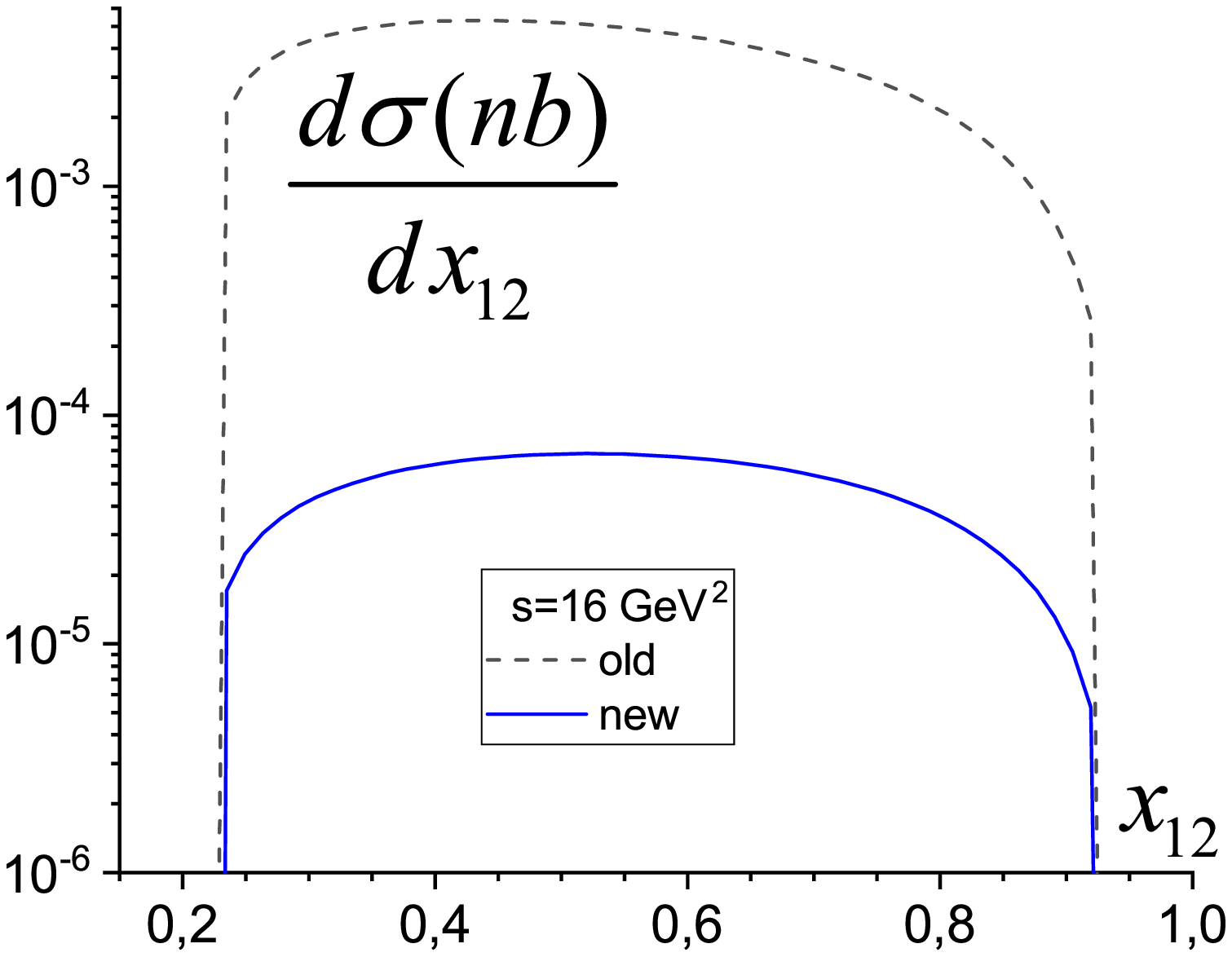}
\parbox[t]{0.9\textwidth}{\caption{The same as in Fig.~7 but for the  $\pi^0 n \bar n$-channel.}\label{fig.8}}
\end{figure}

The total cross sections for the $\pi^0 p \bar p$ and $\pi^0 n \bar n$ channels are shown in Fig.~9.

\begin{figure}
\centering
\includegraphics[width=0.4\textwidth]{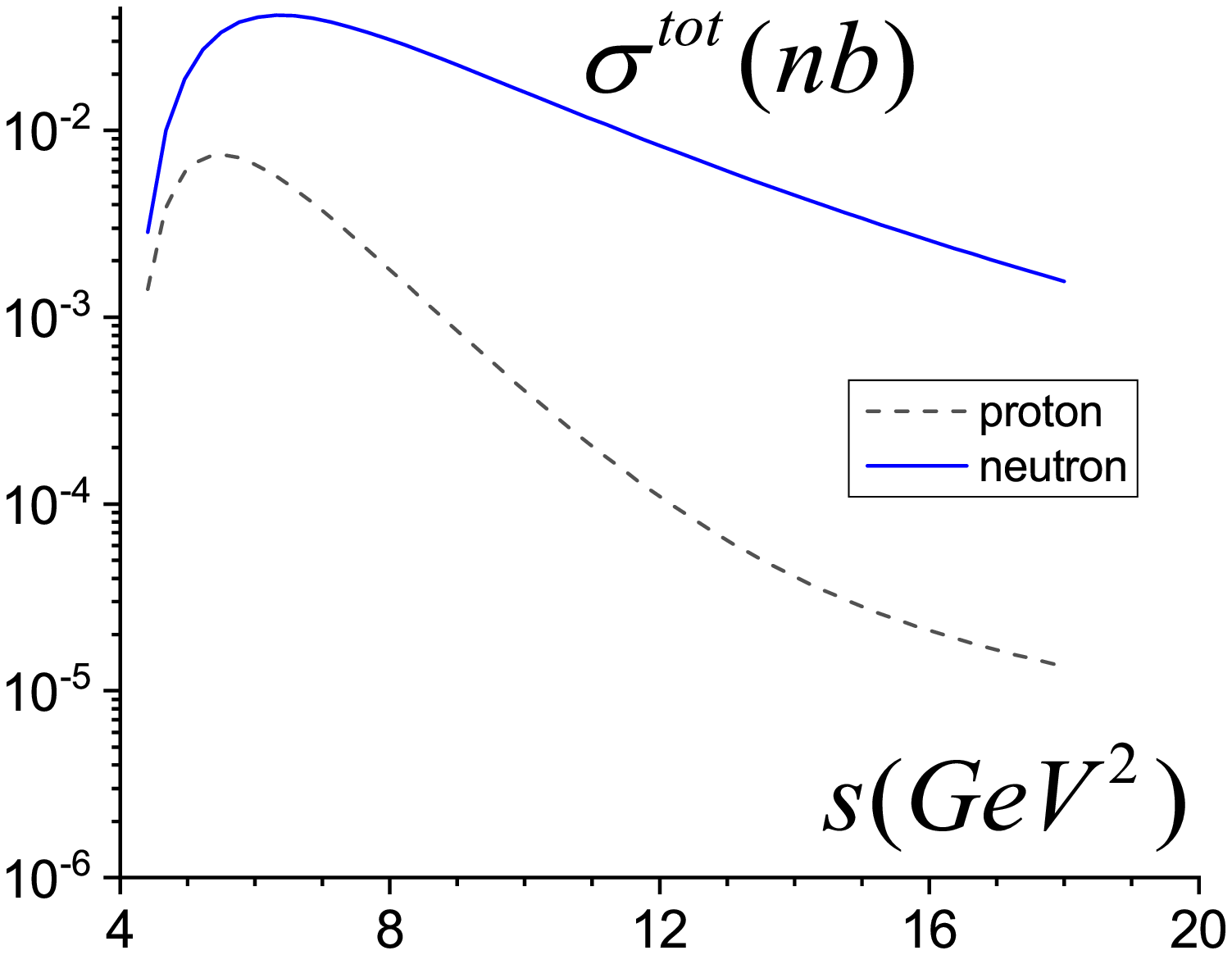}
\hspace{0.1cm}
\includegraphics[width=0.4\textwidth]{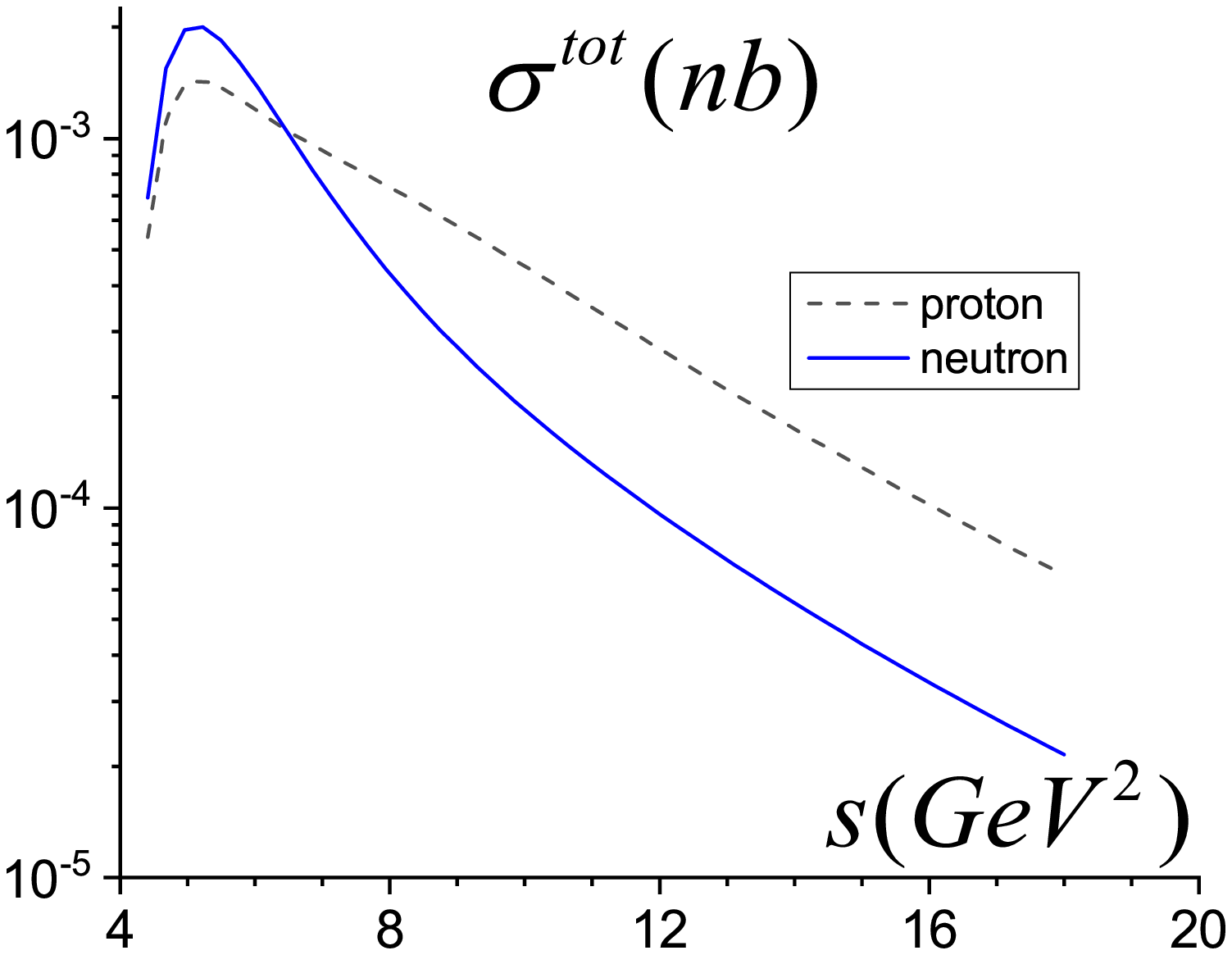}
\parbox[t]{0.9\textwidth}{\caption{The total cross sections for the $\pi^0 p \bar p$ and $\pi^0 n \bar n$ channels: left panel corresponds to the "old" version
and the right panel $-$ to the "new" one. }\label{fig.9}}
\end{figure}
%%%%%%%%%%%%%
\section{Discussion}
%%%%%%%%%%%%%%%%%
About 20 years ago, the BES Collaboration initiated a systematic study of baryon resonances \cite{Zou:2000nu,Li:1999uwc} at Beijing Electron-Positron Collider (BEPC). The major experimental results  obtained on N$^*$ from $e^+e^-$ annihilations and some of their interesting phenomenological implications are reviewed in \cite{Zou:2018lse}. The reaction $e^+ + e^- \to p + \bar p +\pi^0 $ was recently  measured with BESIII detector at the BEPCII collider. In the experiment \cite{Ablikim:2014kxa}, this reaction has been studied in the vicinity of the $\psi(3770)$ resonance. The Born cross section of $e^+ + e^- \to \psi(3770) \to p + \bar p +\pi^0 $ has been extracted allowing the continuum production amplitude to interfere with the resonance production amplitude. Later, the measurement of this reaction was performed at higher energies \cite{Ablikim:2017gtb}, namely at 13 center of mass energies, $\sqrt{s}$, from 4.008 to 4.600 GeV (in the vicinity of the $Y(4260)$ resonance).

The upper limit on the Born cross section of the reaction $e^+e^- \to R\to p\bar p \pi^0$, where $R$ is the $\psi(3770)$ or $Y(4260)$ resonance,
is determined by a least squares fit of
$$\sigma(s)=|\sqrt{\sigma_{con}}+\sqrt{\sigma_R}\frac{m\Gamma}{s-m^2+im\Gamma}exp(i\phi)|^2, $$
where $\sigma_{con}$ and $\sigma_R$ represent the continuum cross section and resonant cross section,
respectively, and $\sigma_{con}$ can be described by a function of s, $\sigma_{con}=C/s^{\lambda}$, where
the exponent $\lambda$ is a priori unknown. The parameter $\phi$ describes the phase between resonant and
continuum production amplitudes. The values of C,$ \lambda, \sigma_R , $ and the interference phase $\phi$
are free parameters of the fit. So, the precision of the determination of the resonance parameters depends on the knowledge
of the continuum cross section.

The total, single and double differential distributions are calculated  for the reactions $e^+e^- \to p\bar p\pi^0 $ and
$e^+e^- \to n\bar n\pi^0 $ using the non-resonant (continuum) contribution which is described by the diagrams given in Fig. 1.

The analytical expressions are calculated for the double differential distributions over ($s_1, s_2$) and ($s_1, s_{12}$)
variables. The integration of these distributions over the corresponding variables allows to obtain analytical expressions
for the single differential distributions over the invariant mass squared of the nucleon-pion and nucleon-antinucleon system.

The numerical estimation of the various differential distributions requires the knowledge of the electromagnetic nucleon
form factors in the time-like region. We use two different parameterizations of two-component model based on the vector
dominance (VDM) at low and intermediate energies and predictions of the perturbative QCD at the large ones. Some features of these
parameterizations are shortly considered in Sec. IV.

The double differential distributions for the $e^+e^- \to p\bar p\pi^0 $ and $e^+e^- \to n\bar n\pi^0 $ channels at different values
of variable $s$ (from 5 to 16 GeV$^2$) are given in Figs. 5,6. At chosen parametrization of the form factors the differential distribution
of the $n\bar n\pi^0$-channel is systematically larger than the $p\bar p\pi^0$-channel. It is hard to say to what extent this
feature depends on the choice of the parameterizations of the nucleon form factors. The BESIII
experiment has collected data samples between $\sqrt{s}$=2 GeV and 3.08 GeV to study baryon cross sections and form factors \cite{Mezzadri:2021yce}.
This lead to the world’s most precise measurement of the $e^+e^- \to n\bar n$ cross section. It is interesting to note that the ratio
R=$\sigma(e^+e^- \to n\bar n)/\sigma(e^+e^- \to p\bar p)$ seems to change at 2.4 GeV. Above this value, the ratio becomes closer to R$\sim$1, that is the expected results predicted by perturbative QCD \cite{Pacetti:2015iqa}. Below this value,  the ratio is flat and smaller, R$\sim$0.25, increasing again at threshold  \cite{CMD-3:2018kql}.

The single differential distribution over the variables $x_1, x_3$ and $x_{12}$ for the $p\bar p\pi^0$ ($n\bar n\pi^0$) channel are
given in Fig. 7(8) for various values of the variable $s$. One can see that for these differential distributions 
strongly differ  for the old and new parameterizations of the nucleon form factors. But at $s$=10 GeV$^2$ this difference is
small for the $p\bar p\pi^0$-channel, as an effect of  these  specific parameterizations.

The total cross sections for the $e^+e^- \to p\bar p\pi^0 $ and $e^+e^- \to n\bar n\pi^0 $ reactions as a functions of the variable $s$
are given in Fig. 9. One can see that the behaviour of the cross sections as a function of  $s$  depends strongly on the nucleon form factor parametrization. For the case of the 'old' parameterization, the cross section of the $e^+e^- \to n\bar n\pi^0 $
reaction is appreciably larger than the cross section of the $e^+e^- \to p\bar p\pi^0 $ reaction. The last cross section, in this
case, decreases more rapidly than the $e^+e^- \to n\bar n\pi^0 $ one. For the case of the 'new' parameterization of the nucleon form
factors, the cross section of the $e^+e^- \to n\bar n\pi^0 $ reaction is smaller than the cross section of the $e^+e^- \to p\bar p\pi^0 $ one, both decreasing rapidly when the variable $s$ increases.

The ongoing physics programme at BESIII is described in the review \cite{Yuan:2019zfo}. One of the goals of this programme is the experimental
study of  hadron spectroscopy, namely to map out all the resonances and determine their properties. This requires a good knowledge of the corresponding background. The non-resonant (continuum) contribution to the $e^+e^- \to N\bar N\pi $
reaction constitutes the  background for the resonances decaying mostly to $N\pi$ state.
%%%%%%%%%%%%%%%%
\section{Conclusion}
%%%%%%%%%%%%%%%%%%
The general analysis of the differential cross section and various polarization observables is performed
for the process $e^+ + e^- \to N + \bar N +\pi^0 $ in the one-photon-annihilation approximation. This analysis is useful
for the description of the continuum (non-resonant) and resonant (with different possible vector mesons or excited baryons
in the intermediate virtual states of the Feynman diagrams) contributions. A number of double differential distributions
is calculated analytically and numerical estimates are given for the $p\bar p\pi^0$ and $n\bar n\pi^0$ channels in the
Born (non-resonant) approximation.

The general structure of the matrix element of the reaction (\ref{eq:1}) has been determined in terms of the 6 independent
invariant amplitudes. The expression of the hadronic tensor is given for the case of the unpolarized final particles or polarized nucleon.
The formalism  is very general, as it is based on fundamental symmetries of the strong and electromagnetic interaction as Parity and Time invariance, and holds for different models of the nucleon structure.

The kinematics of this process is investigated in details. We  introduced useful invariant variables and illustrated the physical kinematical  range. The allowed double invariant variables regions are illustrated for  $s$ = 10 GeV$^2$ in Fig. \ref{fig.2}.

The non-resonant (continuum) contribution to the reaction $e^+ + e^- \to N + \bar N +\pi^0 $ has been
calculated. This contribution is described by two diagrams in\ref{fig.1}, where the pion is emitted by the nucleon or the  antinucleon.

The numerical  results  depend on the choice of the nucleon electromagnetic form factors in the
time-like region. We use two different parameterizations of two-component model based on the vector dominance (VDM) at
low and intermediate energies and predictions of the perturbative QCD at the large ones. The predictions differ, depending on the kinematical region. 

The present calculation can be generalized to other 'inelastic' annihilation processes, with emission of different mesons and can be used to model the background contribution for the experimental study of nucleon resonances.

\section{Appendix A: Invariant Structures }
%\begin{enumerate}{A.1}
\label{section:A.} \setcounter{equation}{0}
\def\theequation{A.\arabic{equation}}

The 13 chosen symmetrical gauge invariant structures are:
\ba
\tilde{S}^{\mu\nu}&=&\tilde{S}^{\mu}(\nu k p q)+\tilde{S}^{\mu}(\nu k p q),\,\,\,(qs)\tilde{k}^{\mu\nu}=(qs)[\tilde{k}^{\mu}(\nu k p q)+\tilde{k}^{\nu}(\mu k p q)]\,,\nn\\
(ks)\tilde{k}^{\mu\nu}&=&(ks)[\tilde{k}^{\mu}(\nu k p q)+\tilde{k}^{\nu}(\mu k p q],\,\,\,(qs)\tilde{p}^{\mu\nu}=(qs)[\tilde{p}^{\mu}(\nu k p q)+\tilde{p}^{\nu}(\mu k p q)]\,,\nn\\
(ks)\tilde{p}^{\mu\nu}&=&(ks)[\tilde{k}^{\mu}(\nu k p q)+\tilde{p}^{\nu}(\mu k p q)],\,\,\,\tilde{k}^{\mu\nu}_k=\tilde{k}^{\mu}(\nu k q s)+\tilde{k}^{\nu}(\mu k q s)\,,\nn\\
\tilde{k}^{\mu\nu}_p&=&\tilde{k}^{\mu}(\nu p q s)+\tilde{k}^{\nu}(\mu p q s),\,\,\,\tilde{p}^{\mu\nu}_k=\tilde{p}^{\mu}(\nu k q s)+\tilde{p}^{\nu}(\mu k q s)\,,\nn\\
\tilde{p}^{\mu\nu}_p&=&\tilde{p}^{\mu}(\nu p q s)+\tilde{p}^{\nu}(\mu p q s),\,\,\,\tilde{G}^{\mu\nu}=\tilde{g}^{\mu\nu}(kpqs),\,\,\,\tilde{K}^{\mu\nu}=\tilde{k}^\mu\tilde{k}^\nu(kpqs)\,, \nn\\
\tilde{P}^{\mu\nu}&=&\tilde{p}^\mu\tilde{p}^\nu(kpqs),\,\,\,\widetilde{KP}^{\mu\nu}=[\tilde{k}^\mu\tilde{p}^\nu +\tilde{p}^\mu\tilde{k}^\nu](kpqs)\,. \label{aq:eqstr}
\ea
Our aim is to show that only eight from them are independent and we choose the eight bottom ones. We use the well known relation
\begin{equation}\label{eq:A1g}
g^{\mu\nu}(\alpha \beta \lambda \rho)=g^{\mu\alpha}(\nu \beta \lambda \rho)-g^{\mu\beta}(\nu \alpha \lambda \rho)+g^{\mu\lambda}(\nu \alpha \beta \rho)-g^{\mu\rho}(\nu \alpha \beta \lambda)\,.
\end{equation}
By contraction of  (Eq.{\ref{eq:A1g}}) with $k_{\alpha} p_{\beta} q_{\lambda} s_\rho$ we obtain after symmetrization
\ba
\label{eq:A2gmn}
g^{\mu\nu}(k p q s)&=&\frac{1}{2}\Big\{\big[k^\mu(\nu p q s)+k^\nu(\mu p q s)\big]-\big[p^\mu(\nu k q s)+p^\nu(\mu k q s)\big] +\nn\\
&&+\big[q^\mu(\nu k p s)+q^\nu(\mu k p s)\big] - \big[s^\mu(\nu k p q)+s^\nu(\mu k p q)\big] \Big\}\,.
\ea
The quantity $q^\mu (kpqs)$ is derived using $\big[g^{\mu\nu} q_\nu=q^\mu \big](kpqs).$ 
Such procedure gives
\begin{equation}\label{eq:A3q}
q^{\mu}(kpqs)=(kq)(\mu p q s)-(pq)(\mu k q s) + q^2(\mu k p s) - (qs)(\mu k p q)\,.
\end{equation}

We then multiply (\ref{eq:A3q}) by $q^{\nu}$ and symmetrize. This leads to
\ba
q^{\mu}q^{\nu}(kpqs)&=&\frac{1}{2}\Big\{(kq)\big[q^\mu (\nu p q s)+q^\nu (\mu p q s)\big]-(pq)\big[q^\mu (\nu k q s)+q^\nu (\mu k q s)\big]+
\nn\\
&&+ q^2\big[q^\mu (\nu k p s)+q^\nu (\mu k p s)\big]-(qs)\big[q^\mu (\nu k p q)+q^\nu (\mu k p q)\big]\Big\}\,.\label{eq:A4qq}
\ea

We have all tools to write the structure $\tilde{G}^{\mu\nu},$ namely
\begin{equation}\label{eq:A5G}
\tilde{G}^{\mu\nu}=\frac{1}{2}\big[\tilde{k}^{\mu\nu}_p - \tilde{p}^{\mu\nu}_k - \tilde{S}^{\mu\nu} \big]\,.
\end{equation}
As one can see, the structure $\tilde{S}^{\mu\nu}$ is not independent (it is expressed in terms of the chosen independent structures).

Now we use (Eq.\ref{eq:A5G}) to write the quantities $\tilde{k}^\mu (kpqs)$ and $\tilde{p}^\mu (kpqs)$ bearing in mind that
$$\tilde{g}^{\mu\nu} k_\nu = \tilde{k}^{\mu}, \ \tilde{g}^{\mu\nu} p_\nu = \tilde{p}^{\mu}.$$
We have
\ba\label{eq:A6kt}
\tilde{k}^{\mu}(kpqs)&=&C_{kk}(\mu p q s)-C_{kp}(\mu k q s) - C_{ks}(\mu k p q)\,,\\
C_{kk}&=&(\tilde{k}k)= m^2-\frac{(kq)^2}{q^2}\,, \ C_{kp} = (\tilde{k}p)=(\tilde{p}k) = (kp)-\frac{(kq)(pq)}{q^2}\,, \nn\\
 C_{ks}&=&(\tilde{s}k)=(sk)-\frac{(qs)(kq)}{q^2}\,,\nn
 \ea
and
\ba\label{eq:A7pt}
\tilde{p}^{\mu}(kpqs)&=&C_{kp}(\mu p q s)-C_{pp}(\mu k q s) - C_{ps}(\mu k p q)\,, 
C_{pp}=(\tilde{p}p)=p^2-\frac{(pq)^2}{q^2}\,, \\
\ C_{ps} &=& (\tilde{s}p)=(ps)-\frac{(qs)(pq)}{q^2}\,, \ (ps) =(ks)-(qs)\,.\nn
\ea
Now, we are ready to write the structures $\tilde{K}^{\mu\nu}$ and $\tilde{P}^{\mu\nu}.$ 
The multiplication of (Eq.\ref{eq:A6kt}) by $\tilde{k}^\nu$ and the symmetrization gives
\begin{equation}\label{eq:A8KK}
\tilde{K}^{\mu\nu}=\frac{1}{2}\big(C_{kk}~\tilde{k}^{\mu\nu}_p -C_{kp}~\tilde{k}^{\mu\nu}_k-C_{ks}~\tilde{k}^{\mu\nu}\big)\,,
\end{equation}

A fully analogous procedure with the use of  Eq. (\ref{eq:A7pt}) and $\tilde{p}^\nu$ leads to
\begin{equation}\label{eq:A9PP}
\tilde{P}^{\mu\nu}=\frac{1}{2}\big(C_{kp}~\tilde{p}^{\mu\nu}_p -C_{pp}~\tilde{p}^{\mu\nu}_k-C_{ps}~\tilde{p}^{\mu\nu}\big)\,.
\end{equation}

As concern the structure $\widetilde{KP}^{\mu\nu},$ it may be expressed by two different equations and both of them  will be applied.
It easy to show that on the one side
\begin{equation}\label{eq:A10KPP}
\widetilde{KP}^{\mu\nu}=C_{kk}~\tilde{p}^{\mu\nu}_p -C_{kp}~\tilde{p}^{\mu\nu}_k-C_{ks}~\tilde{p}^{\mu\nu}\,,
\end{equation}
and on the other one
\begin{equation}\label{eq:A11KPK}
\widetilde{KP}^{\mu\nu}=C_{kp}~\tilde{k}^{\mu\nu}_p -C_{pp}~\tilde{k}^{\mu\nu}_k-C_{ps}~\tilde{k}^{\mu\nu}\,,
\end{equation}
To exclude the structures $(qs)\tilde{k}^{\mu\nu}$ and $(ks)\tilde{k}^{\mu\nu}$, we use the relations (\ref{eq:A8KK}) and (\ref{eq:A11KPK}) and obtain
\begin{equation}\label{eq:A12qsk}
(qs)\tilde{k}^{\mu\nu}=\frac{q^2}{2(q p_1)}\Big[(C_{kk}-C_{kp})\tilde{k}^{\mu\nu}_p - (C_{kp}-C_{pp})\tilde{k}^{\mu\nu}_k + \widetilde{KP}^{\mu\nu}-2\tilde{K}^{\mu\nu}\Big]\,,
\end{equation}
and
\ba\label{eq:A13ksk}
(ks)\tilde{k}^{\mu\nu}&=&\frac{1}{2(q p_1)}\Big\{\big[C_{kk}(q^2+(pq))-C_{kp}(kq)\big]\tilde{k}^{\mu\nu}_p - \big[C_{kp}(q^2+(pq))-C_{pp}(kq)\big]\tilde{k}^{\mu\nu}_k +\nn\\
&& +\widetilde{KP}^{\mu\nu}(kq)-2\tilde{K}^{\mu\nu}(q^2+(pq))\Big\}\,.
\ea
To exclude the structures $(qs)\tilde{p}^{\mu\nu}$ and $(ks)\tilde{p}^{\mu\nu}$, we use the relations (\ref{eq:A9PP}) and (\ref{eq:A10KPP}) and obtain
\ba 
(qs)\tilde{p}^{\mu\nu}&=&\frac{q^2}{2(q p_1)}\Big[(C_{kk}-C_{kp})\tilde{p}^{\mu\nu}_p - (C_{kp}-C_{pp})\tilde{p}^{\mu\nu}_k - \widetilde{KP}^{\mu\nu}+2\tilde{P}^{\mu\nu}\Big]\,,
\label{eq:A14qsp}\\
(ks)\tilde{p}^{\mu\nu}&=&\frac{1}{2(q p_1)}\Big\{\big[C_{kk}(q^2+(pq))-C_{kp}(kq)\big]\tilde{p}^{\mu\nu}_p - \big[C_{kp}(q^2+(pq))-C_{pp}(kq)\big]\tilde{p}^{\mu\nu}_k -
\nn\\
&&-\widetilde{KP}^{\mu\nu}(q^2+(pq))+2\tilde{P}^{\mu\nu}(kq)\Big\}\,.
\ea
Thus, we demonstrated that the five upper structures in  (\ref{aq:eqstr}) are expressed as a function of the bottom eight ones.

%\end{enumerate}{A.1}

\section{Appendix B: the hadronic tensor}
\label{section:B.} \setcounter{equation}{0}
\def\theequation{B.\arabic{equation}}

The structure functions of the symmetrical spin-dependent part of the hadronic tensor (see Eq.~(\ref{eq:7})) read
\ba
T_{kk}&=&\Big[p\cdot q(A_4-2 M A_2)-q^2(2 M A_5 + A_6) +\nn\\
&&+ \frac{1}{2q\cdot p_1}\big[q^2(p^2+q^2-k\cdot q)+p\cdot q(k\cdot q-p\cdot q)\big]A_3\Big]\,A_{14}^*\nn\\
&& +\frac{p^2}{2\,q\cdot p_1}\big(k\cdot q\,A_3+p\cdot q\,A_4 - q^2A_6\big)\big(p\cdot q\,A_2^* + q^2 A_5^*\big) \,,\label{eq:B1}\\
T_{kp}&=& \Big[q^2 A_6-p\cdot q\,A_4 + \frac{(p\cdot q)^2 + m^2 q^2-k\cdot q(p\cdot q+q^2)}{2 q\cdot p_1}A_3  \Big]A_{14}^*+\nn\\
&&
 +2\,M\big(p\cdot q\,A_3\,A_4^*-q^2A_3\,A_6^*\big) +\nn\\
&& +\frac{(p_1+p_2)^2}{2 q\cdot p_1}\big(k\cdot q\,A_3+p\cdot q\,A_4 - q^2A_6\big)\big(p\cdot q\,A_2^* + q^2\,A_5^*\big)\,, \label{eq:B2}
\\
T_{pk}&=&\Big[k\cdot q\,(2 M A_2 + A_3)-q^2 A_6-\big(k\cdot q - q^2 - \frac{p^2 q^2}{2 q\cdot p_1}\big)A_4\Big]A_{14}^*-\nn\\
&&
- \frac{k\cdot q\,p^2}{2 q\cdot p_1}\big(k\cdot q\,A_3+p\cdot q\,A_4 - q^2\,A_6\big)A_2^*\,,\label{eq:B3}
\\
T_{pp}&=&\Big[q^2 A_6 -k\cdot q\,A_3 - \big[p\cdot q-\frac{(p\cdot q)^2+q^2 m^2-(k\cdot q)^2}{2 q\cdot p_1}\big]A_4  \Big]A_{14}^*+\nn\\
&&
+2 M (q^2 A_6 -k\cdot q\,A_3)\,A_4^* - \frac{k\cdot q\,(p_1+p_2)^2}{2 q\cdot p_1}\big(k\cdot q\,A_3+p\cdot q\,A_4 - q^2\,A_6\big)A_2^*\,,\label{eq:B4}\\
T_K&=&\frac{1}{q\cdot p_1}\Big\{\big[ q^2\,A_6 -p\cdot q\,A_4 - (p\cdot q+q^2)A_3 \big]\big(p\cdot q\,A_2^* + q^2 A_5^*\big) -q^2\,A_3\,A_{14}^*\Big\}\,,\label{eq:B5}\\
T_P&=&\frac{1}{q\cdot p_1}\big\{k\cdot q\,[q^2\,A_6-(k\cdot q-q^2)\,A_4-k\cdot q\,A_3]A_2^* + q^2 A_4 A_{14}^* \big\}\,,\label{eq:B6}
\\
T_{KP}&=&-\frac{1}{2 q\cdot p_1}\big\{\big[q^2(p\cdot q +k\cdot q)\,A_6 - k\cdot q\,(2\,p\cdot q+q^2)\,A_3 -p\cdot q\,(2 k\cdot q-q^2)\,A_4 \big]A_2^*+\nn\\
&&
q^2(A_4-A_3)A_{14}^* +q^2[q^2 A_6 - k\cdot q\,A_3 + (q^2-k\cdot q)\,A_4]A_5^* \big\}\,, \label{eq:B7}\\
T_G&=&2\big( k\cdot q\,A_3 + p\cdot q\,A_4 - q^2 A_6\big)A_{14}^*\,. \label{eq:B8}
\ea
Let us remind that according to Eq.~(\ref{eq:7}) one has to take the imaginary part from these structure functions, therefore in Eqs. (\ref{eq:B1}-\ref{eq:B8}) we can use $A_i\,A_j^* = - A_i^*\,A_j.$

The structure functions of the antisymmetrical spin-dependent part of the hadronic tensor, see Eq.~(\ref{eq:8}),  can be written as follows:
\ba
T_s&=&2 M p\cdot q\big(k\cdot q\,|A_3|^2 + p^2\,|A_4|^2 + q^2\,|A_6|^2 + |A_{14}|^2\big) + \label{eq:B9}\\
&&
+\Big[2 M p\cdot q(k\cdot q-m^2)\,A_2 + [k\cdot q\,(p^2-k\cdot q)+(p\cdot q)^2 + m^2 q^2]A_3 +p\cdot q\,(4 M^2 +p^2) A_4 +\nn\\
&&
+2 M[(k\cdot q)^2 - m^2 q^2]A_5 + [(k\cdot q-q^2)^2 - (p\cdot q)^2 - 4 M^2 q^2 ]A_6 \Big]A_{14}^* +\nn\\
&&+
\Big[[(4 M^2 - p^2)(p\cdot q)^2 + p^2 k\cdot q\,(k\cdot q-q^2)] A_2 -\nn\\ 
&&-2 M\big[p\cdot q\,(k\cdot q+q^2)A_3 + ((p\cdot q)^2+p^2 q^2)A_4\big]+\nn\\
&&+(4 M^2- p^2)(p\cdot q)q^2 A_5 \Big] A_6^* + \Big[-(4 M^2 - p^2)p\cdot q\,k\cdot q\, A_3 + p^2[(k\cdot q)^2-m^2 q^2] A_4\Big] A_5^* +\nn\\
&&
+\Big[p^2\, p\cdot q(k\cdot q-m^2) A_2 + 2 M[p^2\, k\cdot q + (p\cdot q)^2] A_3\Big] A_4^* + p^2 \,k\cdot q\,(k\cdot q-m^2) A_2 A_3^*\,, \nn\\
T_{pps}&=&2 M\big(k\cdot q\,|A_3|^2 + p\cdot q\,|A_4|^2\big) + p\cdot q\,(A_3+A_4) A_{14}^* +\label{eq:B10}
\\
&&
+\big[(k\cdot q-p\cdot q)(k\cdot q-q^2) A_2 - 2 M q^2 (A_3 + A_4)\big] A_6^* + \big[(k\cdot q)^2-m^2 q^2 \big]A_4 A_5^* +\nn\\
&&
+\big[(k\cdot q-m^2)\,p\cdot q\, A_2 +2 M (k\cdot q + p\cdot q)A_3\big] A_4^* +
\nn\\
&&
+(k\cdot q-m^2)\,k\cdot q\, A_2 A_3^* +(k\cdot q- q^2)(k\cdot q A_3-q^2 A_6) A_5^*,\nn\\
T_{pqs} &=& 2 M\big(k\cdot q\,|A_3|^2 + q^2 |A_6|^2 + p\cdot q\, A_3 A_4^*\big)+2\big(q\cdot p_2\, A_6 - k\cdot p_2\, A_3\big) A_{14}^* + \label{eq:B11}\\
&&
+\big[(4 M^2-p^2)(p\cdot q\,A_2 + q^2 A_5) - 2 M[(k\cdot q+q^2) A_3 + p\cdot q\,A_4 ]\big] A_6^* -\nn\\
&& 
-(4 M^2-p^2)\,k\cdot q\,A_3 A_5^* \,, \nn\\
T_{kps} &=& \big[2 M[(k\cdot q-p\cdot q)\,A_2-q^2 A_5] + (q^2-k\cdot q)(A_3+A_4) \big] A_{14}^* +\label{eq:B12}\\
&&+p\cdot q \big[\, (p\cdot q-k\cdot q)\,A_2 + q^2\, A_5\big] A_6^* + (p\cdot q-p^2)(k\cdot q\, A_2 A_3^* + p\cdot q\, A_2 A_4^*) +\nn\\
&&
+\big[-k\cdot q\,p\cdot q\,A_3 + [-k\cdot q\,p\cdot q+q^2(p\cdot q-p^2)]A_4\big] A_5^* \,,
\nn\\
T_{kqs} &=& -2 M |A_{14}|^2 +\big[  2 M[(k\cdot q-q^2)A_5 - p\cdot q\,A_2] +2q\cdot p_2\,(A_3-A_6) -p^2 A_4\big] A_{14}^* +
\label{eq:B13}\\
&&
+p\,^2\big[k\cdot q\, (A_2 A_6^* -A_2 A_3^*) - p\cdot q\,A_2 A_4^*)  +(k\cdot q-q^2) A_4 A_5^*   \big] \,,
\nn
\ea
where we bear in mind the real part of the structure functions  (\ref{eq:B9}-\ref{eq:B13}).

\section{Appendix C: form factor parametrizations}
\label{section:C.} \setcounter{equation}{0}
\def\theequation{C.\arabic{equation}}

Here we report the "old" \cite{Iachello:2004aq} and "new"\cite{Bijker:2004yu} parameterizations of the Dirac and Pauli electromagnetic form factors in the case when the intermediate photon couples with
an intrinsic quark-gluon structure and a meson cloud. The interaction with the intrinsic structure is described by the pQCD form factor $g(Q^2),\, Q^2 =-q^2,$ whereas the interaction with the meson cloud $-$ by the vector dominance $(\rho,\,\omega,\,\varphi).$

Firstly these parameterizations were written for the space-like regions and then rules are formulated for the analytical extension to  the time-like region. The analytic form
of the Dirac form factors $F_1^S$ and $F_1^V$ is the same  for both parameterizations, namely
\begin{equation}\label{eq:C1}
F_1^S(Q^2)= g(Q^2)\Big[1-\beta_{\omega}-\beta_{\varphi}+\beta_{\omega}\frac{m^2_{\omega}}{m^2_{\omega}+Q^2}+\beta_{\varphi}\frac{m^2_{\varphi}}{m^2_{\varphi}+Q^2}\Big],
\end{equation}
\begin{equation}\label{eq:C2}
F_1^V(Q^2)= g(Q^2)\Big[1-\beta_{\rho}+\beta_{\rho}\frac{m^2_{\rho}+8\Gamma_\rho\,m/\pi}{m^2_{\rho}+Q^2+(4m^2+Q^2)\Gamma_\rho\,\alpha(Q^2)/m}\Big],
\ \ g(Q^2)=\frac{1}{(1 +\gamma\,Q^2)^2},
\end{equation}
where
$$\alpha(Q^2)=\frac{2}{\pi}\sqrt{\frac{4m^2+Q^2}{Q^2}}\,\ln\Big(\frac{\sqrt{4m^2+Q^2}+\sqrt{Q^2}}{2\,m}\Big).$$

The values of the fitting parameters for the "old" version are
$$\beta_\rho = 0.672, \ \beta_\omega =1.102, \ \beta_\varphi =0.112, \ \gamma=0.25(GeV)^{-2}, \ \Gamma_\rho = 0.112,GeV$$
and for the "new" version
$$\beta_\rho = 0.512, \ \beta_\omega=1.129, \ \beta_\varphi =-0.263, \ \gamma=0.515(GeV)^{-2}.$$
The values of the vector meson masses are $m_\rho = 0.776\,GeV,\,\,m_\omega=0.783\,GeV,\\
m_\varphi = 1.019\,GeV.$

As concerns the Pauli form factors, they are for the "old" version
\begin{equation}\label{eq:C3}
F_2^S= g(Q^2)\Big[(-0.120-\alpha_\varphi)\frac{m^2_{\omega}}{m^2_{\omega}+Q^2}+\alpha_\varphi\,\frac{m^2_{\varphi}}{m^2_{\varphi}+Q^2}\Big],
\end{equation}
$$F_2^V = 3.706\,g(Q^2)\frac{m^2_{\rho}+8\Gamma_\rho\,m/\pi}{m^2_{\rho}+Q^2+(4m^2+Q^2)\Gamma_\rho\,\alpha(q^2)/m}, \ \ \alpha_\varphi = -0.052,$$
and for the "new" version
\begin{equation}\label{eq:C4}
F_2^S= g(Q^2)\Big[(\mu_p +\mu_n-1-\alpha_\varphi)\frac{m^2_{\omega}}{m^2_{\omega}+Q^2}+\alpha_\varphi\,\frac{m^2_{\varphi}}{m^2_{\varphi}+Q^2}\Big], \
\alpha_\varphi = -0.200,
\end{equation}
$$F_2^V = g(Q^2)\Big[\frac{\mu_p-\mu_n-1-\alpha_\rho}{1+\gamma\,Q^2} +\alpha_\rho\frac{m^2_{\rho}+8\Gamma_\rho\,m/\pi}{m^2_{\rho}+Q^2+(4m^2+Q^2)\Gamma_\rho\,\alpha(Q^2)/m}\Big], \ \alpha_\rho = 2.675,$$
where $\mu_p=2.793,\,\,\mu_n=-1.913$ are the magnetic moments of proton and neutron.

The second step consists of the analytic continuation  to the time-like region of the intrinsic and vector meson contributions to form factors.
Due to the complex nature of the $N\,\bar{N}$ interaction, the intrinsic part can be written as
\begin{equation}\label{eq:C5}
g(q^2) = \frac{1}{(1-\tilde{\gamma}\,q^2)^2}, \ \tilde{\gamma} = \gamma\,e^{i\theta}, \ q^2 =-Q^2,
\end{equation}
where the fitting phase is $\theta = 53^o \,(22.7^o)$ for the "old"\,\,("new") version. The vector meson part is obtained with the replacement
\begin{equation}\label{eq:C6}
Q^2\to -q^2, \ \alpha(Q^2) \to \alpha(q^2) - i\,\pi\beta(q^2)/2,  \  q^2 > 4\,m^2,
\end{equation}
$$\alpha(q^2)=\frac{2}{\pi}\sqrt{\frac{q^2-4m^2}{q^2}}\,\ln\Big(\frac{\sqrt{q^2-4m^2}+\sqrt{q^2}}{2\,m}\Big), \ \beta(q^2)= \sqrt{\frac{q^2-4m^2}{q^2}}.$$

%\bibliography{Biblio} 
\input{Born21UA.bbl}
\end{document}